\documentclass[superscriptaddress,groupedaddress,nofootnoteinbib,11pt]{article}
\pdfoutput=1 

\usepackage{jheppub} 

\usepackage{amsmath, amsfonts, amsthm, amssymb, graphicx, color, hyperref}

\usepackage{float, subfig}
\usepackage{pstool}
\usepackage[utf8]{inputenc}
\usepackage[normalem]{ulem}
\usepackage[english]{babel}
\usepackage{blkarray}
\usepackage{dsserif}
\usepackage{mathtools}
\usepackage{makecell}
\usepackage{multicol}
\usepackage{cleveref}

\allowdisplaybreaks[3]

\def \bal#1\eal  {\begin{align} #1 \end{align}}
\def\({\left(}
\def\){\right)}
\def\[{\left[}
\def\]{\right]}
\def\<{\langle}
\def\>{\rangle}
\def\d{\mathrm{d}}

\newcommand{\eref}[1]{Eq.~(\ref{#1})}
\newcommand{\f}[2]{\frac{#1}{#2}}
\newcommand{\bim} {\begin{itemize}[noitemsep]}

\newcommand{\eim}{\end{itemize}}
\newcommand{\be} {\begin{equation}}
\newcommand{\ee} {\end{equation}}
\newcommand{\bc}{\begin{center}}
\newcommand{\ec}{\end{center}}

\newcommand{\nn} {\nonumber\\}

\newcommand{\marrow}{~~\Longrightarrow~~}

\newcommand{\nd} {\nabla}
\newcommand{\pd} {\partial}
\newcommand{\mc} {\mathcal}

\newcommand{\ai}{{\alpha}}
\newcommand{\bi}{{\beta}}
\newcommand{\gi}{{\gamma}}

\newcommand{\ri}{{\rho}}
\newcommand{\si}{{\sigma}}
\newcommand{\li}{{\lambda}}

\newcommand{\oi}{\omega}
\newcommand{\epi}{\epsilon}
\newcommand{\thi}{\theta}

\newcommand{\Li}{\Lambda}
\title{Causality bounds on scalar-tensor EFTs}

\author[a]{Dong-Yu Hong,}
\author[a]{Zhuo-Hui Wang}
\author[a,b]{and Shuang-Yong Zhou}

\affiliation[a]{Interdisciplinary Center for Theoretical Study, University of Science and Technology of China, Hefei, Anhui 230026, China}
\affiliation[b]{Peng Huanwu Center for Fundamental Theory, Hefei, Anhui 230026, China}

\emailAdd{principle@mail.ustc.edu.cn}
\emailAdd{wzh33@mail.ustc.edu.cn}
\emailAdd{zhoushy@ustc.edu.cn}

\preprint{{\small USTC-ICTS/PCFT-23-10}}

\date{\today}

\abstract{
We compute the causality/positivity bounds on the Wilson coefficients of scalar-tensor effective field theories. Two-sided bounds are obtained by extracting IR information from UV physics via dispersion relations of scattering amplitudes, making use of the full crossing symmetry. The graviton $t$-channel pole is carefully treated in the numerical optimization, taking into account the constraints with fixed impact parameters. It is shown that the typical sizes of the Wilson coefficients can be estimated by simply inspecting the dispersion relations. We carve out sharp bounds on the leading coefficients, particularly, the scalar-Gauss-Bonnet couplings, and discuss how some bounds vary with the leading $(\pd\phi)^4$ coefficient and as well as phenomenological implications of the causality bounds. 
}

\begin{document}
\maketitle
\flushbottom

\section{Introduction and summary}

{\bf Causality/Positivity bounds} 
Relativistic causality is a foundational concept that underpins the modern construction of the fundamental models of nature. It is conjectured to imply analyticity and crossing symmetry of the S-matrix \cite{Eden:1966dnq}. Unitarity of the quantum theory, another foundational cornerstone, also plays a vital role in restricting the forms the S-matrix can take. On the other hand, effective field theories (EFTs) are part and parcel of model building in modern particle physics and cosmology. Using merely the low energy field contents and symmetries, an EFT, arising from integrating out heavy degrees of freedom, can parametrize generic effects of possible UV completions at low energies. Interestingly, causality and unitarity, along with locality, can impose strong constraints on the theory space, {\it i.e.,} the space of the Wilson coefficients, of effective field theories, often known as causality or positivity bounds (see \cite{deRham:2022hpx} for a concise review). 

A simple and efficient way to derive these constraints on the Wilson coefficients is via the dispersion relations or dispersive sum rules, which provide a portal to connect the accessible EFT coefficients in the IR with the generic unknown physics in the UV \cite{Adams:2006sv}. They can be derived from analyticity, crossing symmetry and locality of the scattering amplitudes, and causality bounds are precisely the unitarity conditions on the UV amplitudes passed down to the IR via the dispersive sum rules. In the forward-limit of identical particle scattering, a simple positivity bound on the $s^2$ ($s,t,u$ being the Mandelstam) coefficient can be easily seen using the textbook optical theorem \cite{Adams:2006sv}. The $s^2$ bound is usually the most accessible one phenomenologically. For 2-to-2 scattering between multiple species of particles, there are a set of $s^2$ coefficients since the amplitude can have different in and out states. Positivity bounds tell us that these $s^2$ coefficients form a convex cone, whose extremal rays (or kinks from the viewpoint of the cross section of the convex cone) correspond to tree-level UV (irrep) states, which are endowed with the projected-down versions of the UV symmetries \cite{Zhang:2020jyn, Bellazzini:2014waa}. Particularly, this means that one can infer the existence of certain UV states from the causality convex cone, which helps inverse engineer the UV model from the EFT data. Furthermore, the dual cone of this amplitude cone is a spectrahedron, so the optimal causality bounds on the $s^2$ coefficients can also be effectively computed with semi-definite programing (SDP), even for the case of many degrees of freedom with less symmetries \cite{Li:2021lpe}. The Standard Model EFT (SMEFT) contains many degrees of freedom, so its parameter space is vast, especially at higher orders. Positivity bounds have been found to significantly restrict the viable space of dimension-8 operators \cite{Zhang:2020jyn, Li:2021lpe, Zhang:2018shp, Bi:2019phv, Bellazzini:2018paj, Remmen:2019cyz, Yamashita:2020gtt, Trott:2020ebl, Remmen:2020vts, Bonnefoy:2020yee, Davighi:2021osh, Chala:2021wpj, Li:2022tcz,Ghosh:2022qqq,Remmen:2022orj}. One may also reverse the argument and use the positivity bounds to test the fundamental principles of quantum field theory in some seemingly benign parameter regions \cite{Fuks:2020ujk, Gu:2020ldn, Li:2022rag}, or inverse bootstrap the UV from the IR \cite{Alberte:2021dnj, Alberte:2020bdz}.

Highly nonlinear constraints on the coefficients of higher powers of $s$ can also be gleaned once realizing that the forward-limit dispersion relations readily define a Hausdorff moment problem \cite{Arkani-Hamed:2020blm, Bellazzini:2020cot}.  Away from the forward limit, a series of easily-to-use analytic bounds on both $s$ and $t$ derivatives of the amplitudes can be obtained using the Martin extension of analyticity \cite{Martin:1965jj} and the positivity of the derivatives of the Legendre polynomials \cite{deRham:2017avq} (see also \cite{Manohar:2008tc, Nicolis:2009qm, Bellazzini:2016xrt, Wang:2020jxr} for related works). These bounds can be generalized to the case of massive particles with spin utilizing the transversity formalism (as opposed to the helicity formalism) for the external polarizations \cite{deRham:2017zjm}.

However, since the dispersive sum rules used to derive the above bounds are only $su$-symmetric, the full crossing symmetry of the S-matrix has not been used thoroughly, and they usually only constrain the coefficients from one side. Indeed, two-sided bounds can be derived for the coefficients once the full crossing symmetry is used \cite{Tolley:2020gtv, Caron-Huot:2020cmc}. One pathway to achieve the triple crossing symmetry is simply to impose $st$ symmetry on the $su$-symmetric sum rules. For the case of identical scalar scattering, the bounds on the explicitly computed coefficients are consistent with the usual dimensional analysis expectations for EFT coefficients. More importantly, this excludes the possibility that some delicate design of the UV model can lead to arbitrary disparity among different orders of Wilson coefficients --- ``not everything goes for an EFT'' \cite{Vafa:2005ui}. This formalism can be easily extended to the case of multi-field theories using the generalized optical theorem for partial waves \cite{Du:2021byy}. Compared to linear programing for the case of a single scalar, the optimization scheme now needs to be promoted to be a SDP problem with a continuous variable, which parametrizes the scales of the UV states. Both of them can be efficiently solved by the {\tt SDPB} package \cite{Simmons-Duffin:2015qma}. Alternative methods, also based on dispersive relations, have been developed for obtaining the fully crossing symmetric causality bounds. These include directly using triple crossing symmetric dispersive relations \cite{Sinha:2020win}, and formulating the (non-forward) dispersion relations as a double moment problem and slicing out the triple crossing bounds towards the end \cite{Chiang:2021ziz}. Triple crossing positivity bounds have also been used to constrain EFTs with spinning particles  \cite{Bern:2021ppb, Henriksson:2021ymi, Chowdhury:2021ynh, Caron-Huot:2022ugt, Chiang:2022jep, Caron-Huot:2022jli, Henriksson:2022oeu}, and extra causality constraints using the upper bounds on the spectral functions can be found in \cite{Caron-Huot:2020cmc, Chiang:2022jep,Chiang:2022ltp}. Moreover, the powerful primal approach of S-matrix bootstrap has also been developed to chart the space of EFTs; see, {\it e.g.,} \cite{Guerrieri:2020bto, Guerrieri:2021ivu, EliasMiro:2022xaa, Haring:2022sdp} and for a review \cite{Kruczenski:2022lot}. The primal approach directly parametrizes the crossing symmetric amplitudes themselves and expands viable theory space by imposing unitarity conditions. In this language, the above methods that rule out unphysical parameter regions is referred to as the dual approach, which parallels the difference between the cone and dual cone of the $s^2$ coefficients above.

In the presence of graviton exchanges in the scattering, a $t$-channel pole appears in the left hand side of the sum rules, because a spin-2 particle $t$-channel exchange term, different from the cases of lower spins, can survive the twice subtractions in deriving the sum rules. While we can still Taylor expand in terms of $s$, the existence of the $t$-channel pole prevents us from Taylor expanding in terms of $t$. Indeed, this $t$-channel pole must be balanced by a divergence in the dispersive integral on the right hand side as $t\to0$.
Apart from balancing the pole, the dispersive integral also gives rise to extra terms which can be negative and violate the would-be strict positivity in theories without the gravitons \cite{Alberte:2020jsk, Alberte:2020bdz, Tokuda:2020mlf}. Nevertheless, each of the $s$-expanded sum rules can be viewed as a one-parameter ($t$) family of IR-UV relations, and one can effectively use them by optimizing over a set of continuous functions for the range that $t$ can take within the EFT \cite{Caron-Huot:2021rmr}. It turns out that the strongest constraints come from when $t$ is far away from the forward limit and close to the cutoff. (A similar phenomenon was also seen in the earlier non-forward-limit bounds without full crossing symmetry \cite{deRham:2017imi, deRham:2018qqo}.) Physically, this means that some important constraints arise from when the impact parameter is small \cite{Caron-Huot:2021rmr}. This approach has been used to constrain the Wilson coefficients of Einstein gravitational EFTs \cite{Caron-Huot:2022ugt, Caron-Huot:2022jli} and Einstein-Maxwell EFTs \cite{Henriksson:2022oeu}.

Besides using the dispersion relations, causality bounds can also be derived from within the EFT by requiring information not propagating faster the speed of light. Although less algorithmic than the optimized dispersion relation approach, this approach is more intuitive and can sometimes produce very strong constraints with less efforts. In flat space, subluminality can usually be directly imposed on the dynamical modes of theory in a nontrivial background, which leads to conditions consistent with the positivity bounds obtained above \cite{Adams:2006sv}. In a gravitational EFT, the situation is more subtle, as the definition of speed is frame-dependent. So one resorts to observables such as the time delay in a classical scattering. An often used causality condition is that the Eisenbud-Wigner time advance be not resolvable for the scattering wave, which is called asymptotic causality \cite{Camanho:2016opx}. However, a more refined criterion for an EFT, called infrared causality, may be imposed that the time advance with the GR part subtracted should be non-resolvable for the scattering wave \cite{Chen:2021bvg, deRham:2020zyh}. Applications of the infrared causality can be found in \cite{Chen:2021bvg, deRham:2021bll}, and those of the asymptotic causality can be found in \cite{Camanho:2016opx,Goon:2016une,Hinterbichler:2017qyt,AccettulliHuber:2020oou,Bellazzini:2021shn}. A few other interesting applications of positivity bounds on gravitational and cosmological EFTs can be found in for example \cite{Bellazzini:2015cra, Cheung:2016yqr, Bonifacio:2016wcb, Bellazzini:2017fep, Bonifacio:2018vzv, Melville:2019wyy, deRham:2019ctd, Alberte:2019xfh, Chen:2019qvr, Huang:2020nqy,  Wang:2020xlt, Herrero-Valea:2019hde, Herrero-Valea:2020wxz, deRham:2021fpu, Arkani-Hamed:2021ajd, Bellazzini:2022wzv}.\\

\noindent{\bf Scalar-tensor theory}
General relativity (GR), with only the Einstein-Hilbert term, has been extensively tested in the solar system where it is relatively convenient for us to carry out gravitational experiments and where gravity is weak and velocities are small compared to the speed of light \cite{Will:2014kxa,Berti:2015itd}. The development of the Parameterized Post-Newtonian formalism has put severe constraints on possible deviations from GR in the weak gravity limit. The formalism is quite systematic, as it thoroughly parameterizes all possible deviations directly at the level of the metric. The discovery of binary pulsars has allowed us to confirm viability of GR in stronger gravity environments, with somewhat less accuracy, but those environments are still well approximated by the linearized GR. Therefore, the lesson is that, to be a viable alternative or extended gravity theory, it first needs to very precisely reduce to GR in the weak field limit.

However, this does not necessarily mean that sizable beyond GR effects have been completely ruled out in astrophysics, an intriguing possibility being that they are hidden in the highly dynamical and strong-field regimes, such as near black holes and neutron stars. Indeed, we are just starting to probe these regimes with the new observational tools such as LIGO-Virgo-KAGRA gravitational wave detectors \cite{LIGOScientific:2016aoc} and the Event Horizon Telescope \cite{EventHorizonTelescope:2019dse}. While GR can still pass the tests from these experiments to date, the accuracy is still quite low. Since interpolating between the weak gravity GR regime and the strong gravity regime with non-GR effects requires some degrees of ``dynamical'' nonlinearity, one of the simplest ways is to introduce new field degrees of freedom. Scalar-tensor theory is a simple extension of GR in this direction which only adds one extra field degree of freedom. Brans-Dicke theory \cite{Brans:1961sx}, which give rises to a ``variable gravitational constant'', is one of the earliest such models. It is currently tightly constrained by observations \cite{Will:2014kxa}. However, its extensions such as Horndeski theory/Generalized Galieon \cite{Horndeski:1974wa, Deffayet:2011gz} and Degenerate Higher-Order Scalar-Tensor theories \cite{Langlois:2015cwa} are being intensively investigated to fit astronomical and cosmological data \cite{Berti:2015itd}.
 Another motivation for scalar-tensor theory comes from string/M theory, where a dilaton naturally arises as a low energy degree of freedom from compactification \cite{Green:2012oqa}.  The scalar degree of freedom is natural to consider also because fermions, due to the Pauli exclusion principle, can not form classical configurations, which need high occupation numbers at a range of momentum modes, while long-distance vector fields, endowed with a direction, are difficult to be compatible with the cosmological principle.

There is a growing body of research dedicated to understanding scalar-tensor theory in the strong regimes. The class of models involving the Gauss-Bonnet invariant $\mc{G}=R_{\mu\nu\ri\si}R^{\mu\nu\ri\si} -  4R_{\mu\nu} R^{\mu\nu}+ R^2 $ stand out, as they are low orders in the EFTs and can give rise to hairy black holes \cite{Kanti:1995vq, Sotiriou:2013qea, Sotiriou:2014pfa, Yagi:2011xp} and the phenomenon of (spontaneous) scalarization \cite{Silva:2017uqg, Doneva:2017bvd}. These operators have been confronted with gravitational wave observations and beyond \cite{Yagi:2012gp, Witek:2018dmd, Carson:2019fxr, Wang:2021jfc, Perkins:2021mhb,  Pani:2011xm, Saffer:2021gak, Antoniou:2022dre, Lyu:2022gdr, Wong:2022wni}. Wheeler famously coined the phrase that a black hole has no hair \cite{Bekenstein:1996pn}. More precisely, due to the uniqueness theorems in GR, a (non-charged) black hole in GR can be solely described by its mass and angular momentum, and a bunch of no-hair theorems generally prevent a black hole from having other parameters/pieces of hair \cite{Bekenstein:1996pn, Herdeiro:2015waa}. A few exceptions include the presence of the scalar-Gauss-Bonnet couplings. In fact, assuming shift symmetry for the scalar and the equations of motion being second order, the linear scalar-Gauss-Bonnet coupling $\phi \mc{G}$ is necessary to sustain hairy solutions in Horndeski theory \cite{Sotiriou:2013qea, Sotiriou:2014pfa}. Furthermore, the $\phi \mc{G}$ term leads to the same parametrized post-Newtonian parameters as in GR \cite{Sotiriou:2006pq}, and in particular it does not lead to nontrivial scalar charges for neutron stars or other extended objects \cite{Yagi:2011xp}. Therefore, the current gravitational wave experiments are an ideal place to test this leading quadratic curvature term. 

On the other hand, the Damour-Esposito-Farese model \cite{Damour:1993hw} is the first model of scalarization, which was proposed when the weak field gravity tests had reached an unprecedented accuracy such that viable deviations from GR was seemingly impracticable. It was also when binary pulsar observations became available, ushering in a new arena to test GR with the compact stars. In the Damour-Esposito-Farese model, the scalar field obtains a nontrivial profile once the density/curvature within the star exceeds a threshold, and this can be the case for a neutron star, resulting in strong deviations from GR, but not for the Sun.  With the arrival of gravitational wave astronomy, another new window has been opened up to test GR in stronger and more dynamical gravity environments. Recently, a new class of scalarization models involving the Gauss-Bonnet invariant and black holes have been proposed, in which the black hole becomes hairy if the curvature outside the horizon exceeds a threshold \cite{Silva:2017uqg, Doneva:2017bvd} (see \cite{Doneva:2022ewd} for a review). The underlying reason for the scalarization to happen is because in these models the strong gravity environment induces tachyonic instabilities for the unscalarized configuration. In the inspiral phase of a binary black hole coalescence, a dynamical de-scalarization can occur, which can give rise to extra scalar radiation and thus observational constraints \cite{Silva:2020omi}.

With the arrival of the gravitational wave astronomy and advances of more traditional observational means, it is becoming increasingly accessible to test gravity, along with possible accompanied extra degrees of freedom, in the strong and dynamical regimes \cite{Berti:2015itd}. As we shall see, the causality bounds can strongly constrain the parameter spaces of gravitational EFTs, which may help orient current and future experiments to more theoretically favorable directions. On the flip side, one may also use the new observational data to test the fundamental principles of quantum field theory or the S-matrix theory. \\

\noindent {\bf Summary}
In this paper, we investigate how causality bounds constrain the parameter space of scalar-tensor theory by means of dispersive sum rules of the scattering amplitudes. To fully utilize the crossing symmetry of amplitudes, we start with dispersive sum rules that are only $su$-symmetric and then impose the $st$ symmetry on these sum rules. In a multi-field theory such as scalar-tensor theory here, only a few amplitudes are truly symmetric in full permutations of $s,t,u$ in the strict sense, some being not even strictly symmetric in $s$ and $u$, so the $su$ or $st$ crossing symmetry is used loosely in this context, with the understanding that some crossings actually link distinct amplitudes. Nevertheless, the working mechanism of improving the bounds with crossing symmetry is exactly the same as in the single scalar case. In the presence of massless gravitons, the $t$-channel pole prevents us from Taylor-expanding some sum rules in the forward limit, so the decision variables for the optimization involve a set of weight functions of $t$, which numerically will be evaluated with a finite dimensional truncation. In this setup, some important constraint space can be effectively sampled using the impact parameter \cite{Caron-Huot:2021rmr}. Various causality bounds without full crossing symmetry and/or neglecting the $t$-channel pole have previously been used to constrain scalar-tensor models \cite{Melville:2019wyy, deRham:2021fpu, Tokuda:2020mlf, Herrero-Valea:2021dry,  Bellazzini:2022wzv, Serra:2022pzl, Hertzberg:2022bsb}.

While the Froissart-Martin bound \cite{Froissart:1961ux, Martin:1962rt} for the high energy behaviors of amplitudes is rigorously established for massive particles, which suggests that only two subtractions are needed to derive the dispersive sum rules, it is more subtle for massless particles especially in the presence of gravitons. We will make the usual assumption that only two subtractions are needed when $t<0$ and three subtractions when $t\leq 0$ \cite{Alberte:2021dnj, Caron-Huot:2022ugt}. We will also assume that the EFT is weakly coupled in the IR so that we can use tree-level amplitudes at low energies, but we are agnostic about the attributes of the UV theory, as manifest in our exclusive use of the dispersive sum rules in deriving the bounds. We will only make use of positivity of partial wave unitarity, which leads to the semi-positive conditions on the $B_{P_X,\ell}(\mu)$ matrices (see \eref{BposCon}). Nevertheless, with full crossing symmetry incorporated, we find that the Wilson coefficients projected to the gravitational coupling $1/M_P^2$ are already bounded to finite regions. This is of course except for the $(\pd\phi)^4$ coefficient (and consequently some correlated coefficients), for which the upper bound of partial wave unitarity is needed to cap from the above. 

We find that a simple method can be devised to estimate the sizes/scalings of the Wilson coefficients via the dispersive sum rules, without the need for heavy numerical calculations. This proceeds by first normalizing the Mandelstam variables in the dispersive sum rules with the cutoff of the EFT. Then, from some simple sum rules that only contain the gravitational coupling $1/M^2_P$, we can establish correspondences between the UV spectral functions and the hierarchy between the cutoff and the Planck mass. A scaling correspondence can not be uniquely assigned in this way to the UV spectral function $c^{00\to X}_{\ell,\mu}$ (the partial amplitude from two scalars to a heavy state $X$, cf.~\eref{gotAcc}), for which we can either let it saturate the unitarity upper bound or assign a desired correspondence, the latter of which will lead to an {\it ad hoc} class of theories with reduced scalings for the relevant terms. These correspondences can then be used to infer the dimensions of the Wilson coefficients by simple inspection of available sum rules. The scalings of the coefficients extracted in this way are consistent with the sharp numerical bounds obtained by SDP.

The causality bounds on some Wilson coefficients are intimately correlated with each other, while others are quite independent. This can be often inspected from the $B_{P_X,\ell}(\mu)$ matrices that are constructed from dispersive sum rules. If the relevant quantities are in different diagonal blocks, then the corresponding coefficients are insensitive to each other. However, even if the relevant quantities overlap in the $B_{P_X,\ell}(\mu)$ matrix, a strong correlation between the corresponding coefficients is not guaranteed. At the practical level, the bounds on certain coefficients can not be numerically optimized unless we specify the value of the coefficient of the scalar self-interaction operator $(\pd\phi)^4$. These are the coefficients that only appear in the sum rules involving the UV spectral function $c^{00\to X}_{\ell,\mu}$.  

We also derive the causality bounds on some fine-tuned EFTs. The bounds on a set of Wilson coefficients in the fine-tuned EFT can be considered as taking an appropriate crossing section in the Wilson coefficient space, while the bounds on a given set of Wilson coefficients in a generic EFT amounts to projecting the causality spectrahedron down to an appropriate subspace. We show that some phenomenological models such as the $f(\phi)\mc{G}$ model should not be taken at its face value, because only adding exactly $f(\phi)\mc{G}$ but no other terms inevitably violates causality bounds. Indeed, in a model where the operators essential for causality bounds to uphold are turned on but highly suppressed compared to the usual EFT power counting, we can see that the Wilson coefficients of concern are also highly constrained by causality bounds. We give a simple criterion to test whether a given/fine-tuned scalar-tensor model will run into contradictions with causality bounds.   

Particular attention has been given to the scalar Gauss-Bonnet couplings, which can give rise to hairy black holes and scalarization and are currently undergoing intense scrutiny in astrophysics by gravitational wave and other observations. We carve out the 2D bounds on the leading order $\phi\mc{G}$ coefficient together with the coefficient of the Riemann cubed operator, which is independent of the coefficient $\ai$ of $(\pd\phi)^4$. On the other hand, the bounds on the coefficient of $\phi^2\mc{G}$, which is essential for scalarization, strongly depend on $\ai$. We also compare the causality bounds with the observational bounds for the coefficients of $\phi\mc{G}$ and $\phi^2\mc{G}$, which allows us to impose bounds on the cutoffs for these EFTs and reduce the viable parameter space, thanks to the fact that for a capped $\ai$ these fully crossing symmetric bounds have restricted the viable parameters to an enclosed region.

If the scalar interacts with the heavy modes weakly in the UV theory, {\it i.e.,} if the UV spectral function $c^{00\to X}_{\ell,\mu}$ is suppressed by $\mc{O}({\Lambda}/{M_P})$, the scalar  will interact with the graviton with the usual gravitational strength in the low energy scalar-tensor EFT. This will lead to the scaling of \eref{action22}. For the terms involving the Gauss-Bonnet invariant, this gives rise to the usual scaling implicitly used in most literature: $\mc{L}\supset M_P^2\sqrt{-g} (\f{\mc{O}(1)}{\Li^2}  \varphi {\cal G}    +\frac{\mc{O}(1) }{\Li^2}  \varphi^2 {\cal G})$, where $\varphi\equiv\phi/M_P$.
However, for a generic UV completion, as we see in \eref{action21}, the couplings for terms like $\varphi^2 {\cal G}$ are allowed to be much larger, without running into the trouble with causality bounds: $ \mc{L}\supset M_P^2\sqrt{-g} ( \f{\mc{O}(1)}{\Li^2}  \varphi {\cal G}    +\frac{\mc{O}(1) M_P}{\Li^3}  \varphi^2 {\cal G})$. This arises when the low energy scalar interacts the heavy modes more strongly than the gravitational force, a scenario aligned with the weak gravity conjecture. Incidentally, in this scenario, the spontaneous scalarization models are natural where a vanishing $\varphi {\cal G}$ term is usually assumed and a sizable $\varphi^2 {\cal G}$ is required for tachyonic instabilities to take place. We have confirmed the above scalings with the numerical causality bounds in Section \ref{sec:bounds}.

We have focused on the parity conserving sector in this paper. Once the parity violating operators are involved, the complexity of numerics will increase significantly, as we have to augment the dimension of the vector $\mc{C}_{P_X,\ell,\mu}$ and consequently the matrix $B_{P_X,\ell}(\mu)$ (see \eref{BposCon}). There has also been a growing interest in examining the observational implications of parity-violating operators in scalar-tensor theories (see for example \cite{Yagi:2011xp, Berti:2015itd}). We defer the extraction of causality bounds on these terms to future work \cite{pvpaper}.

The paper is organized as follows. In Section \ref{sec: stEFT}, we present the scalar-tensor EFT both at the level of Lagrangian, with independent operators, and at the level of the amplitudes that will be needed to derive the dispersive sum rules. The sum rules will be derived in a couple of steps in Section \ref{sec:disprules}. In Section \ref{sec:dimAna}, we propose a method to perform dimensional analysis of the Wilson coefficients with the dispersive sum rules. In Section \ref{sec:numSet}, we outline the optimization scheme to obtain the optimal bounds with positivity from unitarity, and explain its numerical implementation in details. In Section \ref{sec:bounds}, we present the results of the numerical causality bounds and discuss their implications. In Appendix \ref{sec:4pointAmp}, we show how to construct generic 4-leg EFT amplitudes from scratch. In Appendix \ref{sec:expSum}, we explicitly list all the sum rules used to perform analyses and computations in this paper. In Appendix \ref{sec:Bec_exa}, we show an explicit example exhibiting how the SDP optimization is performed. \\

\noindent {\bf Notation and conventions} The (reduced) Planck mass is $M_P=1/\sqrt{8\pi G_N}$. Our metric signature is  $g_{\mu\nu}=\{-,+,+,+\}$. We choose all momenta to be in-going, so the Mandelstam variables are $s = - (p_1+p_2)^2,~t = - (p_1+p_3)^2,~u = - (p_1+p_4)^2$. A generic four-point helicity amplitude is denoted as $\mc{M}^{\mathbb{1234}}= \mathcal{M}\left(1^{h_{1}} 2^{h_{2}} 3^{h_{3}} 4^{h_{4}}\right)$, where $h_i$ is the helicity for particle $i$, while a specific four-point helicity amplitude is denoted as, say, $\mc{M}^{++ 0-}  = \mathcal{M}\left(1^{+2} 2^{+2} 3^{0} 4^{-2}\right)$. Our convention for he partial wave expansion of the four-point amplitude is $\mc{M}^{\mathbb{1234}}(s,t,u) =16\pi\sum_{\ell} (2\ell+1)d^{\ell}_{h_{12},h_{43}}\(1+\frac{2t}{s}\)A^{\mathbb{1234}}_{\ell}(s)$, where $A^{\mathbb{1234}}_{\ell}(s)\equiv A_{\ell}^{h_1h_2h_3h_4}(s)$ is the partial wave amplitude, $d_{h_{12}, h_{43}}^{\ell}(z)$ is the Wigner (small) d-matrices and $h_{ij}\equiv h_i-h_j$ . The dimensionful scalar field $\phi$ is related to the dimensionless one $\varphi$ by $\phi=M_P\varphi$.

\section{Scalar-tensor EFT}
\label{sec: stEFT}

Scalar-tensor theory is a popular extension of Einstein's metric tensor theory. It augments gravity by coupling the massless spin-2 field to a scalar, arguably the simplest kind of fields that can form classical configurations which may affect local or large-scale gravitational physics. The scalar can minimally couple to the metric with possible potential self-interactions. However, from an EFT point of view, non-minimal and derivative interactions are generically present in the theory. For example, these couplings are also ubiquitous in EFTs from string/M theory which generally predicts existence of scalars due to compactification from higher dimensions \cite{Green:2012oqa}. Indeed, the effects of these non-minimal and derivative couplings have been extensively studied in astrophysics and cosmology \cite{Clifton:2011jh}. 

We will be interested in 4D scalar-tensor theory where the mass of the scalar is negligible, and also assume that the theory is weakly coupled below the cutoff $\Lambda$ so that we can take the tree-level approximation in the IR. We are agnostic about the UV theory, in particular, not assuming it to be weakly coupled. Up to six derivatives and including only terms that can give rise to tree-level 2-to-2 amplitudes, the lowest order terms of such a theory are given by
\bal
\label{action0}
S  &=   \int \d^4 x \sqrt{-g} \bigg(  \f{M_P^2}2 R -\f12 \nd_\mu \phi \nd^\mu \phi   - \f{\li_3}{3!} \phi^3 -  \f{\li_4}{4!}\phi^4 +  \f{\ai}{2} (\nd_\mu \phi \nd^\mu \phi)^2    + \f{\bi_1}{2!}  \phi {\cal G}    +\frac{\bi_2}{4}  \phi^2 {\cal G}
 \nn
 &\hskip 70pt + \f{\gi_0}{3!} {\cal R}^{(3)} + \f{\gi_1}{3!}  \phi{\cal R}^{(3)}  +  \frac{\gi_2}{2} \nd_\mu\phi\nd^\mu\phi {\cal R}^{(2)}  -\frac{4\gi_3}{3}  \nd_\mu\phi\nd_\ri\phi \nd_\nu\nd_\si \phi  R^{\mu\nu\ri\si}
 \nn
 &\hskip 70pt  +  \f{\gi_4}{3}  \nd_\mu\phi \nd^\mu  \phi   \nd_\ri \nd_\si \phi \nd^\ri \nd^\si \phi
 + \cdots \bigg)\,,
\eal
where  $M_P=1/\sqrt{8\pi G_N}$ is the (reduced) Planck mass and we have defined ${\cal R}^{(2)}$, ${\cal R}^{(3)}$ and the Gauss-Bonnet invariant ${\cal G}$,
\be
{\cal R}^{(2)}=R_{\mu \nu \rho \sigma} R^{\mu\nu\ri\si}\,,~~{\cal R}^{(3)}=R_{\mu \nu}{ }^{\rho \sigma} R_{\rho \sigma}{ }^{\alpha \beta} R_{\alpha \beta}{ }^{\mu \nu}\,,~~ {\cal G}= R_{\mu\nu\ri\si}R^{\mu\nu\ri\si} -  4R_{\mu\nu} R^{\mu\nu}+ R^2 \,.
\ee
We have focused on a scalar-tensor theory that conserves parity, so Lagrangian terms with odd numbers of the Levi-Civita tensor such as the Chern-Simons term $R_{\mu\nu\ri\si}  \epsilon^{\mu \nu \alpha \beta} R_{\alpha \beta}{}^{\rho \sigma}$ are absent from the Lagrangian. Naively, there are several other terms that can be written down in the Lagrangian, but those terms can be reduced to the above terms by field redefinitions and integration by parts \cite{Solomon:2017nlh,Ruhdorfer:2019qmk}. This can be partially checked by explicit scattering amplitudes computed in the following, since amplitudes are free of ambiguities of field redefinitions and integration by parts.

As mentioned in the introduction, the scalar coupled quadratic curvature terms are being actively looked at phenomenologically, in search of/to rule out possible deviations from Einstein's gravity in strong and/or dynamical gravity environments near compact stars. In principle, a couple of scalar self-interaction operators are of lower orders in terms of the EFT cutoff, but they are only minimally coupled to gravity, which by themselves would not give rise to significant modifications to the gravitational force. More practically, for the positivity bounds that will be extracted later, since we make use of the generic twice subtracted dispersion relations, the scalar potential terms are unconstrained, while, say, the scalar four-derivative self-coupling can be bounded. In fact, the coefficient of the dim-8 contact interaction being bounded to be positive in flat space has inspired the name of these bounds.

Particular attention has been paid to the operators involving the Gauss-Bonnet invariant, as these operators can give rise to hairy black holes \cite{Kanti:1995vq, Sotiriou:2013qea, Sotiriou:2014pfa, Yagi:2011xp} and the interesting phenomenon of spontaneous scalarization \cite{Silva:2017uqg,Doneva:2017bvd}, which is the reason why we have chosen to parametrize the Lagrangian terms with the Gauss-Bonnet invariant, instead of the Riemann tensor squared. The linear scalar-Gauss-Bonnet term $\phi \mc{G}$ \cite{Sotiriou:2014pfa, Sotiriou:2013qea, Yagi:2011xp} is special in the sense that it is shift-symmetric $\phi\to \phi +const$, as $\mc{G}$ is famously a total derivative. Significant efforts have been put into constraining the Wilson coefficient of this operator with the gravitational wave and X-ray data from binary compact stars \cite{Yagi:2012gp, Witek:2018dmd, Carson:2019fxr, Wang:2021jfc, Perkins:2021mhb,  Pani:2011xm, Saffer:2021gak, Antoniou:2022dre, Lyu:2022gdr, Wong:2022wni}. These observations capitalize on the fact that the scalar-Gauss-Bonnet coupling alters the star configurations and as well as induces significant dipole radiation in binaries, thanks to the scalar degree of freedom.  In Section \ref{sec:bounds}, we shall use these data to infer observational bounds on the EFT cutoff. Furthermore, the $\phi^2 \mc{G}$ operator has also attracted a lot of interest lately, due to its ability to generate tachyonic instabilities to make the scalar field nontrivial for black holes and neutron stars \cite{Silva:2017uqg, Doneva:2017bvd}.

Since we will be constraining the Wilson coefficients with the dispersion relations of the scattering amplitudes, we may as well parametrize the EFT at the level of amplitudes. General EFT amplitudes can be parametrize by considering little group scalings and crossing symmetries. After factoring out the helicity structures, the amplitudes can be written as scalar functions of Mandelstam variables $s,t,u$. Crossing symmetries dictate the symmetries of these functions, and also allow us to focus on a few independent amplitudes to extract all available information. For the lowest orders of the amplitudes with double 3-leg insertions, one can simply calculate them explicitly from the EFT Lagrangian. Contributions from the 4-leg contact interactions can be constructed based on some simple principles. For our purposes, we choose a representation for the helicity spinors to also convert the helicity structures into expressions in terms of $s,t,u$. After these considerations (see more details in Appendix \ref{sec:4pointAmp}), the independent amplitudes can thus be parametrized as follows
\bal
\label{gstartdef}
\mathcal{M}^{0000}&= f_{S}(s,t,u) =g^{S}_{-1,1}\frac{x}{y}+g^{S}_{0,0}+g^{S}_{-1,2}\frac{x^2}{y}+g^{S}_{0,1}x+g^{S}_{1,0}y+g^{S}_{0,2}x^2+\cdots
\\
\mathcal{M}^{++--}&=([12]\<34\>)^{4} f_{T_1}(t,u)
  =g^{T_1}_{3,-1}\frac{s^3}{t u}+g^{T_1}_{3,0}s^3+g^{T_1}_{4,0}s^4+g^{T_1}_{3,1}s^3t u+g^{T_1}_{5,0}s^5+\cdots \\
\mathcal{M}^{+++-}&=([12][13]\langle 14\rangle)^{4} f_{T_2}(s,t,u)
  =g^{T_2}_{1,0}y+g^{T_2}_{2,0}y^2+g^{T_2}_{2,1}y^2x+g^{T_2}_{3,0}y^3+\cdots \\
\mathcal{M}^{++++}&=\frac{\([12][34]\)^2}{\(\langle 12\rangle\langle 34\rangle\)^2} f_{T_3}(s,t,u)
  =g^{T_3}_{1,0}y+g^{T_3}_{0,2}x^2+g^{T_3}_{1,1}yx+g^{T_3}_{2,0}y^2+g^{T_3}_{0,3}x^3
  +\cdots
\\
\mc{M}^{+++0} &= ([12][23][31])^2 f_{M_1}(s,t,u)
=g^{M_1}_{1,0}y+g^{M_1}_{1,1}yx+g^{M_1}_{2,0}y^2+\cdots \\
\mc{M}^{++0-} &= [12]^6 (\<14\>\<24\>)^2  f_{M_2}(t,u)
 =g^{M_2}_{2,0}s^2+g^{M_2}_{2,1}s^2t u+g^{M_2}_{3,1}s^3t u+\cdots \\
\mc{M}^{++00}  &=[12]^4 f_{M_3}(t,u)
 =g^{M_3}_{1,0}s+g^{M_3}_{2,0}s^2+g^{M_3}_{1,1}s t u+g^{M_3}_{3,0}s^3+g^{M_3}_{2,1}s^2tu+\cdots \\
\mc{M}^{+-00} &= \([13]\<23\>[14]\<24\>\)^2  f_{M_4}(t,u)
 =g^{M_4}_{-1,1}\frac{t u}{s}+g^{M_4}_{1,1}s t u+g^{M_4}_{0,2}(t u)^2+\cdots \\
\mc{M}^{+000} &= ([12]\<23\>[31])^2 f_{M_5}(s,t,u)
 =g^{M_5}_{0,1}x+g^{M_5}_{1,0}y+g^{M_5}_{1,1}yx+g^{M_5}_{2,0}y^2+\cdots
 \label{gfinaldef}
\eal
where we have defined the shorthand for the amplitudes, say, $\mc{M}^{++0-}=\mc{M}(1^{+2} 2^{+2} 3^{0} 4^{-2} )$ (particle 1 having helicity $+2$, etc.) and the basic symmetric polynomials of the Mandelstam variables
\bal
x=s^2+t^2+u^2\,,~~y=s t u\,.
\eal
 The $f_i(t,u)$ functions are $tu$ symmetric, while the $f_i(s,t,u)$ functions are $stu$ symmetric. Thus, in scalar-tensor theory, a  whole amplitude is either symmetric under the full permutations of $s,t,u$ or symmetric under the exchange of two of $s,t,u$, accompanied by exchanges of the helicities accordingly. Explicitly, the ones with full $stu$ permutation symmetries are given by
\bal
	\label{fullcrossS1}
		\mc{M}^{0000}(s,t,u)&=\mc{M}^{0000}(u,t,s)=\mc{M}^{0000}(t,s,u)\,,\\
		\mc{M}^{+000}(s,t,u)&=\mc{M}^{+000}(u,t,s)=\mc{M}^{+000}(t,s,u)\,,\\
		\mc{M}^{+++0}(s,t,u)&=\mc{M}^{+++0}(u,t,s)=\mc{M}^{+++0}(t,s,u)\,,\\
		\mc{M}^{+++-}(s,t,u)&=\mc{M}^{+++-}(u,t,s)=\mc{M}^{+++-}(t,s,u)\,,\\
	\label{fullcrossS5}
		\mc{M}^{++++}(s,t,u)&=\mc{M}^{++++}(u,t,s)=\mc{M}^{++++}(t,s,u) \,,
\eal
and the ones with only one exchange symmetry are
\bal
	\label{onecrossS1}
		\mc{M}^{++00}(s,t,u)&=\mc{M}^{++00}(s,u,t)=\mc{M}^{+00+}(u,t,s)=\mc{M}^{+00+}(t,u,s)\nn
		&=\mc{M}^{+0+0}(t,s,u)=\mc{M}^{+0+0}(u,s,t)\,,\\
		\mc{M}^{+-00}(s,t,u)&=\mc{M}^{+-00}(s,u,t)=\mc{M}^{+00-}(u,t,s)=\mc{M}^{+00-}(t,u,s)\nn
		&=\mc{M}^{+0-0}(t,s,u)=\mc{M}^{+0-0}(u,s,t)\,,\\
		\mc{M}^{0-++}(s,t,u)&=\mc{M}^{0-++}(s,u,t)=\mc{M}^{0++-}(u,t,s)=\mc{M}^{0++-}(t,u,s)\nn
		&=\mc{M}^{0+-+}(t,s,u)=\mc{M}^{0+-+}(u,s,t)\,,\\
	\label{onecrossS4}	
		\mc{M}^{++--}(s,t,u)&=\mc{M}^{++--}(s,u,t)=\mc{M}^{+--+}(u,t,s)=\mc{M}^{+--+}(t,u,s)\nn
		&=\mc{M}^{+-+-}(t,s,u)=\mc{M}^{+-+-}(u,s,t)\,.
\eal
Note that for particles with spin the crossing symmetry is generally highly non-trivial except for the massless case we are considering. We see that some of the above equalities are more appropriately called crossing {\it relations} rather than crossing {\it symmetries}, as they link different amplitudes rather than reflect symmetries within an amplitude. We shall adapt the standard terminology that crossing symmetry refers to the collection of all crossing symmetries and relations. The amplitudes with the remaining helicities are not independent and can be obtained by using the relation $\mc{M}^{\overline{\mathbb{1234}}}(s,t,u)=(\mc{M}^{\mathbb{1234}}(s^*,t^*,u^*))^*$. So we will only need to use the dispersion relations for the amplitudes above in Eqs.~(\ref{fullcrossS1}-\ref{onecrossS4}) to constrain the Wilson coefficients.

By an explicit computation of the amplitudes from Lagrangian (\ref{action0}) with Feynman diagrams, we find that to the lowest orders the $g^i_{a,b}$ coefficients above are related to the Lagrangian Wilson coefficients as follows
\bal
\label{Mwcstart}
\mc{M}^{0000} ~~&=\li_3^2\(\f1s+\f1t+\f1u\)-\li_4+\frac{1}{M_P^2} \(\frac{su}{t}+\frac{st}{u}+\frac{ut}{s}\)\nn
&~~~~+\ai(s^2+t^2+u^2)+\gi_4 stu +...
\\
\mc {M}^{++--} & = \frac {1} {M_P^2}\frac{s^3}{tu} - \frac {\beta_1^2} {M_P^4} s^3 + \frac {\gi_0^2} {M_P^6} s^3 tu +...
\\
\mc {M}^{+++-} & = \frac {\gi_ 0} {M_P^4} stu +...
\\
\mc {M}^{++++} & = \(\frac {10\gi_ 0} {M_P^4}  - \frac{3\beta_1^2}{M_P^4}\) stu + \frac{\gi_0^2}{M_P^6} stu (s^2 + t^2 + u^2) +...
\\
\mc {M}^{+++0} & = \f{\gi_1}{M_P^3}stu +...
\\
\mc {M}^{++0-} & = \frac {\beta_1} {M_P^3} s^2 - \frac {\gi_0\beta_1} {M_P^5} s^2 tu +...
\\
\mc {M}^{++00} & =\f{\li_3 \bi_1}{M_P^2}s+\f{\bi_2}{M_P^2}s^2+\frac {\gi_0} {M_P^4} stu + \frac {\beta_1^2+\gi_2 M_P^2}{M_P^4} s^3 +...
\\
\mc {M}^{+-00} & = \frac {1} {M_P^2}\frac {tu} {s} + \frac {\beta_1^2} {M_P^4} stu +...
\\
\label{Mwcfinal}
\mc {M}^{+000} & = \frac {\beta_1} {2 M_P^3} (s^2 + t^2 + u^2)+\f{\gi_3}{M_P}stu +...
\eal

\section{Dispersive sum rules}
\label{sec:disprules}

In constructing the EFT Lagrangian or parameterizing the EFT scattering amplitudes in the last section, it would seem that the Wilson coefficients are allowed to take arbitrary values. The existence of causality/positivity bounds suggests that this would be an approach that sometimes leads to erroneous results. In particular, the consistency of the UV physics can actually impart many constraints on these EFT couplings. These UV consistency conditions include fundamental principles of S-matrix theory such as causality and unitarity, and can be utilized in the form of a series of dispersive sum rules or dispersion relations. In this section, we shall derive these dispersion relations and discuss how to effectively use them for scalar-tensor theory.

\subsection{Dispersion relations}
\label{sec:pse}

Before introducing the dispersion relations, let us first briefly recall partial wave unitarity that will be used shortly.   General 2-to-2 amplitudes for particles with spin in the helicity basis can be decomposed into partial wave amplitudes in terms of the Wigner (small) d-matrices
\bal
\mc{M}^{\mathbb{1234}}(s,t,u) =16\pi\sum_{\ell} (2\ell+1)d^{\ell}_{h_{12},h_{43}}\bigg(1+\frac{2t}{s}\bigg)A^{\mathbb{1234}}_{\ell}(s) \,,
\label{MtodA}
\eal
where $A^{\mathbb{1234}}_{\ell}(s)\equiv A_{\ell}^{h_1h_2h_3h_4}(s)$ is the spin-$\ell$ partial wave amplitude and $d_{h_{12}, h_{43}}^{\ell}(z)$ is the Wigner (small) d-matrices with $h_{ij}\equiv h_i-h_j$ (see,{\it e.g.,} Appendix F of \cite{deRham:2017zjm} for properties of the Wigner d-matrices). Note that $A^{\mathbb{1234}}_{\ell}(s)$ is a function of $s$ only, while $\mc{M}^{\mathbb{1234}}(s,t,u)\equiv \mc{M}^{\mathbb{1234}}(s,t)$ is a function of $s$ and $t$ because of the constraint $s+t+u=0$. The argument of the Wigner d-matrix is $\cos\thi=1+{2t}/{s}$, where scattering angle $\thi$ is the angle between the physical momenta of particle 1 and 3. Since the angular momentum is conserved in a scattering, the S-matrix is block-diagonal for different spin-$\ell$, so unitarity of the $\mc{M}^{\mathbb{1234}}$ amplitudes implies that the partial wave amplitudes $A^{\mathbb{1234}}_{\ell}$ are also unitary. This means that we can split the absorptive part of $A^{\mathbb{1234}}_{\ell}(s)$ into
\begin{equation}
	{\rm Abs} \, A^{\mathbb{1234}}_{\ell}(s)=\sum_{X} c^{\mathbb{12} \to X}_{\ell,s} c^{*\bar{\mathbb{3}}\bar{\mathbb{4}}\to { X }}_{\ell,s}\,,
	\label{gotAcc}
\end{equation}
where the sum over $X$ is for a complete basis of the Hilbert space, $c^{\mathbb{12} \to X}_{\ell,s}$ denotes the partial wave amplitude from particle 1 and 2 to the intermediate state $X$ with center of mass energy $s$, and $c^{*\bar{\mathbb{3}}\bar{\mathbb{4}}\to { X }}_{\ell,s} \equiv (c^{\bar{\mathbb{3}}\bar{\mathbb{4}}\to { X }}_{\ell,s} )^*$ with $\bar{\mathbb{3}}$ and $\bar{\mathbb{4}}$ denoting that particle 3 and 4 carry helicity $-h_3$ and $-h_4$ respectively. The reason for the extra minus signs for the helicities of particle 3 and 4 is that we are using the convention where all external particles are in-going. The absorptive part of $A^{\mathbb{1234}}_{\ell}$ is defined as
\be
\label{absEqdisc}
{\rm Abs} \, A^{\mathbb{1234}}_{\ell}(s) \equiv  \f{1}{2i} \( A^{\mathbb{1234}}_{\ell}(s+i\epi)  - (A^{\bar{\mathbb{3}} \bar{\mathbb{4}}\bar{\mathbb{1}}\bar{\mathbb{2}}}_{\ell}(s+i\epi))^* \)
={\rm Disc}A^{\mathbb{1234}}_{\ell}(s)  \,,
\ee
where the last equality is because the S-matrix is Hermitian analytic $(A^{\bar{\mathbb{3}} \bar{\mathbb{4}}\bar{\mathbb{1}}\bar{\mathbb{2}}}_{\ell}(s+i\epi))^*=A^{\mathbb{1234}}_{\ell}(s-i\epi)$.
For a time reversal invariant theory, as we are focusing on in this paper, we have $A^{\mathbb{\bar3\bar4\bar1\bar2}}_{\ell}(s+i\epi)=A^{\mathbb{1234}}_{\ell}(s+i\epi)$, in which case the absorptive part is simply the imaginary part: ${\rm A bs} \, A^{\mathbb{1234}}_{\ell}(s)={\rm Im} \, A^{\mathbb{1234}}_{\ell}(s)$.

Now, let us derive the dispersion relations we will use later. The most important ingredient in deriving the dispersion relations is the analyticity of the amplitudes when $s$ is analytically continued to the complex plane. While analyticity has not been rigorously proven, it is conjectured to be implied by causality of the UV theory (see \cite{deRham:2017zjm} for a brief account and \cite{Mizera:2021ujs} for a recent discussion), justifying the name of causality bounds, and we shall take it as a fundamental assumption. More precisely, we will make use of the analyticity condition that for fixed $t$ the amplitude $\mc{M}^{\mathbb{1234}}(s,t)$ is analytic in the complex $s$ plane except for singularities on the real $s$ axis that can be readily inferred from unitarity. Additionally, we shall assume that our EFT is weakly coupled in the IR and take the leading tree level approximation below the EFT cutoff $\Lambda$. This means that we can take the approximation that the amplitudes do not have branch cuts on the real $s$ axis in the low energy EFT region. That is, when $-t-\Lambda^2<s<\Lambda^2$, the only singularities in the low energy amplitude $\mc{M}^{\mathbb{1234}}$ are poles from exchange diagrams calculable within the EFT. Beyond the cutoff, unknown UV poles and branch cuts can appear.

\begin{figure}[tbp]
	\centering
	\includegraphics[scale=0.28]{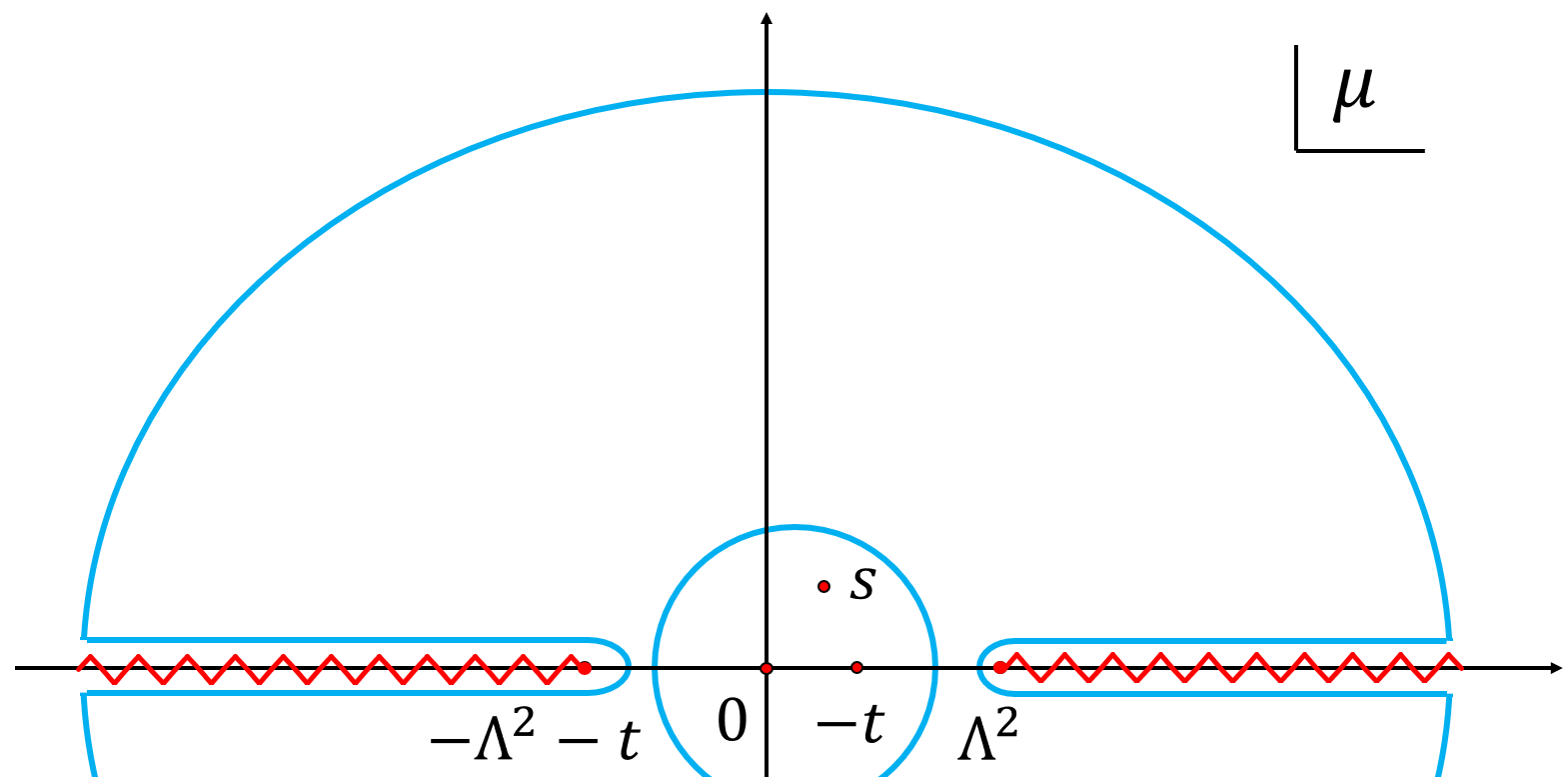}
	\caption{Analytic structure of ${\mc{M}^{\mathbb{1234}}(\mu,t)}/({\mu-s})$ in the complex $\mu$ (center-of-mass energy squared) plane. The pole at $\mu=s$ and the low energy poles $\mc{M}^{\mathbb{1234}}(\mu,t)$ are referred to as the ``EFT poles'', with $\Lambda$ being the EFT cutoff. The small (closed) contour is valid within the EFT, while the equivalent big (closed) contour encodes the UV information.}
	\label{fig:analyticity}
\end{figure}

Then we can look at the quantity ${\mc{M}^{\mathbb{1234}}(\mu,t)}/(\mu-s)$ in the complex $\mu$ plane for fixed $s$ and $t$ which are chosen to be in the EFT region $|s|<\Lambda^2, 0\leq -t<\Lambda^2$ . The analytic structure of this quantity is shown in Figure \ref{fig:analyticity}, which allows us to perform the contour integration as depicted. Due to analyticity, the integration with the small (closed) contour, which is valid in the EFT, is equivalent to the big (closed) contour that goes around the UV branch cut and the infinity. We will refer to the pole at $\mu=s$ as well as other low energy poles of ${\mc{M}^{\mathbb{1234}}(\mu,t)}$ as the ``EFT poles''. For massless scalar-tensor theory we are considering, the only low energy poles of ${\mc{M}^{\mathbb{1234}}(\mu,t)}$ for fixed $t$ are at $\mu=0$ and $\mu=-t$.  By the residue theorem, the big contour integral gives rise to
\bal
\label{4.1.3}
	\sum_{\rm EFT\,poles}\!\! {\rm Res}\frac{\mc{M}^{\mathbb{1234}}(\mu,t)}{\mu-s} &=\int_{\Lambda^2}^{+\infty}\frac{\d\mu}{\pi}\frac{{\rm Abs}\mc{M}^{\mathbb{1234}}(\mu,t)}{\mu -s} +\int_{-\infty}^{-t-\Lambda^2}\frac{\d\mu}{\pi}\frac{{\rm Abs}\mc{M}^{\mathbb{1234}}(\mu,t)}{\mu -s} \nn
	& ~~~~ +\int_{C_{\infty}^{\pm}} \frac{\d\mu}{2\pi i}\frac{\mc{M}^{\mathbb{1234}}(\mu,t)}{\mu-s}\,,
\eal
where we have made use of \eref{absEqdisc} and $C_{\infty}^{\pm}$ denotes the upper and lower semi-circles at infinity. The second term on the right hand side can be written in a form similar to the first term by the $su$ crossing of the amplitude and a change of the integration variable, so we get
\bal
	\sum_{\rm EFT\,poles}\!\!{\rm Res}\frac{\mc{M}^{\mathbb{1234}}(\mu,t)}{\mu-s}&=\int_{\Lambda^2}^{+\infty} \frac{\d\mu}{\pi}\bigg(\frac{{\rm Abs}\mc{M}^{\mathbb{1234}}(\mu,t)}{\mu -s}+\frac{{\rm Abs}\mc{M}^{\mathbb{1432}}(\mu,t)}{\mu -u}\bigg)
	\nn
	&~~~~ +\int_{C_{\infty}^{\pm}}  \frac{\d\mu}{2\pi i}\frac{\mc{M}^{\mathbb{1234}}(\mu,t)}{\mu-s}\,.
	\label{disprel1}
\eal
The aforementioned equation in its current form is not particularly useful, as the two integrals on the right-hand side may not converge due to the UV behavior of the amplitude. Typically, in order to respect locality, momentum space scattering amplitudes are polynomially bounded in terms of the Mandelstam variables so that Fourier transforms to real space amplitudes are well-defined. However, the case for a theory with the massless graviton can be more delicate, as will be discussed shortly. Nevertheless, we shall assume that the UV theory is polynomially bounded such that for fixed $t$ we have
\begin{equation}
	\lim_{|s|\to \infty}\mc{M}(s,t)/s^N = 0 \,,
	\label{MstotheN}
\end{equation}
where $N$ is a positive integer that depends on the value of $t$, as will be explained shortly. To render \eref{disprel1} useful, the standard remedy is to make ``subtractions''. For an $N$ subtraction, we can simply utilize the following algebraic identity
\begin{equation}
	\frac{{\rm Abs}\mc{M}(\mu,t)}{\mu-s}=\sum^N_{i=0}\binom{N}{i}\frac{(s-\mu_p)^{N-i}(\mu-s)^{i}}{(\mu-\mu_p)^N}\frac{{\rm Abs}\mc{M}(\mu,t)}{\mu-s} \,,
	\label{4.1.5}
\end{equation}
where $\mu_p$ is the subtraction point that can be arbitrarily chosen and $\binom{N}{i}\equiv N!/(i!(N-i)!)$. Notice that, except for the $i=0$ term, all the other terms in \eref{4.1.5} are just $(N-1)$-th degree polynomials of $s$. Since the left hand side of \eref{disprel1} is finite except for $t=0$, the divergences on the right hand must cancel. So all the $i\neq 0$ terms on the right hand side \eref{4.1.5} must group into an $(N-1)$-th-degree polynomial of $s$ whose coefficients are finite functions of $t$, while the $i=0$ term converges thanks to the high energy bound (\ref{MstotheN}). Thus, \eref{disprel1} can be re-written as an $N$-th subtracted dispersion relation:
\bal
\label{4.1.6}
		\sum_{\rm EFT\,poles}{\rm Res}&\frac{\mc{M}^{\mathbb{1234}}(\mu,t)}{\mu-s}=\sum_{m=0}^{N-1}b_{(N)m}^{\mathbb{1234}}(t)s^m
\\
	& \hskip 40pt	+\int_{\Lambda^2}^{+\infty}\frac{\d\mu}{\pi}\bigg(\frac{(s-\mu_s)^N}{(\mu-\mu_s)^N}\frac{{\rm Abs}\mc{M}^{\mathbb{1234}}(\mu,t)}{\mu -s}+\frac{(u-\mu_u)^N}{(\mu-\mu_u)^N}\frac{{\rm Abs}\mc{M}^{\mathbb{1432}}(\mu,t)}{\mu -u}\bigg) \,, \nonumber
\eal
where we have allowed the $s$ and $u$ channel subtraction points $\mu_{s}$ and $\mu_{u}$ to be different. Then, by the partial wave expansion (\ref{MtodA}) and the generalized optical theorem for the partial waves (\ref{gotAcc}), we can get
\bal
	\label{4.1.7}
		\sum_{\rm EFT\,poles}{\rm Res}\frac{\mc{M}^{\mathbb{1234}}(\mu,t)}{\mu-s}&= \sum_{m=0}^{N-1}b_{(N)m}^{\mathbb{1234}}(t)s^m
		\\
		&~~~~ +\bigg \langle \frac{(s-\mu_s)^N}{(\mu-\mu_s)^N}\frac{d^{\ell,\mu,t}_{h_{12},h_{43}}c^{\mathbb{12}}_{\ell,\mu}c^{*\bar{\mathbb{3}}\bar{\mathbb{4}}}_{\ell,\mu}}{\mu -s}+\frac{(u-\mu_u)^N}{(\mu-\mu_u)^N}\frac{{d^{\ell,\mu,t}_{h_{14},h_{23}}c^{\mathbb{14}}_{\ell,\mu}c^{*\bar{\mathbb3}\bar{\mathbb2}}_{\ell,\mu}}}{\mu -u}\bigg \rangle \,, \nonumber
\eal
where we have defined the shorthands
\be
\label{shortHandn}
\Big \langle \cdots \Big \rangle := 16\pi\sum_{\ell,X}(2\ell+1)\int_{\Lambda^2}^{\infty}\frac{\d\mu}{\pi}(\cdots) \,,~
c^{\mathbb{12}}_{\ell,s} := c^{\mathbb{12} \to X}_{\ell,s}\,,~
d^{\ell,\mu,t}_{h_{12},h_{43}}:= d^{\ell}_{h_{12},h_{43}}\(1+\frac{2t}{\mu}\) .
\ee
Note that each of the dispersion relations is actually a one-parameter family of relations parametrized by the momentum transfer $t$.

To determine the number of subtractions $N$, we need to have a better understanding of the Regge behavior of the amplitudes. Let us recall that for a non-gravitational massive field theory, the rigorous results of Froissart \cite{Froissart:1961ux} and Martin \cite{Martin:1962rt} suggest that two subtractions are sufficient: $\lim_{|s| \to \infty}\mc{M}(s, t)/s^2=0$ for a range of physical $t\leq 0$ and even for a range of non-physical $t>0$. For massless fields, especially when gravitons are included in the low energy spectrum, it is more subtle, not the least for the presence of the spin-2 $t$ channel pole. Generically, one expects that for a gravitational theory the Regge behavior of the amplitude may change for different fixed $t$ (see, {\it e.g.}, \cite{Alberte:2021dnj, Herrero-Valea:2022lfd})
\begin{equation}
\label{4.2.2}
\begin{dcases}
\lim_{|s|\to \infty}\mc{M}(s, t)/s^2 =0 \,, & t<0 \,,\\
 \lim_{|s|\to \infty}\mc{M}(s, t)/s^3 =0 \,, & 0\leq t \leq \xi \,,
\end{dcases}
\end{equation}
where $\xi$ is a small positive number. While string theory gives rise to this behavior, it is believed to be generically valid for a theory with a spin-2 $t$-channel pole. Although the original Froissart bound does not apply for massless particles, twice subtracted dispersion relations in the physical region $t<0$ is implied at least in the weak coupling limit by causality considerations for impact parameter amplitudes \cite{Alberte:2021dnj}. In any case, we shall assume that twice subtractions are sufficient for $t<0$. Then, from twice-subtracted dispersion relations, say, $\mc{M}^{++--}$, in the $t\to 0^{-}$ limit
\begin{equation}
	 \lim_{t\to 0^{-}}\bigg( \int_{\Lambda^2}^{\infty} \frac{d\mu}{\pi}\frac{\text{Disc}\mc{M}^{++--}(s,t)}{\mu^2 (\mu-s)}+s\leftrightarrow u \bigg)\sim -\frac{1}{t}\,,\nonumber
\end{equation}
we can infer that the dispersive integral on the left hand side must diverge as $t\to 0^{-}$, since the integrand does not give rise to any negative power of $t$. 
However, a thrice subtraction eliminates the spin-2 $t$-channel pole $s^2/t$, and therefore, we have $\lim_{|s|\to \infty}\mc{M}(s, t)/s^3 =0$ for $0\leq t \leq \xi$. In this paper, we shall simply assume the Regge bounds of Eq.~(\ref{4.2.2}) to hold. Since we will use the dispersion relations for the range of $t\leq 0$, $N$ is chosen to be 2 for $t<0$ and 3 for $t\leq 0$.

Therefore, for $t<0$, choosing $\mu_s=\mu_u=0$,  we can get twice subtracted dispersion relations
\begin{equation}\label{4.3.1}
		\sum_{{\rm EFT\,poles}}\!\!\!\!\!{\rm Res}\frac{\mc{M}^{\mathbb{1234}}(\mu,t)}{\mu-s}=
		b_{(2)0}^{\mathbb{1234}}(t)+b_{(2)1}^{\mathbb{1234}}(t)s+\!\bigg \langle  \!\frac{s^2d^{\ell,\mu,t}_{h_{12},h_{43}}c^{\mathbb{12}}_{\ell,\mu}c^{*\bar{\mathbb{3}}\bar{\mathbb{4}}}_{\ell,\mu}}{\mu^2(\mu -s)}+\frac{{u^2 d^{\ell,\mu,t}_{h_{14},h_{23}}c^{\mathbb{14}}_{\ell,\mu}c^{*\bar{\mathbb3}\bar{\mathbb2}}_{\ell,\mu}}}{\mu^2(\mu -u)} \bigg \rangle  .
\end{equation}
For a $su$-symmetric amplitude, we additionally have $b^{\mathbb{1234}}_{(2)1}(t)=0$. Later, we will also use thrice subtracted dispersive relations at $t=0$, which helps impose the $st$ crossing symmetry of the amplitudes to get more useful dispersion relations. The use of forward-limit dispersive relations also helps harvest effective constraints numerically in the finite $\mu$ and large $\ell$ region. A remarkable feature of the dispersion relations (\ref{4.3.1}) is that they link the EFT couplings in the IR (on the left hand side) to the unknown UV behaviors of the amplitudes (on the right hand side) via dispersive integrals. To see this more clearly, let us parametrize the residues of the EFT poles on the left hand side of \eref{4.3.1}  as follows
\begin{equation}\label{4.3.2}
\sum_{{\rm EFT\,poles}}{\rm Res}\frac{\mc{M}^{\mathbb{1234}}(\mu, t)}{\mu-s}=a_{2,-1}^{\mathbb{1234}}\frac{s^2}{t}+\sum_{k,n\ge0}a_{k,n}^{\mathbb{1234}}s^kt^n \,.
\end{equation}
The $a^{\mathbb{1234}}_{k, n}$ coefficients can be easily expressed in terms of the independent $g$ coefficients introduced in Eqs.~(\ref{gstartdef}-\ref{gfinaldef}) or in terms of the Lagrangian Wilson coefficients via Eqs.~(\ref{Mwcstart}-\ref{Mwcfinal}). For a particular EFT amplitude, some of the $a^{\mathbb{1234}}$ coefficients can vanish. The term ${s^2}/{t}$ comes from a $t$-channel exchange of the massless graviton. This prevents a Taylor expansion in terms of $t$ in the forward limit $t=0$ for the two sides of these dispersion relations. For some of the twice-subtracted dispersion relations that do not contain $t$-channel pole, this pathology also manifests as the fact that the expansions at $t=0$ on the two sides can not be matched without imposing unphysical restrictions on the Wilson coefficients. {(For the twice-subtracted dispersion relations listed in Appendix \ref{sec:expSum}, those of $\mc{M}^{0000}$, $\mc{M}^{+0-0}$ and $\mc{M}^{++--}$ contain the $s^2/t$ pole, while expanding those of $\mc{M}^{+000}$, $\mc{M}^{+++-}$, $\mc{M}^{+0+0}$, $\mc{M}^{+-00}$, $\mc{M}^{0-++}$ and $\mc{M}^{0+-+}$ will impose unphysical constraints on the Wilson coefficients.) For example, if we expand the right hand side of the  $\mc{M}^{+000}$ dispersion relation around $t=0$, the series of $t$ within $\langle ~ \rangle$ begins with $t^2$ because of the structure of $d^{\ell}_{2,0}$, which implies that the coefficient of the $st$ term on the left hand side must be zero, {\it i.e.}, $\beta_1=0$. This clearly is an unphysical constraint, meaning that it is invalid to expand around $t=0$ even for those dispersion relations. Even if the two sides of a twice-subtracted dispersion relation could be matched for the expansion around $t=0$, we might still not use its forward limit simply because of the Regge behavior Eq.~(\ref{4.2.2}) of the amplitude. Nevertheless, since $\mc{M}^{\mathbb{1234}}(\mu,t)$ only contains simple poles in the EFT region, the left hand side of \eref{4.3.1} is analytic around $s=0$, as shown explicitly in \eref{4.3.2}. We can Taylor-expand both sides of \eref{4.3.1} in the neighborhood of $s=0$, and matching coefficients of $s^k$ gives
\begin{equation}
\label{4.3.3}
\delta_{k, 2}a^{\mathbb{1234}}_{k, -1}\frac{1}{t}+\!\sum_{n=0}a^{\mathbb{1234}}_{k,n}t^n= \bigg \langle  \frac{\partial_s^k}{k!}\bigg[\frac{s^2d^{\ell,\mu,t}_{h_{12}, h_{43}}c^{\mathbb{12}}_{\ell, \mu}c^{*\bar{\mathbb{3}}\bar{\mathbb{4}}}_{\ell, \mu}}{\mu^2(\mu-s)}+\frac{(-s-t)^2d^{\ell,\mu,t}_{h_{14}, h_{23}}c^{\mathbb{14}}_{\ell, \mu}c^{*, \bar{\mathbb3}\bar{\mathbb2}}_{\ell, \mu}}{\mu^2(\mu +s+t)}\bigg]\bigg|_{s\to 0} \bigg \rangle \,,
\end{equation}
which for fixed $k$ and $n$ is a one-parameter ($t$) family of sum rules. If $\mc{M}^{\mathbb{1234}}(s,t)$ is $su$-symmetric, \eref{4.3.3} is valid for $k\geq 1$, because in this case we have $b^{\mathbb{1234}}_{(2)1}(t)=0$; if $\mc{M}^{\mathbb{1234}}$ is not $su$-symmetric, \eref{4.3.3} is valid for $k\geq 2$, remembering that $b^{\mathbb{1234}}_{(2)1}(t)$ is then generically nonzero and unknown. That is, for $su$-symmetric amplitudes, we have some extra sum rules. These extra low order sum rules are constraining in bounding the Wilson coefficients, so it is important to make use of them effectively.

\subsection{Imposing $st$ crossing symmetry}
\label{sec:ids}

\begin{figure}[http]
 \centering
 \includegraphics[scale=0.19]{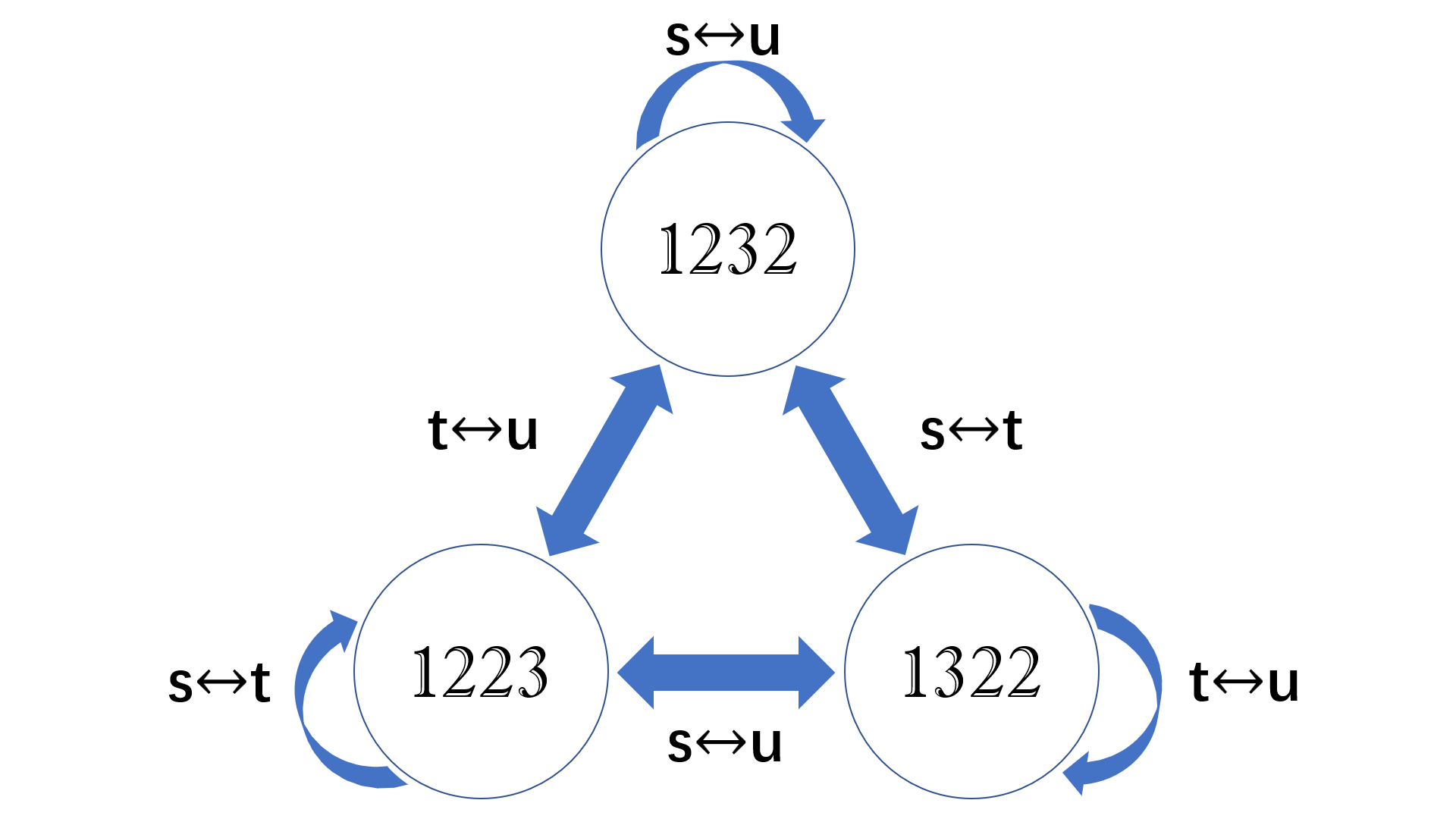}
 \caption{Crossing relations for the amplitudes with only one crossing symmetry. The $su$ crossing symmetry for $\mc{M}^{\mathbb{1234}}(s,t,u)$ is built-in in our dispersion relations.}
 \label{fig:crossing}
\end{figure}

In deriving the sum rules (\ref{4.3.3}), we have already used the $su$ crossing symmetry of the amplitudes. However, that is not the full crossing symmetry that the amplitudes have. We also have the $st$ crossing symmetry, whose information is not contained in the sum rules (\ref{4.3.3}). It has been realized recently that imposing the $st$ crossing symmetry on the $su$ dispersion relations is very potent in improving causality bounds on the Wilson coefficients \cite{Tolley:2020gtv,Caron-Huot:2020cmc}.

As an aside, note that in the absence of gravitational interactions, dispersion relations can be expanded in the forward limit as well as around $s=0$, and one can express individual amplitude coefficients in terms of UV dispersive integrals. In that case, the $st$ crossing symmetry directly links different amplitude coefficients, giving rise to vanishing dispersive integrals, known as null constraints. For a theory with multiple degrees of freedom, the coefficient sum rules and the null constraints can be combined to define a SDP with one continuous decision variable \cite{Du:2021byy}, solvable by the powerful ${\tt SDPB}$ package. In the presence of the massless graviton, the expansion in the forward limit is invalid, and we need to be content with sum rules where the left hand sides generally contain the momentum transfer $t$. This will also be usually true after imposing the $st$ crossing symmetry, as shown in Appendix \ref{sec:expSum}.

The $tu$ crossing symmetry is implied by the $su$ plus $st$ crossing symmetry, so we do not need additionally impose the $tu$ crossing. Let us see how to implement this concretely in our case. In the massless scalar-tensor theory, there are two kinds of amplitudes: the ones that are fully $stu$ symmetric, whose crossing symmetries have been listed in Eqs.~(\ref{fullcrossS1}-\ref{fullcrossS5}), and the ones with only one of the $su$, $st$ and $tu$ symmetries, whose crossing symmetries and relations have been listed in Eqs.~(\ref{onecrossS1}-\ref{onecrossS4}). For the fully crossing symmetric cases, after imposing the $st$ crossing symmetry, we can easily see that the $tu$ crossing symmetry is redundant. For the amplitudes with only one crossing symmetry, there are three different types: $\mc{M}^{\mathbb{1232}}$, $\mc{M}^{\mathbb{1322}}$ and $\mc{M}^{\mathbb{1223}}$. Crossing then either maps one amplitude into itself or into anther amplitude, see Figure \ref{fig:crossing}. Again, since we have used the $su$ crossing symmetry, it is sufficient to impose the $st$ crossing symmetry, $\mc{M}^{\mathbb{1232}}(s,t,u)=\mc{M}^{\mathbb{1322}}(t,s,u)$, to extract the full crossing information. We would like to remind the reader that we use the terminology that crossing symmetry refers to the collection of the crossing symmetries that map one amplitude to itself and crossing relations that map one amplitude to another.

To impose the $st$ crossing symmetry, we first note that the amplitudes with full $stu$ symmetry separate into 5 groups and the amplitudes with only $su$, $st$ or $ut$ crossing symmetry separate into 4 groups. The $st$ crossing relations are imposed separately for each of these groups, which can be done by equating the following EFT coefficients in the expansion (\ref{4.3.2}):
\begin{equation}
	\label{4.3.6}
	a^{\mathbb{1234}}_{k,n}=a^{\mathbb{1324}}_{n,k}\,,~~n\geq 3 \,,
\end{equation}
where $k\geq 1$ if $\mc{M}^{\mathbb{1234}}$ is $su$-symmetric in the narrow sense and $k\geq 2$ if $\mc{M}^{\mathbb{1234}}$ is not $su$-symmetric.

Later, for technical reasons, we shall try to access dispersion relations when $t$ is close to the cutoff $-\Lambda^2$, for which \eref{4.3.3} is not suitable. This is simply because the left hand side of \eref{4.3.3} contains an infinite number of powers of $t$, which all become important when $t$ approaches $-\Lambda^2$. However, this can be overcome by combining different dispersion relations. To this end, we shall also make use of thrice subtracted dispersion relations. In \eref{4.1.7},  we can choose the subtraction points to be $\mu_s=0$ and $\mu_u=-t$, and get
\begin{equation}
	\label{4.3.7}
				\sum_{{\rm EFT\,poles}}\!\!\!\!{\rm Res}\frac{\mc{M}^{\mathbb{1324}}(\mu,t)}{\mu-s}=
				\sum^2_{m=0}b_{(3)m}^{\mathbb{1324}}(t)s^m \!+\! \bigg \langle  \frac{s^3d^{\ell,\mu,t}_{h_{13},h_{42}}c^{\mathbb{13}}_{\ell,\mu}c^{*\bar{\mathbb2}\bar{\mathbb4}}_{\ell,\mu}}{\mu^3(\mu -s)}+\frac{{(-s)^3 d^{\ell,\mu,t}_{h_{14},h_{32}}c^{\mathbb{14}}_{\ell,\mu}c^{*\bar{\mathbb2}\bar{\mathbb3}}_{\ell,\mu}}}{(\mu+t)^3(\mu -u)} \bigg \rangle \,.
\end{equation}
Since these are thrice subtracted dispersive sum rules, which are free of the $t$-channel pole issue, we can then express both sides of \eref{4.3.7} as a Taylor series of $t$ and match the expansion coefficients. The choice of $\mu_s=0$ and $\mu_u=-t$ makes sure that the part within $\langle...\rangle$ only contains terms with $s^{3}$ and higher orders. This leads to
\bal
		\sum_{k=3}a^{\mathbb{1324}}_{k,n}s^k=
		\bigg\langle  \frac{\partial_t^n}{n!} \bigg(
			\frac{s^3d^{\ell,\mu,t}_{h_{13},h_{42}}c^{\mathbb{13}}_{\ell,\mu}c^{*\bar{\mathbb2}\bar{\mathbb4}}_{\ell,\mu}}{\mu^3(\mu-s)}+
			\frac{(-s)^3d^{\ell,\mu,t}_{h_{14},h_{32}}c^{\mathbb{14}}_{\ell,\mu}c^{*\bar{\mathbb2}\bar{\mathbb3}}_{\ell,\mu}}{(\mu+t)^3(\mu+s+t)}
		\bigg)\bigg|_{t\to 0} \bigg \rangle \,,
	\label{4.3.8}	
\eal
where $n\geq 0$. Then, we can relabel $s$ as $t$ in \eref{4.3.8}, and subtract \eref{4.3.3} with this $s$ and $t$ swapped equation. This gives the final $st$ crossing imposed sum rules that we will use in a SDP problem to get the causality bounds
\be
\label{mainsumRule}
\delta_{k,2}a^{\mathbb{1234}}_{2,-1}\frac{1}{t}+a^{\mathbb{1234}}_{k,0}+a^{\mathbb{1234}}_{k,1}t+a^{\mathbb{1234}}_{k,2}t^2
=\bigg \langle F^{\mathbb{1234}}_{k,\ell}(\mu,t) \bigg \rangle \,,
\ee
with $F^{\mathbb{1234}}_{k,\ell}(\mu,t)$ defined as
\bal
\label{4.3.9}
		F^{\mathbb{1234}}_{k,\ell}(\mu,t)  &= \frac{\partial_s^k}{k!}\bigg(\frac{s^2}{\mu^2(\mu-s)}d^{\ell,\mu,t}_{h_{12}, h_{43}}c^{\mathbb{12}}_{\ell, \mu}c^{*\bar{\mathbb{3}}\bar{\mathbb{4}}}_{\ell, \mu}+\frac{(-s-t)^2}{\mu^2(\mu +s+t)}d^{\ell,\mu,t}_{h_{14}, h_{23}}c^{\mathbb{14}}_{\ell, \mu}c^{*, \bar{\mathbb3}\bar{\mathbb2}}_{\ell, \mu}\bigg)\bigg|_{s\to 0}
		\\
		&~~~ -\frac{\partial_t^k}{k!} \bigg(
		\frac{s^3}{\mu^3(\mu-s)}d^{\ell,\mu,t}_{h_{13},h_{42}}c^{\mathbb{13}}_{\ell,\mu}c^{*\bar{\mathbb2}\bar{\mathbb4}}_{\ell,\mu}+
		\frac{(-s)^3}{(\mu+t)^3(\mu+s+t)}d^{\ell,\mu,t}_{h_{14},h_{32}} c^{\mathbb{14}}_{\ell,\mu}c^{*\bar{\mathbb2}\bar{\mathbb3}}_{\ell,\mu}
		\bigg)\bigg|_{t\to 0,s\to t} \,,  \nonumber	
\eal
where we have used the $st$ crossing symmetry $a^{\mathbb{1234}}_{k,n}=a^{\mathbb{1324}}_{n,k}$ to cancel all the $t^{n}$ terms with $n\geq 3$ and $k\geq 1$ if $\mc{M}^{\mathbb{1234}}$ is $su$-symmetric and $k\geq 2$ if $\mc{M}^{\mathbb{1234}}$ is not $su$-symmetric. These sum rules are under control even if $t$ is close to $-\Lambda^2$. These explicit independent sum rules are listed in Appendix \ref{sec:expSum}.

A few comments are in order. In \eref{4.3.6}, we have only imposed $st$ crossing relations $a^{\mathbb{1234}}_{k,n}=a^{\mathbb{1324}}_{n,k}$ for $n\geq 3$. In principle, we could also impose the condition $a^{\mathbb{1234}}_{2,1}=a^{\mathbb{1324}}_{1,2}$. However, for an $stu$ symmetric amplitude, this is redundant, because we have already enforced $a^{\mathbb{1222}}_{2,1}=a^{\mathbb{1222}}_{1,2}$ when deriving the dispersion relation with the $su$ crossing symmetry --- the only $su$-symmetric terms at that order are $stu$ and $t^3$. For an amplitude with only one crossing symmetry, the crossing relation $a^{\mathbb{1232}}_{1,2}=a^{\mathbb{1322}}_{2,1}$ does provide some new information. However, since we will for our convenience use both the sum rules involving $a^{\mathbb{1232}}_{1,2}$ and $a^{\mathbb{1322}}_{2,1}$, it is equivalent to imposing crossing relation $a^{\mathbb{1232}}_{1,2}=a^{\mathbb{1322}}_{2,1}$. Using two different expressions for one Wilson coefficient is the same as using one expression for the coefficient plus one $st$ crossing relation.

Note that sometimes the requirement of $a^{\mathbb{1234}}_{k,n}=a^{\mathbb{1324}}_{n,k}$ for an amplitude with $stu$ symmetry can be redundant, since the $st$ symmetry is occasionally guaranteed by the $su$ symmetry already. To find redundant relations at the $N$-th order ($N=k+n$), we can first expand an amplitude at the $N$-th order as $\mc{M}_N= \sum_{m=0}^{\lfloor N/2\rfloor} q_m (su)^m t^{N-2m}$, where $\lfloor~ \rfloor$ denotes taking the flooring integer. Further expanding $u$ as $-s-t$, we get $\mc{M}_N= \sum_{k=0}^{N} c_{k,n}s^k t^n$, which allows us to write $c_{k,n}$ in terms of $q_m$. Then, requiring $c_{k,n}=c_{n,k}$ gives a set of linear equations in terms of $q_m$, and the redundancy of the $st$ symmetry can be obtained by examining the linear dependence of these $q_m$ equations. Let us take the case of scalar scattering for an example, whose amplitude is $stu$ symmetric. When $k+n=3$, the $su$ symmetry requires that the terms of the amplitude must be $stu$ or $t^3$, which means that, without further imposing the $st$ symmetry, we can already have $a^{0000}_{2, 1}=a^{0000}_{1, 2}$. So in this case the $st$ symmetry is redundant. In fact, since the $st$ symmetry results in $\lfloor (N+1)/2\rfloor$ equations and there are only $\lfloor N/2\rfloor+1$ distinct values of $q_m$, redundancy always exists.
	
In principle, the sum rules in the form of \eref{mainsumRule} are all one needs to extract the strongest causality bounds in an ideal optimization scheme. However, to have a scheme that is numerically more tractable, we find that it is beneficial to add some forward-limit sum rules, as will be discussed in Section \ref{sec:numdet}. The forward-limit sum rules can be obtained from \eref{mainsumRule} by simply matching the coefficients in front of $t^n$ on both sides of the equation for the cases of $k\geq 3$:
\be
\begin{aligned}
\label{forlimSum}
a^{\mathbb{1234}}_{k,0}
=\bigg \langle F^{\mathbb{1234}}_{k,\ell}(\mu,0) \bigg \rangle \,,~~
a^{\mathbb{1234}}_{k,1}
&=\bigg \langle \pd_t F^{\mathbb{1234}}_{k,\ell}(\mu,0) \bigg \rangle \,,~~
a^{\mathbb{1234}}_{k,2}
=\f12 \bigg \langle \pd_t^2 F^{\mathbb{1234}}_{k,\ell}(\mu,0) \bigg \rangle \,,
\\
0&=\bigg \langle \pd_t^{n} F^{\mathbb{1234}}_{k,\ell}(\mu,0) \bigg \rangle \,,~~\text{ for $n\geq 3$}\,.
\end{aligned}
\ee

\section{Power counting via dispersion relations}
\label{sec:dimAna}

The dispersive sum rules we have derived can be used to constrain Wilson coefficients of the low energy EFT via an optimization procedure. Before doing that numerically in the next sections, we will see here that these sum rules can be used to do a dimensional analysis on the Wilson coefficients. That is, we will show how schematic estimates on the dimensions of the coefficients can be inspected from the structure of the dispersion relations.

Recall that in the absence of gravity the dimensional analysis of a scalar EFT is usually fairly simple. One just power-counts the mass dimension of an operator and suppresses it with appropriate powers of the cutoff:
\be
\label{Ophiscaling}
\widehat{\cal O}_\phi  \sim \Lambda^{4}\left[\frac{\pd}{\Lambda}\right]^{N_{\pd}}\left[\frac{\phi}{\Lambda}\right]^{N_{\phi}} \,,
\ee
where $N_{\pd}$ is the number of partial derivatives and $N_{\phi}$ is the number of $\phi$ fields in the operator.
A slightly more refined version of this analysis which takes care of loops and factors of $4\pi$, called naive dimensional analysis, can be extended to include spin-1 and spin-1/2 fields \cite{Gavela:2016bzc}. In the presence of gravity, an extra mass scale comes in at the (reduced) Planck mass $M_P= 1/\sqrt{8\pi G_N}$. Then, an important question is how many powers of $M_P$ there are in each of the Wilson coefficients. In the literature, there are a few seemingly plausible arguments supporting different scalings of the Wilson coefficients in terms of $M_P$. In the case of pure gravity that is weakly coupled in the IR, the numerical bounds from causality imply \cite{Caron-Huot:2022ugt} that the typical scalings for generic gravitational EFT operators are given by
\be
\widehat{\cal O}_R  \sim  {M_P^2\Li^2} \left[\frac{\nd}{\Lambda}\right]^{N_{\nd}} \left[ \f{R}{\Li^2} \right]^{N_R} \,,
\ee
where $N_{\nd}$ is the number of covariant derivatives, $R$ stands for a curvature tensor and $N_R$ is the number of curvature tensors. In the following, we shall argue that, in scalar-tensor theory, if the scaling of \eref{Ophiscaling} is recovered in the decoupling limit, the typical scalings of the EFT operators are given by
\be
\label{OphiRscale}
\widehat{\cal O}_{\phi R}  \sim  {M_P^2\Li^2} \left[\frac{\nd}{\Lambda}\right]^{N_{\nd}} \left[ \f{R}{\Li^2} \right]^{N_R} \left[ \f{\phi}{M_P} \right]^{N_\phi}  \[\f{M_P}{\Li}\]^{\tilde N_\phi} \,,
\ee
where the power of the enhancement factor ${\tilde N_\phi}$ can be determined by counting the number of $c^{00}_{\ell,\mu}$ in the most constraining sum rule available. For the lowest orders in \eref{action21}, it happens that $\tilde N_\phi=\lfloor N_\phi/2\rfloor$, where $\lfloor~ \rfloor$ denotes taking the flooring integer, but this has to be modified for higher orders (see Section \ref{sec:xcheck}). On the other hand, for the scenario where the scalar interactions are of the gravitational strength, a typical scalar-tensor operator then has the following scaling
\be\label{little_scale}
\widehat{\cal O}^{\rm st}_{\phi R}  \sim  {M_P^2\Li^2} \left[\frac{\nd}{\Lambda}\right]^{N_{\nd}} \left[ \f{R}{\Li^2} \right]^{N_R} \left[ \f{\phi}{M_P} \right]^{N_\phi} \,.
\ee
Of course, a caveat is that the above scalings have only been explicitly verified for EFT operators of the lowest orders with four fields in a weakly coupled EFT; see \eref{action21} and \eref{action22}.

To see how this schematic method works, we shall first use the sum rules without the $st$ crossing symmetry imposed, {\it i.e.,} \eref{4.3.3}, to infer the typical behaviors of the UV spectral functions $c^{\mathbb{12}}_{\ell,\mu}$. Let us first look at the $++--$ sum rule with $k=2$, which happens to be the same as sum rule (\ref{F22112}). That is, the $st$ crossing does not alter this sum rule. Its explicit form is given by
\be
\label{tMPdis0}
    -\frac{1}{M^2_{P}t}=  \sum_{\ell,X}16\pi(2\ell+1) \int_{\Lambda^2}^{\infty}\frac{\d\mu}{\pi}\bigg[\frac{d^{\ell,\mu,t}_{0,0}|c^{++}_{\ell,\mu}|^2}{\mu^3}+\frac{d^{\ell,\mu,t}_{4,4}|c^{+-}_{\ell,\mu}|^2}{(\mu+t)^3}\bigg] \,.
\ee
The left hand side comes from a $t$-channel exchange, and this sum rule is valid for a range of $|t|$ below the cutoff. When $|t|$ is small, the left hand side is large, which means that the integral over $\mu$ red and/or the sum on the right hand side converges very slowly. A quicker convergence can be achieved by choosing a large $|t|$, so for our estimates we shall choose $|t|\sim \Lambda^2$. Also, this choice does not introduce any extra scale that is not already in the problem. Introducing dimensionless variables $\hat t$ and $\hat \mu$ and normalized $\hat c^{\mathbb{12}}_{\ell,\mu}$:
\be
\hat t = \f{t}{\Li^2}\,,~~\hat \mu = \f{\mu}{\Li^2}\,,~~\hat c^{\mathbb 12}_{\ell,\mu} = \sqrt{16(2\ell+1)} c^{\mathbb 12}_{\ell,\mu}\,,
\ee
we get
\be
\label{tchannelscaled}
    \frac{\Lambda^2}{M^2_{P}}= -\hat t\sum_{\ell,X}\int_{1}^{\infty}\d\hat \mu\bigg[\frac{d^{\ell,\hat \mu,\hat t}_{0,0}|\hat c^{++}_{\ell,\mu}|^2}{\hat \mu^3}+\frac{d^{\ell,\hat\mu,\hat t}_{4,4}|\hat c^{+-}_{\ell,\mu}|^2}{(\hat \mu+\hat t)^3}\bigg] \,.
\ee
Since the quantities on the right hand side are mostly $\mc{O}(1)$ numerically except for $\hat c^{++}_{\ell,\mu}$ and $\hat c^{+-}_{\ell,\mu}$, this means that $\hat c^{++}_{\ell,\mu}$ and $\hat c^{+-}_{\ell,\mu}$ must behave appropriately to make the integral and summation converge to the left hand side. That is, the spectral functions $c^{++}_{\ell,\mu}$ and  $c^{+-}_{\ell,\mu}$ have to conspire to reproduce the hierarchy between $\Lambda$ and $M_P$ in the theory. Thus, we can schematically assign the following correspondences
\be
\label{LiMPscale}
\frac{\Li}{M_P}  \Leftrightarrow \hat c^{++}_{\ell,\mu} \,,~ \hat c^{+-}_{\ell,\mu}\,,~ \hat c^{-+}_{\ell,\mu}\,,~\hat c^{--}_{\ell,\mu} \,,
\ee
which can be used to estimate the sizes of the Wilson coefficients momentarily. Note that we have also added $\hat c^{-+}_{\ell,\mu}$ and $\hat c^{--}_{\ell,\mu}$ because they are related to $\hat c^{+-}_{\ell,\mu}$ and $\hat c^{++}_{\ell,\mu}$ by crossing or parity, and thus they must have the same scaling. In establishing the correspondences such as (\ref{LiMPscale}), the reason for not using the sum rules with the $st$ crossing symmetry imposed is obvious: the $st$ crossing introduces quantities that are cancelable among themselves. For example, the null sum rule (\ref{F21211}) would not tell us any scaling in terms of $\Li$ and $M_P$; it only tells us that there are intricate cancellation among the terms with $\hat c^{++}_{\ell,\mu}$,  $\hat c^{+-}_{\ell,\mu}$,  $\hat c^{-+}_{\ell,\mu}$ and $\hat c^{--}_{\ell,\mu}$. Similarly, even though the sum rule (\ref{F20201}) is not null on the left hand side, its right hand side contains terms that cancel among themselves, so it would be inappropriate to use it to estimate the behavior of $c^{\mathbb{12}}_{\ell,\mu}$.

With these established, we can estimate the sizes of the Wilson coefficients $\gi_0$ and $ \bi_1$ via the improved sum rules in Appendix \ref{sec:expSum}. { Specifically, we can expand \eref{F22113} around the forward limit and match the coefficients to get}
\bal
\label{alpha2scaled}
   \! -\frac{ \gi_0^2}{M^6_{P}}&\!=\!\f{1}{\Li^{10}}\!\sum_{\ell,X}\!\int_{1}^{\infty}\! \!\! \d\hat \mu \!\bigg[\frac{\pd_{\hat t}^2 d^{\ell,\hat\mu,0}_{0,0}|\hat c^{++}_{\ell,\mu}|^2\!}{2\hat \mu^4}
   \! -\!\frac{\pd_{\hat t}^2d^{\ell,\hat\mu,0}_{4,4}|\hat c^{+-}_{\ell,\mu}|^2\!}{2\hat \mu^4}
    \!+\!\frac{4\pd_{\hat t}d^{\ell,\hat\mu,0}_{4,4}|\hat c^{+-}_{\ell,\mu}|^2\!}{\hat \mu^5}
    \!-\!\frac{10d^{\ell,\hat\mu,0}_{4,4}|\hat c^{+-}_{\ell,\mu}|^2 }{\hat \mu^6} \bigg] \!,
\\
\label{beta2scaled}
    -\frac{\beta_1^2}{M^4_{P}}&=\f{1}{\Li^6}\sum_{\ell,X}\int_{1}^{\infty}\d\hat \mu \left[\frac{d^{\ell,\hat\mu,0}_{0,0}|\hat c^{++}_{\ell,\mu}|^2}{\hat \mu^4}
    -\frac{d^{\ell,\hat\mu,0}_{4,4}|\hat c^{+-}_{\ell,\mu}|^2}{\hat \mu^4}\right] \,.
\eal
Making use of the scaling correspondences (\ref{LiMPscale}), we can infer that the typical dimensional scaling of these two Wilson coefficients must be\,\footnote{By the typical scaling of, say, $\gi_0$, we mean that the upper bound of $|\gi_0|$ is around $|\gi_0| \sim {M_P^2}/{\Li^4}$.}
\be
\label{g0b0}
\gi_0 \sim \f{M_P^2}{\Li^4} \,,~~~\beta_1 \sim \f{M_P}{\Li^2}\,.
\ee
As we will see in Section \ref{sec:bounds}, this is consistent with the rigorous numerical results, that is, the upper limits of the causality bounds.

One caveat is in order. Since the sum rules in Appendix \ref{sec:expSum} are with the $st$ crossing symmetry imposed, sometimes a coefficient's dimensional scaling from one sum rule may differ from another. In this case, one should survey all available sum rules and take the smallest dimensional scaling as the {\it bona fide} one. The reason for the difference from different sum rules is that these sum rules are with $st$ crossing imposed so as to pick out a finite number of Wilson coefficients on the left hand side, but this procedure also introduces null constraints in the sum rules. That is, there are $\hat c^{\mathbb{12}}_{\ell,\mu}$ terms that cancel among themselves on the right hand side of the sum rule without affecting the Wilson coefficients, and these terms may have an unusually larger scale, pessimistically overestimating the scaling of the coefficient.

To estimate the sizes of other Wilson coefficients, we also want to establish scale correspondences for the rest of the UV spectral functions $\hat c^{+0}_{\ell,\mu}$, $\hat c^{-0}_{\ell,\mu}$ and $\hat c^{00}_{\ell,\mu}$ that involve the scalar. For $\hat c^{+0}_{\ell,\mu}$, we can use the $+\,0\!-\!0$ sum rule of \eref{4.3.3} with $k=1$, which happens to be \eref{F20101}. {Making use of the correspondences (\ref{LiMPscale}) and the scaling (\ref{g0b0})}, we get
\be\label{hpdisp}
\frac{\Li^2}{M_P^2}\sim  \sum_{\ell,X}\int_{1}^{\infty} \d\hat \mu \frac{\hat t(2\hat \mu+\hat t)d_{2,2}^{\ell,\hat \mu,\hat t}}{\hat \mu^2(\hat \mu+\hat t)^2}|\hat c^{+0}_{\ell,\mu}|^2 \,.
\ee
Thus, we see that $\hat c^{+0}_{\ell,\mu}$ (and hence $\hat c^{-0}_{\ell,\mu}$) leads to the same scale correspondence as those only involving the graviton:
\be
\label{LiMPscale1}
\frac{\Li}{M_P}  \Leftrightarrow \hat c^{+0}_{\ell,\mu} \,,~ \hat c^{-0}_{\ell,\mu}\,,~ c^{0+}_{\ell,\mu} \,,~ \hat c^{0-}_{\ell,\mu} \,.
\ee
For $\hat c^{00}_{\ell,\mu}$, \eref{4.3.3} does not give any readily usable dispersion relation to infer its size in terms of the hierarchy between $\Li$ and $M_P$. This is of course not surprising, as we should be able to define a scalar theory in the decoupling limit of the graviton where $M_P\to \infty$ and $\Lambda$ is held fixed. So in principle $c^{00}_{\ell,\mu}$ should be able to reach its partial wave unitarity limit $c^{00}_{\ell,\mu}\sim 1$. With a mild assumption in the spirit of lower spin dominance $c^{00}_{\ell,\mu}\sim \ell^{-1/2}$, we can then have the scaling correspondence $1  \Leftrightarrow \hat c^{00}_{\ell,\mu}$. This correspondence is also consistent with the pure scalar sum rules in the decoupling limit, which can be expanded in the forward limit and schematically goes like
\be
 a^{0000}_{k,n}\Li^{2k+2n}  =  \sum_{\ell,X}\int_{1}^{\infty} \d\hat\mu (\cdots) |\hat c^{00}_{\ell,\mu}|^2 \,,
\ee
leading to the usual dimensional analysis in the pure scalar theory: $a^{0000}_{k,n}\sim \Li^{-2k-2n}$. Away from the decoupling limit, the 0000 sum rule schematically goes like
\be
\label{sum0000s}
\frac{\Lambda^2}{M_P^2\hat t}+ \sum_n a^{0000}_{k,n}\Li^{2k+2n} \hat t^n =  \sum_{\ell,X}\int_{1}^{\infty} \d\hat\mu  (\cdots) |\hat c^{00}_{\ell,\mu}|^2 \,,
\ee
which contains an extra subdominant $1/M_P^2$ term when $\Li\ll M_P$, so it is also consistent with the $1  \Leftrightarrow \hat c^{00}_{\ell,\mu}$ scaling. For the lowest order terms, from sum rule (\ref{F00001}) or (\ref{F00002}), we see that the scalar self-couplings $\ai$ and $\gi_4$ must scale as
\be
\ai\sim1/\Li^4\,, ~~ \gi_4\sim 1/\Li^6\,.
\ee
On the other hand, in scalar-tensor theory, an interesting parameter regime is when the interactions involving the scalar are comparable with those of the pure gravity, in which case one may view the scalar more as part of gravity rather than some non-minimally coupled matter field. This occurs when the first term is comparable with the rest of the terms on the left hand side of \eref{sum0000s}, which implies a suppressed UV spectral function and the correspondence $\Lambda/M_P  \Leftrightarrow \hat c^{00}_{\ell,\mu}$. In this case, we then have $\ai\sim1/(M_P^2\Li^2)$ and $\gi_4\sim 1/(M_P^2\Li^4)$. Thus, for $\hat c^{00}_{\ell,\mu}$, we may consider the following two scenarios
\be
\label{LiMPscale2}
\begin{dcases}
1  \Leftrightarrow \hat c^{00}_{\ell,\mu} ~~~&\marrow \ai\sim\f1{\Li^4}\,,~ \gi_4\sim \f1{\Li^6}\,,
\\
\f{\Lambda}{M_P}  \Leftrightarrow \hat c^{00}_{\ell,\mu}~~~&\marrow \ai\sim \f{1}{M_P^2\Li^2}\,,~ \gi_4\sim \f{1}{M_P^2\Li^4}\,.
\end{dcases}
\ee
While the first scenario gives the boundary of the causality bounds, the second scenario is more relevant when the scalar plays a significant role in the dynamics, which is phenomenologically more interesting. In the following, we shall discuss the typical scales of the other Wilson coefficients with both the two scenarios in mind.

Now, we are ready to deduce the dimensional scalings of the other Wilson coefficients from the scalings of $\hat c^{\mathbb{12}}_{\ell,\mu}$ from the sum rules in Appendix \ref{sec:expSum}.
Let us now look at the $\gi_1$ coefficient. From the $F^{+++0}_{1,\ell}$ sum rule (\ref{F22201}) (using \eref{F22202} would be similar), we get
\be
\label{gi1disp}
-\frac{\Li^6\gamma_1}{M_P^3}= \sum_{\ell,X}\int_{1}^{\infty}\!\!\d\hat \mu\bigg[ \frac{(2\hat\mu-3\hat t)d_{2,0}^{\ell,\hat \mu,\hat t} \hat c^{+0}_{\ell,\mu} \hat c^{*,--}_{\ell,\mu}}{\hat t \hat\mu^4}
-\frac{\hat t\partial_{\hat t} d_{0,-2}^{\ell,\hat \mu,0} \hat c^{++}_{\ell,\mu} \hat c^{*,-0}_{\ell,\mu}}{\hat \mu^3 (\hat \mu-\hat t)}
+\frac{\hat t\partial_{\hat t} d_{2,0}^{\ell,\hat \mu,0} \hat c^{+0}_{\ell,\mu} \hat c^{*,--}_{\ell,\mu}}{\hat \mu^3 (\hat \mu+\hat t)} \bigg] \,.
\ee
By the scale correspondences (\ref{LiMPscale}) and (\ref{LiMPscale1}), we infer that the typical scale of $\gi_1$ is
\be
\gi_1\sim \f{M_P}{\Lambda^4}\,.
\ee
Note that this is independent of the value of $\ai$, which is consistent with the numerical result in Section \ref{sec:bounds}. Next, we look at $\gi_2$, for which we can use the $F^{++00}_{3,\ell}$ sum rule, whose explicit form in the forward limit is given by
\be
\frac{\Li^6\bi_1^2}{M_P^4} + \frac{\Li^6\gamma_2}{M_P^2}=
\sum_{\ell,X} \int_{1}^{\infty}\d \hat\mu \bigg[\frac{d_{0,0}^{\ell,\hat \mu, 0}}{\hat \mu^4} \hat c^{++}_{\ell,\hat\mu} \hat c^{*,00}_{\ell,\hat\mu}
-\frac{d_{2,2}^{\ell,\hat \mu, 0}}{\hat \mu^4}c^{+0}_{\ell,\hat\mu}c^{*,0-}_{\ell,\hat\mu}\bigg] \,.
\ee
By the scale correspondences (\ref{LiMPscale}), (\ref{LiMPscale1}) and (\ref{LiMPscale2}), we can infer that
\bal
\gamma_2&\sim \frac{M_P}{\Lambda^5}\text{~~~when~~}\alpha\sim \frac{1}{\Lambda^4} \,,
\\
\gamma_2 &\sim \frac{1}{\Lambda^4}\text{~~~~when~~} \alpha\sim \frac{1}{M_P^2\Lambda^2} \,.
\eal
Again, this is consistent with the numerical results in the next sections, and the dependence on $\ai$ is also observed there. Then, we look at the $\gi_3$ coefficient, for which we can use the $F^{+000}_{1,\ell}$ sum rule (\ref{F20001}),
\be
\frac{\beta_1 \Lambda^4}{M_P^3} \hat t-\frac{\gamma_3\Lambda^6}{M_P} \hat t^2=
\sum_{\ell,X}\int_{1}^{\infty}\d\hat \mu\bigg(\frac{\hat t (2\hat \mu-3\hat t)d_{2,0}^{\ell,\hat \mu,\hat t}}{\hat \mu ^4} \hat c^{+0}_{\ell,\mu} \hat c^{*,00}_{\ell,\mu}
+\frac{2 \hat t^4\partial_{\hat t}d_{2,0}^{\ell,\hat \mu,0}}{\hat \mu ^3 \left(\hat t^2-\hat \mu ^2\right)} \hat c^{+0}_{\ell,\mu} \hat c^{*,00}_{\ell,\mu}\bigg) \,.
\ee
We already know that $\bi_1\sim M_P/\Li^2$, so by the scale correspondences (\ref{LiMPscale1}) and (\ref{LiMPscale2}), we can infer that
\bal
\gamma_3&\sim \frac{1}{\Lambda^5}\text{~~~when~~}\alpha\sim \frac{1}{\Lambda^4} \,,
\\
\gamma_3 &\sim \frac{1}{M_P\Lambda^4}\text{~~~when~~} \alpha\sim \frac{1}{M_P^2\Lambda^2}\,.
\eal
We also want to look at the typical size of the coefficient $\bi_2$, which can be inferred from the $F^{++00}_{2,\ell}$ sum rule (\ref{F22002})
\bal
		\frac{\beta_2\Li^4}{M_P^2} -\frac{\gamma_0\Li^6}{M_P^4}\hat t &-g_{2,1}^{M_3}\Lambda^8 \hat t^2 =  \sum_{\ell,X}\int_{1}^{\infty}\! \d\hat \mu\bigg(\frac{d_{0,0}^{\ell,\hat \mu,\hat t} \hat c^{++}_{\ell,\mu} \hat c^{*,00}_{\ell,\mu}}{\hat \mu^3}
		+\frac{d_{2,2}^{\ell,\hat \mu,\hat t} \hat c^{+0}_{\ell,\mu} \hat c^{*,0-}_{\ell,\mu}}{(\hat \mu +\hat t)^3}
		+\!\frac{\hat t^4\partial_{\hat t}^2 d_{2,-2}^{\ell,\hat \mu,\hat t}  \hat c^{+0}_{\ell,\mu} \hat c^{*,-0}_{\ell,\mu}}{2 \hat \mu ^3\! \left(\hat t^2-\hat \mu ^2\right)} \nn
&-\frac{\hat t^3 (4 \hat \mu +3 \hat t) \partial_{\hat t} d_{2,-2}^{\ell,\hat \mu,\hat t} \hat c^{+0}_{\ell,\mu} \hat c^{*,-0}_{\ell,\mu}}{\hat \mu ^4\! \left(\hat t +\hat \mu \right)^2}
+\frac{\hat t^3 (10 \hat \mu^2 +15\hat \mu \hat t +6 \hat t^2) d_{2,-2}^{\ell,\hat \mu,\hat t} \hat c^{+0}_{\ell,\mu}\hat c^{*,-0}_{\ell,\mu} }{\hat \mu ^5\! \left(\hat t + \hat \mu \right)^3}
		\bigg)\,.
\eal
By the scale correspondences (\ref{LiMPscale}), (\ref{LiMPscale1}) and (\ref{LiMPscale2}), this gives us
\bal \label{beta2:1}
\bi_2&\sim \frac{M_P}{\Lambda^3}\text{~~~when~~}\alpha\sim \frac{1}{\Lambda^4} \,,
\\
\bi_2 &\sim \frac{1}{\Lambda^2}\text{~~~when~~} \alpha\sim \frac{1}{M_P^2\Lambda^2} \,.
\eal
As mentioned, all of these will be confirmed with the rigorous numerical results in Section \ref{sec:bounds}. Nevertheless, the scaling exercises above guide us to perform the numerical optimizations as they outline the rough boundaries of the causality bounds.

In summary, by simply inspecting the dispersive sum rules, one can estimate the typical sizes of the Wilson coefficients in the Lagrangian. Without imposing any {\it a priori} constraint on the UV spectral function $c^{00}_{\ell,\mu}$, apart from partial wave unitarity, we find that the scalar-tensor Lagrangian can be parametrized as follows
\bal
\label{action21}
 S  &=  M_P^2\!\! \int\!\! \d^4 x \sqrt{-g} \bigg(  \f{1}2 R -\f12 \nd_\mu \varphi \nd^\mu \varphi  +  \f{\hat\ai M_P^2}{2\Li^4} (\nd_\mu \varphi \nd^\mu \varphi)^2    + \f{\hat\beta_1}{2\Li^2}  \varphi {\cal G}    +\frac{\hat\bi_2M_P}{4\Li^3}  \varphi^2 {\cal G}   + \f{\hat\gi_0}{3!\Li^4} {\cal R}^{(3)}
 \nn
 &\hskip 80pt + \f{\hat\gi_1 }{3!\Li^4}  \varphi{\cal R}^{(3)}  +  \frac{\hat\gi_2M_P}{2\Li^5} \nd_\mu\varphi\nd^\mu\varphi {\cal R}^{(2)}  -\frac{4\hat\gi_3 M_P  }{3\Li^5}\nd_\mu\varphi\nd_\ri\varphi \nd_\nu\nd_\si \varphi  R^{\mu\nu\ri\si}
 \nn
 &\hskip 80pt  +  \f{\hat\gi_4M_P^2}{3\Li^6}  \nd_\mu\varphi \nd^\mu  \varphi   \nd_\ri \nd_\si \varphi \nd^\ri \nd^\si \varphi
 + \cdots \bigg) \,,
\eal
where we have used the dimensionless field $\varphi=\phi/M_P$ and $\hat\ai,\hat \bi_i,\hat\gi_i$ are dimensionless coefficients and are parametrically $\mc{O}(1)$. In this scenario, the scalar self-couplings such as $\alpha$ go like $\sim {1}/{\Lambda^p}$, and these scalings remain the same in the decoupling limit of the graviton where $M_P\to\infty$ and $\Li$ is held fixed. The scalings of the Lagrangian terms in \eref{action21} have been summarized in \eref{OphiRscale}, which for the terms in \eref{action21} has an intriguing integer flooring operation for the power of the $M_P/\Li$ factor, $\tilde N_\phi=\lfloor N_\phi/2\rfloor$. Having gone through the power counting with the sum rules, we can see that the flooring operation originates from the fact that, in the scaling argument above, $\hat c^{\mathbb{12}}_{\ell,\mu}$ with either no or one scalar helicity corresponds to $\Li/M_P$ (see \eref{LiMPscale} and \eref{LiMPscale1}) while $\hat c^{\mathbb{12}}_{\ell,\mu}$ with two scalar helicities corresponds to $1$ (see \eref{LiMPscale2}). Also, given that each term on the right hand side of a sum rule only contains two factors of $\hat c^{\mathbb{12}}_{\ell,\mu}$, there will be a $\hat c^{00}_{\ell,\mu}$ in the sum rule for the lowest orders as long as there are two 0 helicities in the 2-to-2 scattering (except for the case of $F^{+0-0}_{1,\ell}$, which however does not affect our argument). Thus, in these cases, the power of $M_P/\Li$ in \eref{OphiRscale} is determined by the number of 0 helicities in the most constraining 2-to-2 scattering amplitude, upon taking the flooring operation $\lfloor N_\phi/2\rfloor$. We emphasize that the $\tilde N_\phi=\lfloor N_\phi/2\rfloor$ rule is an coincidence, valid only for the lowest orders of the EFT operators. For higher orders, our method precisely predicts the breakdown of this rule, which will be numerically verified in Section \ref{sec:xcheck}. The correct way to get $\tilde{N}_\phi$ for any orders is to count the number of $c^{00}_{\ell,\mu}$ in appropriate dispersion relations, as discussed through this section.

On the other hand, if the scalar interactions are constrained to be comparable with the gravitational interactions,  that is, we assume the scalar UV spectral function is relatively weak and has the correspondence $\hat c^{00}_{\ell,\mu}  \Leftrightarrow {\Lambda}/{M_P}$, then the scalar-tensor Lagrangian can be parametrized as follows
\bal
\label{action22}
 S  &=  M_P^2 \int \d^4 x \sqrt{-g} \bigg(  \f{1}2 R -\f12 \nd_\mu \varphi \nd^\mu \varphi +  \f{\hat\ai}{2\Li^2} (\nd_\mu \varphi \nd^\mu \varphi)^2    + \f{\hat\beta_1}{2\Li^2}  \varphi {\cal G}    +\frac{\hat\bi_2}{4\Li^2}  \varphi^2 {\cal G}   + \f{\hat\gi_0}{3!\Li^4} {\cal R}^{(3)}
 \nn
 &\hskip 80pt + \f{\hat\gi_1}{3!\Li^4}  \varphi{\cal R}^{(3)}  +  \frac{\hat\gi_2}{2\Li^4} \nd_\mu\varphi\nd^\mu\varphi {\cal R}^{(2)}  -\frac{4\hat\gi_3}{3\Li^4}  \nd_\mu\varphi\nd_\ri\varphi \nd_\nu\nd_\si \varphi  R^{\mu\nu\ri\si}
 \nn
 &\hskip 80pt  +  \f{\hat\gi_4}{3\Li^4}  \nd_\mu\varphi \nd^\mu  \varphi   \nd_\ri \nd_\si \varphi \nd^\ri \nd^\si \varphi
 + \cdots \bigg)  \,,
\eal
where again $\hat\ai,\hat \bi_i,\hat\gi_i$ are dimensionless coefficients and are parametrically $\mc{O}(1)$. In this case, we have, for example, $\alpha\sim 1/(M_P^2\Lambda^2)$. Note that the typical size of the coefficient of $\varphi {\cal G}$, a leading operator that gives rise to hairy black holes, is not affected by the constraints on the scalar self-couplings. This surprising fact can be easily spotted in the dispersive sum rules. Our goal in Section \ref{sec:bounds} is to use all available sum rules to numerically compute the bounds on the coefficients $\hat\ai,\hat \bi_i,\hat\gi_i$ and so on, confirming the rough estimates in this section.

\section{Optimization scheme}

\label{sec:numSet}

In this section, we will set up a numerical optimization scheme that effectively utilizes the dispersive sum rules to constrain the Wilson coefficients of scalar-tensor theory in the following section. Recall that the dispersive sum rules establish a remarkable set of relations between the IR coefficients of the EFT and the amplitudes of the unknown UV completion. These relations can be fed into a semi-definite program (SDP) that can be solved numerically. This will confirm the rough estimates in the previous section and, more importantly, lead to ``sharp'' bounds on the coefficients in the next section. Readers uninterested in the detailed numerical setup and methods can go through Section \ref{sec:genStra} and skip Section \ref{sec:numdet}.

\subsection{General strategy}
\label{sec:genStra}

While estimating the scaling rules for the Wilson coefficients, the sum rules (\ref{4.3.3}) are sometimes sufficient and preferred. To numerically obtain the optimal bounds, we shall always use the $st$-improved sum rules (\ref{mainsumRule}). Each of the sum rules (\ref{mainsumRule}) is actually a one-parameter family of dispersive equalities, parametrized by the momentum transfer $t$, connecting the Wilson coefficients and the integrals of the UV amplitudes. To effectively use all of these dispersive equalities, following the approach of \cite{Caron-Huot:2021rmr} and \cite{Caron-Huot:2022ugt}, we integrate the dispersive sum rule against a weight function $\phi^{\mathbb{1234}}_{k}(p)$ over the interval $0\leq p \leq \Lambda$ and as well as sum over several sum rules:
\bal
\label{nnd2}
\sum_{\mathbb{1234},k} \int_{0}^{\Lambda} \! \d p\,  \phi^{\mathbb{1234}}_{k}(p)  \bigg[ \delta_{k,2}a^{\mathbb{1234}}_{k,-1}\f{-1}{p^2}  & +a^{\mathbb{1234}}_{k,0}+a^{\mathbb{1234}}_{k,1}\(-p^2\)+a^{\mathbb{1234}}_{k,2}p^4\bigg]
\nn
&=\bigg\langle \sum_{\mathbb{1234},k}\int_{0}^{\Lambda}\! \d p\, \phi^{\mathbb{1234}}_{k}(p) F^{\mathbb{1234}}_{k,\ell}(\mu ,-p^2) \bigg \rangle \,,
\eal
where we have, for later convenience, introduced a positive real number $p$ such that
\be
t:=-p^2 \,.
\ee
The weight functions $\phi^{\mathbb{1234}}_{k}(p)$ will be the decision variables we optimize over to get the best causality bounds. (For the forward-limit sum rules that will also be used, it is suffice to use normal weight parameters; see Appendix \ref{sec:Bec_exa}.) By the integration and summation in \eref{nnd2}, we can make use of as much information as possible from the dispersive sum rules in extracting the causality bounds. If an appropriate $\phi^{\mathbb{1234}}_{k}(p)$ makes the right hand side of \eref{nnd2} positive, we can then obtain a condition on the Wilson coefficients
\be
\label{WilsonCon2}
\sum_{\mathbb{1234},k}\int_{0}^{\Lambda} \d p \phi^{\mathbb{1234}}_{k}(p)\bigg (\delta_{k,2}a^{\mathbb{1234}}_{k,-1}\frac{-1}{p^2}+a^{\mathbb{1234}}_{k,0}+a^{\mathbb{1234}}_{k,1}(-p^2)+a^{\mathbb{1234}}_{k,2}(-p^2)^2\bigg ) \geq 0 \,.
\ee
Going through all possible $\phi^{\mathbb{1234}}_{k}(p)$, we can find the tightest constraints on these coefficients. The problem of finding the best bounds can be formulated as an SDP with an infinite number of constraints, enumerated by the discrete variable $\ell$ and the continuous variable $\mu$. Also, the functional space of all possible $\phi^{\mathbb{1234}}_{k}(p)$ is parametrized by an infinite number of parameters, so numerically we also need to approximate this functional space, which will be explained shortly in Section \ref{sec:numdet}.

To see how this optimization is implemented, notice that $F^{\mathbb{1234}}_{k,\ell}(\mu ,-p^2)$ contains an infinite number of UV partial amplitudes $c^{\mathbb{12}}_{\ell,\mu}$ and their complex conjugates, which we are agnostic about from the point view of bootstrapping from low energies. In order to proceed, we need to eliminate them in the optimization problem, which naturally turns this into an SDP problem.

Before that, let us isolate the minimal set of $c^{\mathbb{12}}_{\ell,\mu}$ that are necessarily involved when performing this SDP. First, note that in a theory with parity conservation, we can divide the sum over all possible intermediate states in $\<...\>$ (see \eref{shortHandn}) into two parts, one being summation over parity-even $X$ states and the other summation over parity-odd states.  Denoting the parity of state $X$ by $P_X$, we have the following relations for the partial wave amplitudes
\begin{align}
\label{nnd4}
c^{\mathbb{12}}_{P_X,\ell,\mu}&=P_X c^{\mathbb{\bar{2}\bar{1}}}_{P_X,\ell,\mu} \,,
\\
\label{nnd41}
c^{\mathbb{12}}_{P_X,\ell,\mu}&=(-1)^\ell c^{\mathbb{21}}_{P_X,\ell,\mu} \,.
\end{align}
Because of time reversal invariance that we assume, we have $\mathcal{M}^{\mathbb{1234}}=\mathcal{M}^{\bar{\mathbb{3}}\bar{\mathbb{4}}\bar{\mathbb{1}}\bar{\mathbb{2}}}$, which implies that ${\rm  Im} (\sum_{P_X}c^{\mathbb{12}}_{P_X,\ell,\mu}c^{*,\mathbb{\bar{3}\bar{4}}}_{P_X,\ell,\mu} )=0$. Denoting $c^{\mathbb{12}}_{P_X,\ell,\mu}=c^{\mathbb{12},\Re}_{P_X,\ell,\mu}+i\, c^{\mathbb{12},\Im}_{P_X,\ell,\mu}$, we then have
\begin{equation}
\sum_{P_X}c^{\mathbb{12}}_{P_X,\ell,\mu}c^{*,\mathbb{\bar{3}\bar{4}}}_{P_X,\ell,\mu} =\sum_{\mc{I}=\Re,\Im}\sum_{P_X} c^{\mathbb{12},\mc{I}}_{P_X,\ell,\mu}c^{\mathbb{\bar{3}\bar{4}},\mc{I}}_{P_X,\ell,\mu} =\sum_{P_X} \bigg(c^{\mathbb{12},\Re}_{P_X,\ell,\mu}c^{\mathbb{\bar{3}\bar{4}},\Re}_{P_X,\ell,\mu}+c^{\mathbb{12},\Im}_{P_X,\ell,\mu}c^{\mathbb{\bar{3}\bar{4}},\Im}_{P_X,\ell,\mu}\bigg) \,.
\end{equation}
So the real and imaginary parts of $c^{\mathbb{12}}_{P_X,\ell,\mu}$ are separated and play a similar role in the dispersive sum rules. From the perspective of imposing positivity bounds, this extra summation over the real and imaginary part is essentially redundant, since, as mentioned above, we are agnostic about the values of $c^{\mathbb{12}}_{P_X,\ell,\mu}$. Following \cite{Zhang:2020jyn,Du:2021byy}, we will simply absorb the summation over $\mc{I}=\Re,\Im$ into the summation over $X$ and take $c^{\mathbb{12}}_{P_X,\ell,\mu}$ as real functions in the following. Using these separations, we can express a generic quantity obtained by mixing different helicities of $F^{\mathbb{1234}}_{k,\ell}$ and integrating over $p$ in the following form:
\begin{equation}\label{nnd5}
\sum_{\mathbb{1234},k}\int_{0}^{\Lambda} \d p \phi^{\mathbb{1234}}_{k}(p) F^{\mathbb{1234}}_{k,\ell}(\mu ,-p^2):=\sum_{P_X=\pm 1}\sum_{\mathbb{A,B}}B_{P_X,\ell}^{\mathbb{A,B}}(\mu )c^{\mathbb{A}}_{P_X,\ell,\mu }c^{\mathbb{B}}_{P_X,\ell,\mu } \,,
\end{equation}
where the summation of $\mathbb{A}$ and $\mathbb{B}$ is over $00,+0,++,+-$ and $B_{P_X,\ell}^{\mathbb{A,B}}$ is independent of $p$ and can be extracted from \eref{4.3.9}. The reason why $\mathbb{A}$ and $\mathbb{B}$ only run over $00,+0,++,+-$ is that we can use Eqs.~(\ref{nnd4}) and (\ref{nnd41}) to convert other helicities to these four. According to parity $P_X$ and whether $\ell$ is odd, the summation on the right hand side of \eref{nnd5} splits into four independent parts, $(P_X,\ell)=(+1,{\rm even}),(+1,{\rm odd}),(-1,{\rm even}),(-1,{\rm odd})$, each of which can be written in the following form
\begin{equation}\label{nnd6}
\(\mathcal{C}_{P_X,\ell,\mu}\)^T
B_{P_X,\ell}(\mu ) \,
\mathcal{C}_{P_X,\ell,\mu}\,,
\end{equation}
where $B_{P_X,\ell}(\mu )$ is a $4\times 4$ matrix and we have defined that
\begin{equation}\label{nnd_define1}
\mathcal{C}_{P_X,\ell,\mu}=
\begin{pmatrix}
c^{00}_{P_X,\ell,\mu } & c^{+0}_{P_X,\ell,\mu } & c^{++}_{P_X,\ell,\mu } &c^{+-}_{P_X,\ell,\mu }
\end{pmatrix}^T\,.
\end{equation}
The reason why it is beneficial to separate the sum in \eref{nnd5} according to parity $P_X$ and the oddness of $\ell$ is that some of the $c^{\mathbb{12}}_{P_X,\ell,\mu }$ often vanish owing to \eref{nnd4} and  \eref{nnd41}, in which case we can omit the corresponding entries of the $B_{P_X,\ell}(\mu )$ matrix in the SDP. This leads to better bounds and reduces computational costs. Again, the non-vanishing $c^{\mathbb{12}}_{P_X,\ell,\mu }$ depend on the UV model, and for a generic bootstrap program we choose to be agnostic about them.

With these established, we see that the requirement of the right hand side of \eref{nnd5} being positive is equivalent to the conditions that all the $B_{P_X,\ell}(\mu )$ matrices be positive semi-definite
\be
\label{BposCon}
B_{P_X,\ell}(\mu ) \succeq 0\,,~~\text{for  $P_X=\pm 1$, all possible $\ell$ and all $\mu \geq \Lambda^2$}\,.
\ee
These conditions will in turn ensure that the left hand side of \eref{nnd5} is positive, giving rise to a condition for some Wilson coefficients (\ref{WilsonCon2}) for a given set of $\phi^{\mathbb{1234}}_{k}(p)$. To obtain the best bounds, we optimize over all possible $\phi^{\mathbb{1234}}_{k}(p)$. In practice, of course, we can not impose the conditions for all $\ell$ and $\mu$ and go through all possible $\phi^{\mathbb{1234}}_{k}(p)$, and some numerical approximations are needed. Note that the {\tt SDPB} package can deal with an SDP with only one continuous parameter if the entries of the linear matrix inequalities \eref{BposCon} are polynomials of this parameter, but unfortunately this is not the case here. In the following subsection, we shall outline approximations that can be used to overcome this problem, along with how to effectively truncate the $\phi^{\mathbb{1234}}_{k}(p)$ functional space.

\subsection{Numerical details}
\label{sec:numdet}

Having formulated the causality bounds finding as a SDP, we now get to the nitty-gritty of implementing it numerically, largely following the numerical implementation of \cite{Caron-Huot:2021rmr} and \cite{Caron-Huot:2022ugt}. To simplify the expressions, we shall set $\Li=1$ from now on, but restore it in the final results for clarity.

 As mentioned, {\tt SDPB} can directly solve a SDP with a finite number of linear matrix inequalities, and the entries of these matrices can be polynomials of a continuous variable. However, for our current case, entries of $B_{P_X,\ell}(\mu )$ are more complex than polynomials of a continuous variable. To take in as many constraints as possible in the numerical program, we can divide the $\mu$-$\ell$ constraint space into five regions, as shown in Figure \ref{fig:quyuhuafen},  and will make approximations for the five regions separately.\\

\begin{figure}[tbp]
	\centering
	\includegraphics[scale=0.35]{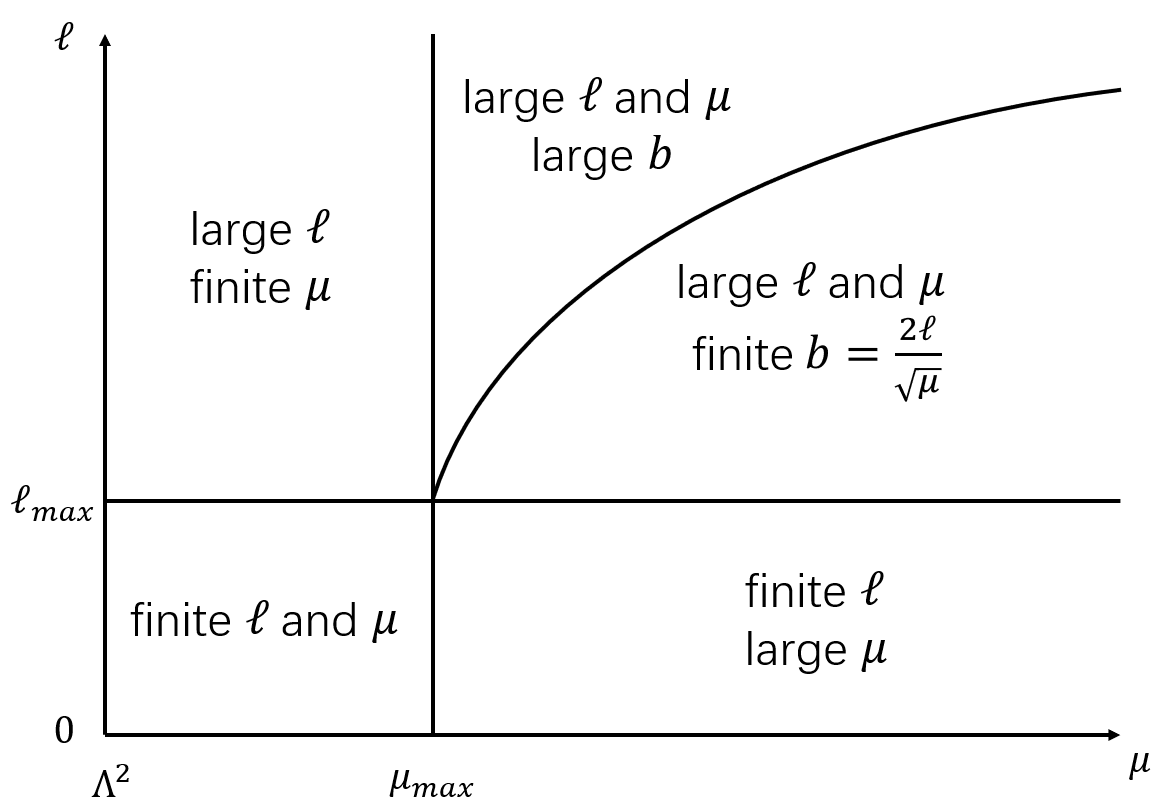}
	\caption{Various regions in the $\mu$-$\ell$ constraint space. The constraints are implemented differently in different regions.}
\label{fig:quyuhuafen}
\end{figure}

\textbf{Finite $\mu$ and finite $\ell$}: In this region, we will simply discretize the continuous parameter $\mu$.  Since the UV scale $\mu\geq 1$, we can choose a discrete set of $\mu$. We find that the point density needed to achieve convergence depends on the dimension of the truncated $\phi^{\mathbb{1234}}_{k}(p)$ functional space, which is the main limiting factor to use a higher dimensional functional space. We also only make use of the partial waves up to $\ell_{\rm max}$.

\textbf{Large $\mu$ and finite $\ell$}: When $\mu$ is large, the entries of the $B_{P_X,\ell}(\mu )$ matrices can be expanded as a Taylor series of $1/\mu$ around $\mu \to \infty$, which allows us to approximate the entries of $B_{P_X,\ell}(\mu )$ by truncating the expansion and retaining the leading few orders. Then, we multiply all the sum rules by an appropriate power of $\mu$ to make entries of the $B_{P_X,\ell}(\mu )$ matrices polynomials of $\mu$, and take $\mu$ as the continuous parameter in $\tt{SDPB}$. Alternatively, when the dimension of the $\phi^{\mathbb{1234}}_{k}(p)$ functional basis is not too large, we find that it is also numerically sufficient to work with the exact dependence on $\mu$ and just take a few discrete large $\mu$ points along with finite $\ell\leq \ell_{\text{max}}$.

\textbf{Finite $\mu$ and large $\ell$}: When $\ell$ is large, the Wigner d-functions (or rather the hypergeometric function) oscillate with $p^2/\mu$ and thus tend to vanish after integrating against the weight functions. This is the reason why we also seemingly redundantly add the forward-limit sum rules (\ref{forlimSum}) in the SDP, in order to effectively use the constraints from this region. That is, in the large $\ell$ limit, with the forward-limit sum rules included, we can neglect the terms with the hypergeometric functions from the non-forward sum rules, since the contributions from the forward-limit sum rules dominate in this limit. In the large $\ell$ limit, we can approximate $\ell$ as a continuous variable; However, the forward-limit sum rules contain square roots of polynomials of $\ell$: $\sqrt{(\ell+c_1)(\ell+c_2)\dots (\ell +c_n)}$, where $c_i$ are real constants, which are not admissible by {\tt SDPB}. To resolve this problem, we shall expand them as a Laurent series in the limit $\ell\to \infty$ and only keep a few leading terms: $\ell^{n/2}+1/2 (c_1+c_2+\dots+c_n) \ell^{n/2-1}+\mc{O}(\ell^{n/2 -2})$. We then make the variable change $\ell\to (y+\sqrt{\ell_{\rm max}})^2$ so that it becomes a polynomial of $y$ where $y\geq 0$. Then, we can again discretize $\mu$, and, for a fixed $\mu$, the entries of the $B_{P_X,\ell}(\mu )$ matrices can be viewed as polynomials of $\ell$ for large $\ell$, the semi-positivity of $B_{P_X,\ell}(\mu )$ then becoming admissible for $\tt SDPB$. Note that while the added forward limit sum rules do technically alter the SDP in this region as well as in the finite $\ell$ regions, they become negligible in other regions.

\textbf{Large $\mu$, $\ell$ and finite $b$}:  This region can be made accessible by using the asymptotic behavior of the Wigner d-functions in $F^{\mathbb{1234}}_{k,\ell}(\mu ,-p^2)$. The Wigner d-functions can be expressed in terms of the hypergeometric function, which has the following asymptotic behavior
\begin{equation}
\label{asym_2f1_1_}
	\lim_{\mu ,\ell \to \infty ;\; {2\ell}/{\!\sqrt{\mu}}=b} \ _2F_1(h_1-\ell,\ell+h_1+1;h_1-h_2+1;{p^2}/{\mu})=\frac{\Gamma(h_1-h_2)}{({bp}/{2})^{h_1-h_2}} J_{h_1-h_2}(bp)\,,
\end{equation}
where $J_{h}(x)$ is the Bessel function of the first kind and the limit is taken with fixed $b= {2\ell}/\!\sqrt{\mu}$. That is, we sample the constraints along lines $b \sqrt{\mu} = {2\ell}$ (with different $b$) in the region of large $\mu$ and large $\ell$, and each of these lines has a natural physical interpretation of scatterings with fixed impact parameter $b= {2\ell}/\!\sqrt{\mu}$ \cite{Caron-Huot:2022ugt}. With these established, we can easily Taylor expand $F^{\mathbb{1234}}_{k,\ell}(\mu ,-p^2)$ around $\mu \to \infty$ with fixed $b$, and only retain the leading terms, namely the ${1}/{\mu^3}$ term in this case. (We do not need to expand $\mu$ in the partial wave amplitudes $c^{\mathbb{12}}_{P_X,\ell,\mu}$, because they are limited in size by partial wave unitarity.) We find that only $F^{\mathbb{1234}}_{(1,2),\ell}(\mu,-p^2)$ have non-vanishing $\mc{O}(1/\mu^{3})$ terms, so only these dispersive sum rules need to be considered in the large $\mu$ and $\ell$ region. For example, the leading term of $F^{+0-0}_{2,\ell}(\mu,-p^2)$ in this limit is given by
\begin{equation}\label{asym_f43}
	F^{+0-0}_{2,\ell}(\mu,-p^2)=\frac{2}{\mu^3}J_0(bp)|c^{+0}_{P_X,\ell,\mu}|^2+\mathcal{O}\(\f1{\mu^4}\) \,.
\end{equation}
Note that in the leading order the $\ell$ dependence is only in $c^{\mathbb{12}}_{P_X,\ell,\mu}$'s, which do not go into the definition of $\tilde{B}_{P_X,\tilde{\ell}}(b)$. However, $\tilde{B}_{P_X,\tilde{\ell}}(b)$ does depend on the oddness of $\ell$, because we need to use $c^{\mathbb{12}}_{P_X,\ell,\mu}=(-1)^\ell c^{\mathbb{21}}_{P_X,\ell,\mu}$ to convert $c^{\mathbb{12}}_{P_X,\ell,\mu}$'s to a standard independent basis. This means that the matrix $\mu^3 B_{P_X,\ell}$ only depends on $b$, $P_X$ and the oddness of $\ell$ at leading order in the large ${\mu}$ and $\ell$ region. Let us define $\tilde{B}_{P_X,\tilde{\ell}}(b):=\mu^3 B_{P_X,\ell}(\mu)$ in this region, where $\tilde{\ell}$ means $\tilde{B}_{P_X,\tilde{\ell}}(b)$ depends on the oddness of $\ell$ rather than its explicit value. Therefore, for large ${\mu}$ and $\ell$, we can simply impose the following linear matrix inequalities as a leading approximation
\begin{equation}\label{Bbcondition}
	\tilde{B}_{P_X,\tilde{\ell}}(b) \succeq 0 , \text{~~for all $b>0$, $P_X=\pm 1$ and $\tilde{\ell}=$ even  or odd} \,.
\end{equation}
To explicitly compute $\tilde{B}_{P_X,\tilde{\ell}}(b)$, we note the following well known integration formula
\begin{equation}\label{damul_jifen1}
	\frac{\Gamma(\nu)}{({b}/{2})^{\nu-1}}\int_{0}^{1}dp p^{n+1-\nu}J_{\nu-1}(bp)=\frac{1}{n+1}\ _1F_2\bigg (\frac{n+1}{2};\frac{n+3}{2},\nu;-\frac{b^2}{4}\bigg )\,.
\end{equation}
So the entries of $\tilde{B}_{P_X,\tilde{\ell}}(b)$ are still not polynomials of $b$, and we need to make further approximations.
For finite $b<b_{\rm max}$, we can discretize it into $b=\{\epsilon_b+k\delta_b|0\leq k <(b_{\rm max}-\epsilon_b)/\delta_b,k\in \mathbb{N}\}$, where $\epsilon_b$ is a very small starting point.

\textbf{Large $\mu$, $\ell$ and large $b$}:  For large $b$, by the asymptotic form of the generalized hypergeometric function,
we can write $\tilde{B}_{P_X,\tilde{\ell}}(b)$ in the following form,
\begin{equation}
	\tilde{B}_{P_X,\tilde{\ell}}(b)=f(b)+g(b)\cos(b)+h(b)\sin(b)\,,
\end{equation}
where $f(b)$, $g(b)$ and $h(b)$ are $4\times 4$ matrices whose entries are polynomials of ${1}/{b}$, truncated to order $(1/b)^{R_b}$. For large $b$, it is a good approximation to replace the semi-positiveness of $\tilde{B}_{P_X,\tilde{\ell}}(b)$ with the following slightly stronger condition
\begin{equation}\label{damul_condi2}
	b^{R_b}
	\begin{pmatrix}
		& f(b)+g(b) & h(b) \\
		&h(b)       & f(b)-g(b)\\
	\end{pmatrix}
	\succeq 0\text{, for $b \geq b_{\rm max}$}\,,
\end{equation}
where the factor $b^{R_b}$ makes $b^{R_b}f(b)$, $b^{R_b}g(b)$ and $b^{R_b}h(b)$ polynomials of $b$.\\

Apart from the approximations in the $\mu$-$\ell$ constraint space, we also need to numerically approximate the functional spaces of all possible $\phi^{\mathbb{1234}}_{k}(p)$. Recall that $\phi^{\mathbb{1234}}_{k}(p)$ are supposed to run over all possible functions within the interval $[0,1]$. By the Weierstrass approximation theorem, a simple functional basis over a finite interval would be power functions $p^n$, and in the numerical approximation we truncate to keep the leading few orders. However, for the technical reasons to be explained below, for some $\phi^{\mathbb{1234}}_{k}(p)$, we will need to choose $(1-p)^2 p^n$.

First, note that, in order to obtain the bounds on the leading order coefficients, the positivity condition (\ref{BposCon}) can not be satisfied without $F^{0000}_{1,\ell}$, $F^{0000}_{2,\ell}$, $F^{+0-0}_{1,\ell}$, $F^{+0-0}_{2,\ell}$ and $F^{++--}_{2,\ell}$. This is because all other leading $F^{\mathbb{1234}}_{k,\ell}$ in the large $\mu$ and large $\ell$ region either lead to a non-diagonal term in $B_{P_X,\ell}$ or contribute to a term in $B_{P_X,\ell}$ that changes its sign under the parity $P_X$ or the oddness of $\ell$. For $B_{P_X,\ell}$ to be semi-positive, we need the diagonal terms to be semi-positive and we need $B_{P_X,\ell}$ to be semi-positive for both all cases of $P_X$ and $\ell$. Additionally, we aim to derive bounds projected onto ${1/M_P^2}$, and only the above five improved sum rules involve ${1/M_P^2}$.

Let us see what kinds of bases are suitable for $F^{0000}_{1,\ell}$, $F^{0000}_{2,\ell}$, $F^{+0-0}_{1,\ell}$, $F^{+0-0}_{2,\ell}$ and $F^{++--}_{2,\ell}$ for our purposes. The technical requirements come from implementing the constraints in the large $\ell$ and $\mu$ region. We take $F^{++--}_{2,\ell}$ as an example. In this region with fixed $b={2\ell }/{\sqrt{\mu}}$, a necessary condition to satisfy the positivity condition (\ref{Bbcondition}) is
\begin{equation}\label{neccon_f71}
	\int_{0}^{1} \d p \phi^{++--}_{2}(p)J_0(bp)\geq 0\text{ , for all $b>0$}\,.
\end{equation}
This actually implies that the Fourier transform of $\phi^{++--}_{2}(p)/p$ is non-negative and also $\lim_{p\to 0}\phi^{++--}_{2}(p)/p>0$. As a result, the basis for $\phi^{++--}_{2}(p)$ should start at $ p^{n_{\text{min}}}$ with $n_{\text{min}}\leq 1$. On the other hand, this choice necessarily results in an IR divergence from integrating in the low energy region near $p=0$. The best one can do for $F^{++--}_{2,\ell}$ is to choose $n_{\text{min}}=1$, which only leads to a logarithmic divergence. This IR divergence arises from how the scattering amplitudes are defined for massless particles in 4D, and may be resolved using better observables \cite{Caron-Huot:2021rmr}. We will simply regulate it with an IR cutoff scale $m_{\text{IR}}$, which may be taken to be the Hubble scale as a conservative choice. The cases of $F^{0000}_{1,\ell}$, $F^{0000}_{2,\ell}$, $F^{+0-0}_{1,\ell}$ and $F^{+0-0}_{2,\ell}$ are analogous. Going through similar steps, we can see that the basis of $\phi^{0000}_1(p)$, $\phi^{0000}_2(p)$, $\phi^{+0-0}_1(p)$, $\phi^{+0-0}_2(p)$ and $\phi^{++--}_2(p)$ should be chosen to start with $ p^{-1}$, $p$, $p^{-1}$, $p$ and $p$ respectively.

\begin{table}[h!]
	\begin{center}
		\begin{tabular}{l|c} 
			\hline
			$\ell_{\rm max}$ & 42 \\
			\hline
			$b_{\rm max}$ & $10001/250$ \\
			\hline
			$\epsilon_b$ & $1/250$ \\
			\hline
			$R_b$       & 10 \\
			\hline
			$N_{\rm p}$   & 7 \\
			\hline
			Discrete set of $\mu$ for finite $\ell$ & \makecell[c]{$\{1/(1-k/100)|0\leq k \leq 90\,,k\in\mathbb{Z}\}$ \\ $\cup \{1/(1-k/400)|361\leq k\leq 399\,,k\in\mathbb{Z}\}$}  \\
			\hline
			Discrete set of $\mu$ for large $\ell$  &  $\{1/2500+1/(1-k/100)^{1/2}|0\leq k \leq 99, k\in\mathbb{Z} \}$ \\
			\hline
			Discrete set of $b=2\ell/\sqrt{\mu}$   & $\{\epsilon_b + k/32|0\leq k \leq1280,k\in\mathbb{Z} \}$    \\
			\hline
			\hline
			Non-default {\tt SDPB} parameters &\makecell[c]{{\tt --precision=766}\\ {\tt --dualityGapThreshold=1e-11} \\ {\tt --maxComplementarity=1e+80}\\ {\tt --maxIterations=20000}}\\
			\hline
		\end{tabular}
	\end{center}
\caption{Numerical parameters used in the {\tt SDPB} computations.}
\label{tab:num}
\end{table}

There is actually one additional consideration for choosing the suitable basis, namely, the requirement that $g(b)$ or $h(b)$ should not dominate in the large $\ell, \mu, b$ region in order to satisfy condition (\ref{damul_condi2}). Again, take $F^{++--}_{2,\ell}$ as an example. By \eref{asym_2f1_1_}, we can get
\bal
		\mu^3\int_{0}^{1} \d p p^n & F^{++--}_{2,\ell}(\mu,-p^2)
		\xrightarrow{{\rm large~} b} \left(|c^{++}_{P_X,\ell,\mu}|^2+|c^{+-}_{P_X,\ell,\mu}|^2\right) \\
		&\cdot\bigg[ \frac{1}{b^{{n+1}}}\frac{2^n \Gamma(\frac{1+n}{2})}{\Gamma(\frac{1-n}{2})}+\frac{\sqrt{2}\cos(b-\frac{3\pi}{4})}{\sqrt{\pi}b^{\frac{3}{2}}}+\frac{\sqrt{2}(8n-5)\sin(b-\frac{3\pi}{4})}{8\sqrt{\pi}b^{\frac{5}{2}}}+\mathcal{O}\bigg(\frac{1}{b^{\frac{7}{2}}}\bigg)\bigg]\,.\nonumber
\eal
If the oscillating term $\cos(b)$ or $\sin(b)$ dominates in the large $b$ limit, the positivity condition  (\ref{damul_condi2}) can not be satisfied. However, we already require that when $p$ goes to zero, $n_{\text{min}}$ in the basis $p^{n_{\text{min}}}$ should not be less than 1 so as to avoid non-logarithmic IR divergences. To overcome this, we can multiply the corresponding weight function with a factor $(1-p)^2$, which cancel the leading oscillating terms upon integration and make $f(b)$ dominate in \eref{damul_condi2}. Again, the cases of $F^{0000}_{1,\ell}$, $F^{0000}_{2,\ell}$, $F^{+0-0}_{1,\ell}$ and $F^{+0-0}_{2,\ell}$ are analogous. Thus, the final result is that the basis of $\phi^{0000}_1(p)$, $\phi^{0000}_2(p)$, $\phi^{+0-0}_1(p)$, $\phi^{+0-0}_2(p)$ and $\phi^{++--}_2(p)$ should be chosen to start from $(1-p)^2 p^{-1}$, $(1-p)^2 p$, $(1-p)^2 p^{-1}$, $(1-p)^2 p$, $(1-p)^2 p$ respectively.

For other $F^{\mathbb{1234}}_{k,\ell}(\mu ,-p^2)$ that result in leading order contributions in the large $\mu,\ell$ limit, the bases are chosen such that they lead to the same large $b$ behavior in the $\tilde{B}_{P_X,\tilde{\ell}}(b)$ matrix as the above five $F^{\mathbb{1234}}_{k,\ell}(\mu ,-p^2)$. For the rest of the $F^{\mathbb{1234}}_{k,\ell}(\mu ,-p^2)$ that are sub-leading in the large $\mu,\ell$ limit, we can simply choose their bases to be ${1,p,p^2,p^3,...}$.
In our numerical calculations, it is sufficient for our purposes to choose the dimension of the functional space of $\phi^{\mathbb{1234}}_{k}(p)$ to be $N_{\rm p}=7$. The numerical parameters we use to run {\tt SDPB} are listed in Table \ref{tab:num}.

In general, when performing the numerical optimization to obtain bounds on a given set of Wilson coefficients, we hope to utilize as many sum rules as possible so as to derive the strongest bounds. For that, we can often include sum rules that contain Wilson coefficients not in that given set. The reason is that we can choose the weight functions $\phi^{\mathbb{1234}}_{k}(p)$ properly so that the weight-function-smeared sum rules do not contain the extra Wilson coefficients.

\section{Bounds on EFT coefficients and their implications}
\label{sec:bounds}

 In this section, we shall constrain the Wilson coefficients of scalar-tensor theory using the dispersive sum rules obtained in Section \ref{sec:disprules}, via the numerical optimization procedure outlined in Section \ref{sec:numSet}. We will consider  generic scalar-tensor EFTs as well as EFTs with some of the coefficients fine-tuned, the latter being also popular as modified gravity and cosmological models phenomenologically. In Appendix \ref{sec:Bec_exa}, we will give an explicit example to demonstrate how to use the optimization scheme to obtain the causality bounds. We will first derive the bound on $\alpha$, the coefficient of the $(\partial \phi)^4$ term in the Lagrangian. We will show that the value of $\alpha$ will significantly affect the bounds on a coefficient when all the sum rules of the coefficient contain $c^{00}_{P_X,\ell,\mu}$, in agreement with the discussions in Section \ref{sec:dimAna}. Therefore, we will compute the bounds on the other coefficients for various values of $\ai$. Particularly, we will compute the bounds on the Gauss-Bonnet couplings, which give rise to the intriguing phenomena of hairy black holes and scalarization in compact stars. These couplings are currently being intensively probed with gravitational wave and other observational means. We shall discuss the phenomenological implications of our bounds for  these couplings. We will also calculate the causality bounds for large values of $\ai$, which confirms the scaling behaviors that have been estimated in Section \ref{sec:dimAna}. We will also show that some fine-tuned EFTs can not be exact, as they will lead to inconsistencies among the sum rules, so some additional terms must exist. Moreover, some higher dimensional coefficients can significantly affect the bounds on the lower dimensional coefficients.

\subsection{Scalar four-derivative term}
\label{sec:alpha}

Let us first derive the lower bound for the coefficient $\alpha$, the coupling constant of the $(\partial \phi)^4=(\nd_\mu \phi \nd^\mu \phi)^2$ term. In the graviton decoupling limit, the lower bound on this dim-8 coefficient is $\ai>0$. This was one of the earliest causality bounds \cite{Adams:2006sv} and gives rise to the term of ``positivity bounds'', often used synonymously with ``causality bounds'', as we do in this paper. In the presence of gravity, however, it has been predicted that the lower bound slightly dips blow zero, the negativity being suppressed by the Planck mass squared \cite{Alberte:2020jsk}. This has been illustrated explicitly with a string theory example \cite{Tokuda:2020mlf} and also numerically confirmed for generic UV completions \cite{Caron-Huot:2021rmr}. 

We refer the readers to Appendix \ref{sec:Bec_exa} for a more detailed explanation of how to implement the numerical procedure of Section \ref{sec:numSet}. Here we shall simply outline the main steps of this procedure for the case of obtaining the lower bound on $\alpha$. 
\begin{itemize}
	\item First, we collect relevant improved dispersive sum rules. In principle, the sum rules that do not contain $\alpha$ should also be included for deriving the strongest bound, because those sum rules contain the information of full crossing symmetry/null constraints. However, for this particular case, we find that the only relevant sum rules are from $\mc{M}^{0000}$, and hence we only need:
	\begin{align}
	        \label{F00001eq}
		-\frac{1}{M_P^2}+2\ai t- \gi_4 t^2&=\Big\langle F^{0000}_{1, \ell}(\mu, t)\Big\rangle    \,,\\
		\label{F00002eq}
		-\frac{1}{M_P^2}\frac{1}{t}+2\ai -\gi_4 t+12g^S_{0,2}t^2&=\Big \langle F^{0000}_{2, \ell}(\mu, t)\Big \rangle  \,.
	\end{align}
	\item Then, we sum over the sum rules after integrating them against the weight functions and define $B_{P_X,\ell}(\mu )$ via $\sum_{k}\int_{0}^{1}\d p F^{0000}_{k,\ell}(\mu,-p^2)=(\mathcal{C}_{P_X,\ell,\mu})^T B_{P_X,\ell}(\mu)\mathcal{C}_{P_X,\ell,\mu}$. The weight functions are decision variables. Imposing $B_{P_X,\ell}(\mu)\succeq 0$ then gives us inequalities on the Wilson coefficients. Since here we are only concerned about the bound on $\alpha$ and agnostic about all the other Wilson coefficients, we can choose the weight functions such that the combinations in front of all the other Wilson coefficients vanish in the summed-over sum rules. For example, if we want to be agnostic about $\gamma_4$, we can impose the condition on the weight functions: 
	\begin{equation}
		\int_{0}^{1}\d p \big[ \phi^{0000}_{1}(p)(-(-p^2)^2)+\phi^{0000}_{2}(p)(-(-p^2)) \big]=0 \label{xianzhi1}\,.
	\end{equation}
where $\phi^{0000}_{1}(p)$ and $\phi^{0000}_{2}(p)$ are the weight functions multiplying Eq.~(\ref{F00001eq}) and Eq.~(\ref{F00002eq}) respectively. We must impose conditions like this such that the only coupling constants are $1/M_P^2$ and $\alpha$ in the summed-over sum rules, which will be used in the numerical optimizaiton.
	\item Then, we impose $B_{P_X,\ell}(\mu)\succeq 0$ and conditions like Eq.~\eqref{xianzhi1}, which leads to inequalities like 
	\begin{equation}
		\int_{0}^{1}\!\! \d p\(-\phi^{0000}_1(p)+\phi^{0000}_2(p)\frac{1}{p^2} \) \frac{1}{M_P^2} + 2\int_{0}^{1}\!\! \d p\(\phi^{0000}_1(p)(-p^2)+\phi^{0000}_2(p)\) \alpha \geq 0 \,.
	\end{equation}
	To find the strongest lower bound on $\ai$, we can normalize $\int_{0}^{1} \d p\big(-\phi^{0000}_1(p)+\phi^{0000}_2(p)\frac{1}{p^2} \big) = 1$, and maximize
	\begin{equation}
		2\int_{0}^{1}\d p\(\phi^{0000}_1(p)(-p^2)+\phi^{0000}_2(p)\) \label{xianzhi4} \,.
	\end{equation}
	for all possible choices of weight functions $\phi^{0000}_i(p)$.  However, as discussed in Section \ref{sec:numdet}, the finite dimensional expansion of $\phi^{0000}_{1}(p)$ must begin with $p^{-1}(1-p)^2$, and that of $\phi^{0000}_{2}(p)$ must begin with $p(1-p)^2$, so the integral of the normalization condition contain a logarithmic divergence and we need to include an IR cutoff $m_{\rm IR}$. It is a good approximation to only preserve the $\log(\Lambda/m_{\rm IR})$ term in the integration, as will be explained in Section \ref{subsec:xian}. With the IR cutoff, the normalization condition becomes 
	\begin{equation}
		-x^{0000}_{1,-1}+x^{0000}_{2,1}=0\,, \label{xianzhi3} 
	\end{equation}
	where we have parameterized the weight functions as $\phi^{\mathbb{1234}}_{k}(p)=\sum_{n=n_{\rm min}}x^{\mathbb{1234}}_{k,n}p^n (1-p)^2$. 
\item Finally, we solve the following SDP: 
\begin{equation}\label{nr5}
	\begin{aligned}
		\text{maximize: }&~~~2\int_{0}^{1}\d p\(\phi^{0000}_1(p)(-p^2)+\phi^{0000}_2(p)\) \,,\\
		\text{subject to: }&~~~ -x^{0000}_{1,-1}+x^{0000}_{2,1}=0\,,
		\\&~~ \text{conditions like Eq.~\eqref{xianzhi1}}: ~~\int_{0}^{1}\d p \sum_{k}\phi^{0000}_k(p)(-p^2)^{n_k}=0\,,
		\\&~~ \text{positivity condition: }~~ B_{P_X,\ell}(\mu)\succeq 0 \text{, for all $P_X$, $\ell$ and $\mu$}\,.
	\end{aligned}
\end{equation}
\end{itemize} 
where $\Lambda$ is set to be 1. We sum $k$ up to $k=4$ and $n_k$ are chosen to eliminate higher order coefficients in the SDP. In the practical numerical calculations, since we can only optimize over a finite dimensional subspace of the infinite $\phi^{0000}_k(p)$ functional space, we supplement this SDP with some extra forward-limit sum rules. These forward-limit sum rules are redundant, but numerically they help minimize the impact of the uncertainties from implementing the constraints in the large $\ell$ and finite $\mu$ region (see Appendix \ref{sec:Bec_exa}). 
Thus, numerically, this SDP gives rise to the lower bound on $\alpha$:
\begin{equation}\label{nr6}
\alpha \geq -16.091\frac{\log({\Lambda}/{m_{\rm IR}})}{\Lambda^2 M_P^2}\,.
\end{equation}
which is consistent with the analysis in \cite{Caron-Huot:2021rmr, Alberte:2020jsk}.  However, we can not use a similar SDP to derive the upper bound on $\alpha$. This is completely analogous to the pure scalar case where the corresponding $\alpha$ has a lower bound $\alpha\geq 0$, which can be recovered from the above bound by taking $M_P\to \infty$, but can not be bounded from above by the positivity of the spectral function. In the pure scalar case, $\alpha$ can be bounded from above by making use of more information from unitarity, particularly using the upper bound on the partial wave amplitude $|c^{00}_{\ell,\mu}|^2\leq\mathcal{O}(1)$. This produces an upper bound of order $\mc{O}({1}/{\Lambda^4})$, which is very large if $M_P\gg \Lambda$. It is expected that the upper bound is something similar in the presence of gravity, which would be consistent with the estimate in Section \ref{sec:dimAna}.

In the following subsections, we will see that the bounds on some coefficients, particularly the couplings involving the scalar, strongly depend on the value of $\alpha$, while the other coefficients are insensitive to $\alpha$. Specifically, we will see that the bounds on the former coefficients, projected on $1/M_P^2$, become weaker as the value of $\alpha$ increases. The sensitivity/insensitivity of the bounds on $\alpha$ originates from the fact that often being agnostic about $\alpha$ essentially means that we are largely agnostic about $c^{00}_{P_X,\ell,\mu}$. So, if the determination of the bounds on a coefficient requires the $B_{P_X,\ell}(\mu)$ matrices to have nontrivial $(00, *)$ or $(*, 00)$ entries (cf.~Eqs.~(\ref{nnd6}) and (\ref{nnd_define1})), this coefficient will at least weakly depend on $\ai$. These coefficients include $\beta_2$ and $\gamma_{1,2,3,4}$. On the other hand, a coefficient strongly depending on $\alpha$ is when all of its sum rules themselves contain $c^{00}_{P_X,\ell,\mu}$; in this case, of course, the $B_{P_X,\ell}(\mu)$ matrices will have nontrivial $(00, *)$ or $(*, 00)$ entries in the optimization results.
For example, the $\gi_1$ sum rules themselves do not involve $c^{00}_{P_X,\ell,\mu}$, so the bounds on $\gi_1$ do not strongly depend  on $\ai$; Nevertheless, $\gi_1$ weakly depends on $\ai$, because the $B_{P_X,\ell}(\mu)$ matrices contain effective $(00, *)$ or $(*, 00)$ entries that are nonzero when optimizing to get the bounds on $\gi_1$, as we shall see in Section \ref{sec:xcheck}. All of these confirm the rough estimates in Section \ref{sec:dimAna}. 

More explicitly, note that, to satisfy our positivity condition $B_{P_X,\ell}(\mu )\succeq 0$, a necessary condition is that all diagonal entries of the matrix $B_{P_X,\ell}(\mu )$ must be non-negative; Making use of the explicit expressions of $F^{\mathbb{1234}}_{m,\ell}$ in Appendix \ref{sec:expSum} and judicious choices of $\phi^{\mathbb{1234}}_{k}(p)$, we find that it is easy to make the $(++,++)$, $(+-,+-)$ and $(+0,+0)$ entries of $B_{P_X,\ell}(\mu )$ positive, but often this can not be done for the $(00,00)$ entry. To see this, note that, the lowest order dispersive sum rules contributing to the $(00,00)$ entry come from amplitude ${\mc M}^{0000}$, of which the only relevant ones, to the lowest orders, are
\begin{equation}\label{nr3}
\begin{aligned}
-\frac{1}{M_P^2}+2\alpha t-\gamma_4 t^2&=\bigg\langle F^{0000}_{1, \ell}(\mu, t)\bigg\rangle\,,\\
-\frac{1}{M_P^2}\frac{1}{t}+2\alpha -\gamma_4 t+12g^S_{0,2}t^2&=\bigg \langle F^{0000}_{2, \ell}(\mu, t)\bigg \rangle\,.\\
\end{aligned}
\end{equation}
If we wish to get a bound on the coefficients while being agnostic about $\alpha$, we can add the above two sum rules and additionally choose more restricted $\phi^{0000}_{1,2}(p)$ to suppress the $\alpha$ terms on the left hand sides of the sum rules
\begin{equation}\label{nr4}
\begin{aligned}
&\int_{0}^{1}\d p \phi^{0000}_1(p)(-p^2)+\int_{0}^{1}\d p \phi^{0000}_2(p)=0 \,.
\end{aligned}
\end{equation}
However, with these extra constraints, we find that usually the $(00,00)$ entry of $B_{P_X,\ell}(\mu)$ can not be positive for every $\mu$ and $\ell$. In the following, we shall probe how the bounds vary with the scalar dim-8 coupling $\alpha$.

\subsection{Linear Gauss-Bonnet coupling}
 \label{subsec:xian}

\begin{figure}[tbp]
	\centering
	\includegraphics[scale=0.28]{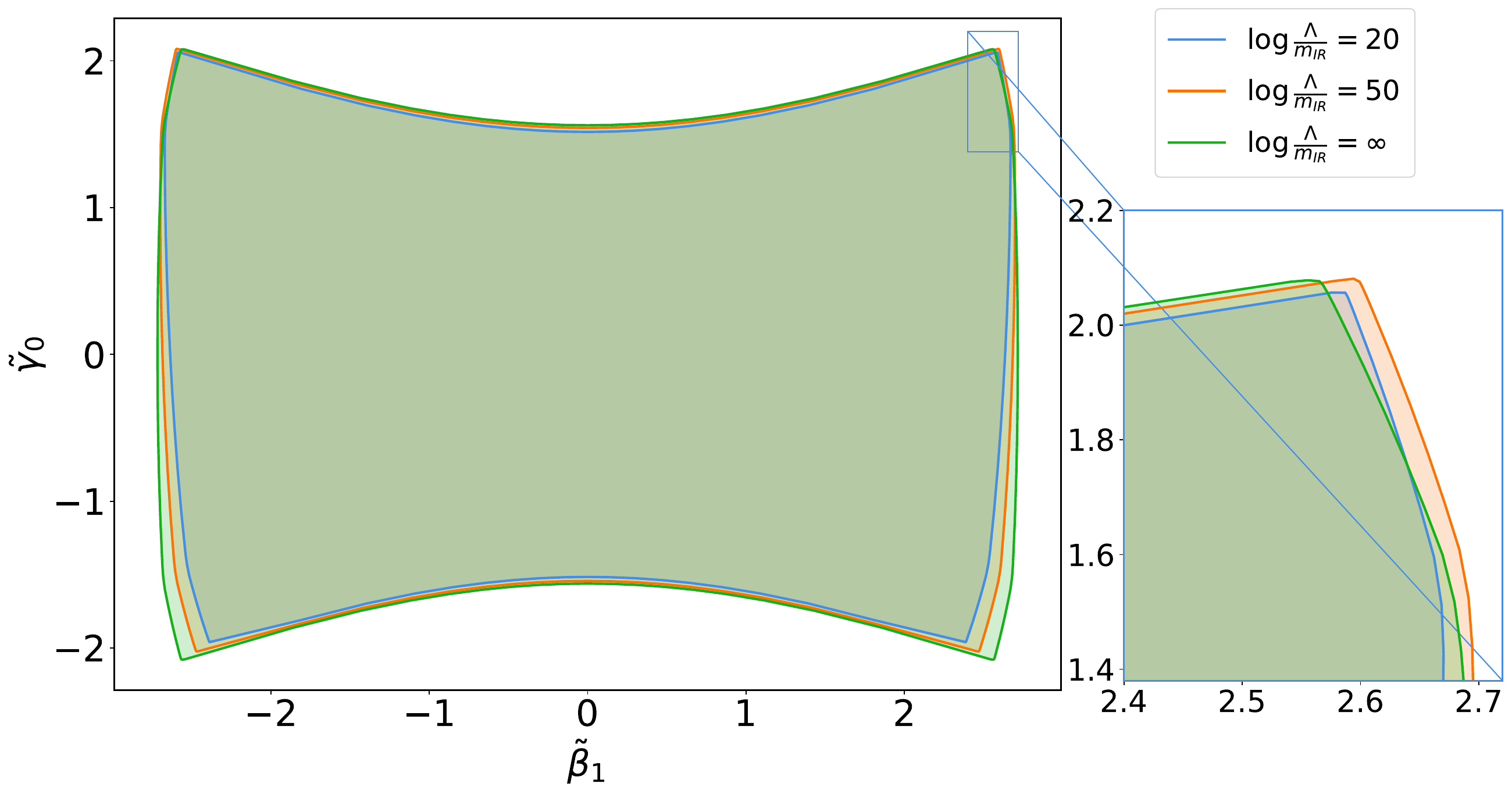}
	\caption{Causality bounds on $\gamma_0$ and $\beta_1$. We have defined $\tilde{\beta_1}=\beta_1 \Lambda^2/\big(M_P \sqrt{\log(\Lambda/m_{\text{IR}})}\big)$ and $\tilde{\gamma}_0=\gamma_0 \Lambda^4/\big(M_P^2 \sqrt{\log(\Lambda/m_{\text{IR}})}\big)$. The $\log({\Lambda}/{m_{\rm IR}})=\infty$ case represents the leading approximation, while the $\log({\Lambda}/{m_{\rm IR}})=50$ and $\log({\Lambda}/{m_{\rm IR}})=20$ (log being the natural logarithm) cases are computed with 2 iteration of linear improvements. The bounds are almost symmetric with respect to $\gi_0\to -\gi_0$ and $\bi_1\to -\bi_1$ because the leading approximation mostly constrains $\gamma_0^2$ and $\beta_1^2$.
 }
 \label{fig:nr2.2}
\end{figure}

The $\phi {\cal G}$ term  has been shown to be one of the very few ways to generate black hole solutions with non-trivial/hairy solutions that are different from those of GR \cite{Sotiriou:2013qea, Sotiriou:2014pfa, Yagi:2011xp}. The no-hair theorems underpin many of our modern understandings of the deep nature of gravity \cite{Bekenstein:1973ur, Bekenstein:1996pn}, and also observational confirmations of black hole solutions are important tests of Einstein's gravity.

In this subsection, we shall compute the positivity bounds on $\beta_1$ in conjunction with the bounds on $\gamma_0$, the coefficient of the ${\cal R}^{(3)}$ term,
\be
\mc{L} \supset  \sqrt{-g} \( \, \f{\bi_1}{2!}  \phi {\cal G}    + \f{\gi_0}{3!} {\cal R}^{(3)} \)\,.
\ee
At times, the specific structure of EFT amplitudes may lead to additional constraints in the sum rules. The case of $\gamma_0$ and $\beta_1$ provides a good example. In the tree level EFT amplitudes, some coefficients are non-negative because they are of the form of $\beta_1^2$ or $\gamma_0^2$. These forms come from squares of 3-leg vertices in the amplitudes, as the Lagrangian terms with coefficient $\beta_1$ and $\gamma_0$ can generate 3-leg vertices. However, the corresponding right hand side terms in the sum rules do not automatically enforce such positivity. So we can take these extra constraints into account when handling the sum rules with $\beta_1^2$ and $\gamma_0^2$. Ignoring them erroneously weakens the bounds on the other coefficients. (For the Newton's constant ${1}/{M_P^2}$, on the other hand, there is no need to impose its positivity in our formalism, as its positivity is implied by the sum rules.)

\begin{figure}[tbp]
	\centering
	\includegraphics[scale=0.28]{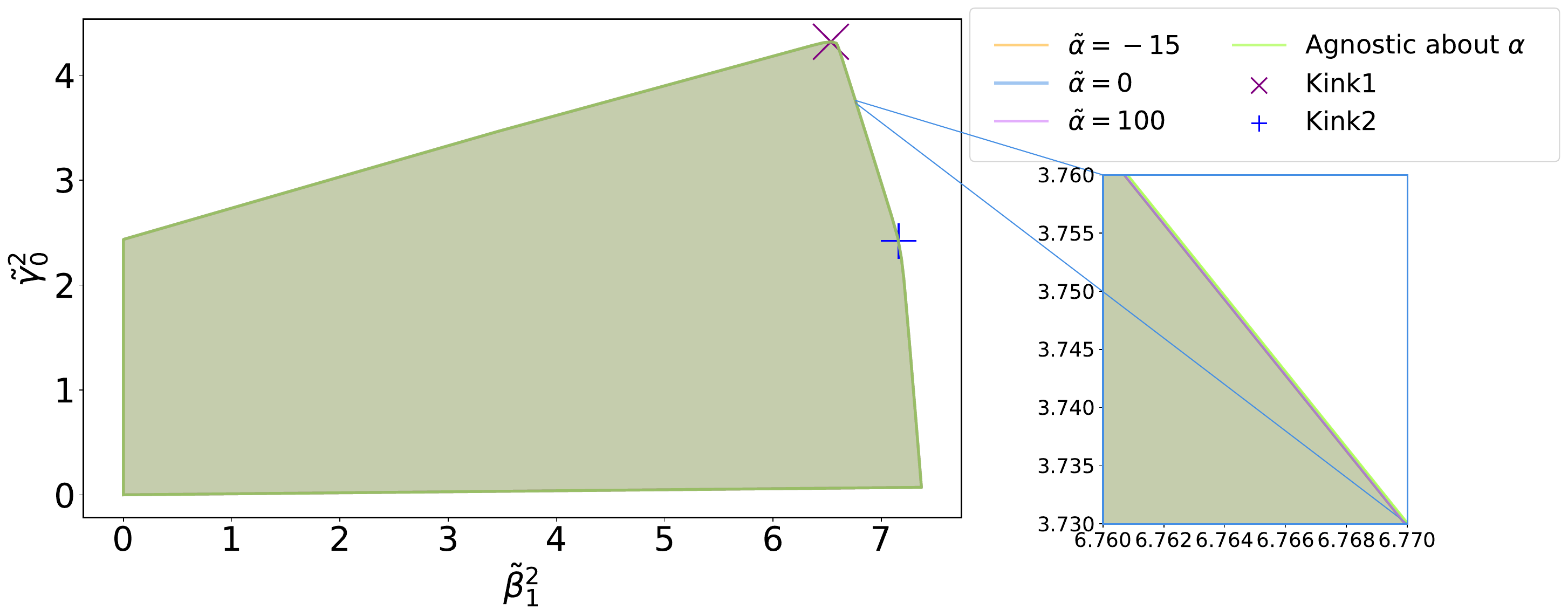}
	\caption{Bounds on $\gamma_0^2$ and $\beta_1^2$ for various $\ai$, where $\tilde{\alpha}={\alpha M_P^2\Lambda^2}/{\log(\Lambda/m_{\text{IR}})}$, $\tilde{\beta}_1^2=\beta_1^2 \Lambda^4/(M_P^2 \log(\Lambda/m_{\text{IR}}))$ and $\tilde{\gamma}_0^2=\gamma_0^2 \Lambda^8/(M_P^4 \log(\Lambda/m_{\text{IR}}))$.
	The main difference from Figure \ref{fig:nr2.2} is that here we also include sum rules involving $c^{00}_{P_X,\ell,\mu}$. The four lines are almost indistinguishable from each other, meaning that the bound on $\gamma_0^2$ and $\beta_1^2$ is insensitive to $\alpha$. There are kinks at $(\tilde \gi_0^2,\tilde \bi_1^2)=(4.32,6.53)$ and $(\tilde \gi_0^2,\tilde \bi_1^2)=(2.42,7.16)$ respectively.}
	\label{fig:yualpha}
\end{figure}

We will be interested in bounds on $\gamma_0$ and $\beta_1$, agnostic about $\alpha$. As discussed in the previous subsection, this means that we can not use the sum rules containing $c^{00}_{P_X,\ell,\mu}$, as well as the sum rules that rely on $c^{00}_{P_X,\ell,\mu}$ to satisfy \eref{BposCon}. This means that we will only use improved dispersive sum rules (\ref{F22211}-\ref{F22224}), (\ref{F21211}-\ref{F21214}) and (\ref{F22112}-\ref{F22114}). Additionally, we also use some forward-limit sum rules to improve the numerical convergence in the large $\ell$ and finite $\mu$ region. Another {\it ad hoc} trick to improve the numerics in this region is to use both sum rules with helicities $\mathbb{1322}$ and $\mathbb{1232}$. This will include more null constraints in the SDP, given that our numerical implementation truncates the sum rules at a finite order of $k$ (see \eref{mainsumRule}).

To determine the boundary of the positivity region, we can make use of angular optimization. To this end, we parametrize $\gamma_0$ and $\beta_1$ as follows
\begin{equation}\label{nr9}
\f{\gamma_0 \Lambda^4}{M_P^2}=r\cos\theta\,,~~
\f{\beta_1 \Lambda^2}{M_P}=r\sin\theta\,.
\end{equation}
Then, for a given $\theta$, the optimization program for bounds on $\gamma_0$ and $\beta_1$ outputs a quadratic inequality of $r$, which gives a bound on $r$; going through sufficiently many $\theta$, we get a 2D bound in the $\gamma_0$-$\beta_1$ plane. However, this is not an optimization problem directly solvable by the {\tt SDPB} package, because both $r$ and $r^2$ are present in the inequality. Nevertheless, for phenomenological interesting cases (for which the IR logarithm $\log({\Lambda}/{m_{\rm IR}})$ is not too small), we can drop the linear term in the $r$ inequality, and then the problem becomes directly solvable by {\tt SDPB} for a given $\theta$. If we want to improve the accuracy of the bound with the linear $r$ terms, we can use the above result as an initial background solution $r_*$ of the quadratic $r$ inequality and seek a linear perturbative improvement $\delta r$ on top of it:
\begin{equation}\label{nr10}
\f{\gamma_0 \Lambda^4}{M_P^2} =(r_*-\delta r)\cos\theta\,,~~
\f{\beta_1 \Lambda^2}{M_P} =(r_*-\delta r)\sin\theta\,.
\end{equation}
This of course can be iterated for further improvements: set $r_*\to r_*-\delta r$ and repeat several times to a desired accuracy. 

In Figure \ref{fig:nr2.2}, we compare the bounds obtained from the leading approximation and its improvements with the above iterations. The leading approximation, where the linear $r$ terms are dropped, can be extracted by the limit $\log({\Lambda}/{m_{\rm IR}})\to\infty$), while for the $\log({\Lambda}/{m_{\rm IR}})=50$ and $\log({\Lambda}/{m_{\rm IR}})=20$ cases we have performed two iterations of linear improvements.  We see that, for a phenomenological interesting $\log({\Lambda}/{m_{\rm IR}})$, the leading approximation is actually rather good. The non-convexity of Figure \ref{fig:nr2.2} results from the fact that the SDP is performed on quadratic functions of $\gi_0$ and $\bi_1$, but Figure \ref{fig:nr2.2} is plotted for $\gi_0$ and $\bi_1$ themselves. Also, from Figure \ref{fig:nr2.2}, we see that the allowed values of the dimensionless coefficients $\hat\gamma_0$ and $\hat\beta_1$, modulo $(\log({\Lambda}/{m_{\rm IR}}))^{1/2}$, are order one, which is consistent with the dimensional analysis in Section \ref{sec:dimAna}. This is also consistent with the parametric bound on $\beta_1$ in Ref \cite{Serra:2022pzl}, estimated from requiring the absence of acausal time advances when the graviton and the scalar scatter off a heavy object in the eikonal regime. Also, the bounds on $\gamma_0$ have previously been computed in Ref \cite{Caron-Huot:2022ugt}, which can be compared with ours by setting $\bi_1=0$. Our bounds on $\gamma_0$ are a few percents stronger than those of Ref \cite{Caron-Huot:2022ugt}, which probably arises from the differences in using dispersion relations and approximations in the large $\mu,\ell$ region.

To obtain the bounds on $\gamma_0$ and $\beta_1$ in Figure \ref{fig:nr2.2}, we only used sum rules that do not relate to the partial wave amplitude $c^{00}_{P_X,\ell,\mu}$. To utilize other sum rules, one needs to have some prior knowledge of $\alpha$ and, potentially, for a given $\alpha$, the bound on $\gamma_0$ and $\beta_1$ could be significantly reduced. However, in Section \ref{sec:dimAna}, we have estimated that this should not happen. Here, with the numerical scheme, we can confirm that the bound on $\gamma_0$ and $\beta_1$ is insensitive to the value of $\ai$; see Figure \ref{fig:yualpha} for how the bound varies with $\ai$ using the leading approximation. Even if the value of $\ai$ has varied from near its lower bound to $\mc{O}(100)$, the impact on the bound on $\gamma_0$ and $\beta_1$ is only about 0.001\%, consistent with a numerical error. Note that in Figure \ref{fig:yualpha} we only plot bounds on $\gamma_0$ and $\beta_1$ (more precisely $\gamma_0^2$ and $\beta_1^2$) in the first quadrant. The bounds on the other quadrants are almost mirror symmetric with respect to the one in the first quadrant, because the sum rules used mostly contain $\gamma_0^2$ and $\beta_1^2$, except for one couple of them which contains $\beta_1\gamma_0$. The effects of the sum rules with $\beta_1\gamma_0$ turn out to be very weak in the optimization.

The $\phi \mc{G}$ operator itself is shift symmetric $\phi\to \phi+const$, but in plotting Figure \ref{fig:yualpha} we are agnostic about the coefficients of non-shift symmetric operators such as $\phi^2 \mc{G}$, that is, Figure \ref{fig:yualpha} is for a generic theory without shift symmetry. Incidentally, one may be interested in how the causality bounds differ if the theory is actually shift-symmetric. This is equivalent to setting the coefficients of terms like $\phi^2 \mc{G}$ to zero, which gives rise to some extra null constraints. However, our numerical results show that the bounds on $\gi_0$ and $\bi_1$ are virtually unchanged if we impose the shift symmetry. The differences between the shift-symmetric bounds and the generic bounds are about $\mc{O}(10^{-5})$, which may well be numerical errors. This is not surprising from the point of view of the dispersion relations: the bounds on $\gamma_0$ and $\beta_1$ mostly come from the constraining powers of the four-graviton dispersion relations, but the latter do not contain $\gamma_1$ and $\beta_2$ at all.

\begin{figure}[tbp]
	\centering
	\includegraphics[scale=0.28]{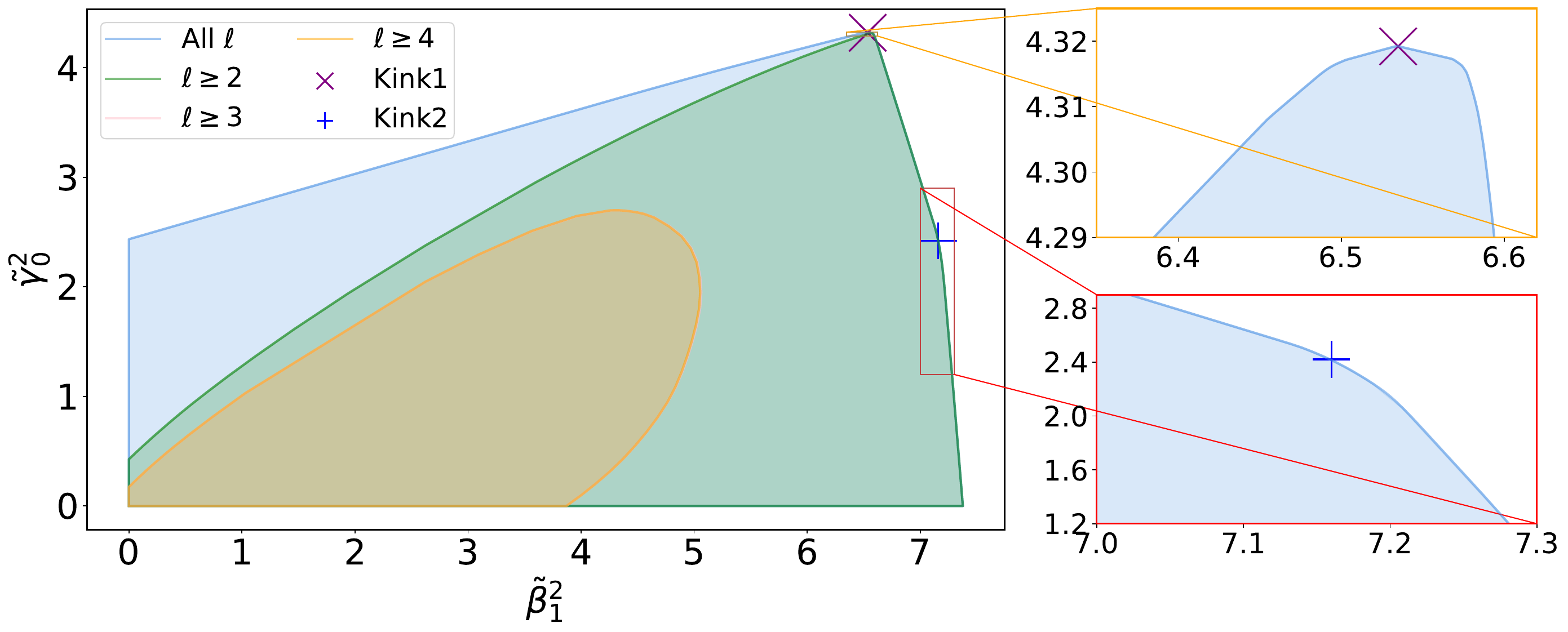}
	\caption{Bounds on $\gamma_0^2$ and $\beta_1^2$ with some low spins removed, where $\tilde{\gi}_0^2={\gi_0^2 \Lambda^8}/{(M_P^4\log(\Lambda/m_{\text{IR}}))}$ and $\tilde{\beta}_1^2=\beta_1^2 \Lambda^4/(M_P^2 \log(\Lambda/m_{\text{IR}}))$. For a line with $\ell\geq\ell_0$, we only impose positivity conditions with UV spin $\ell\geq\ell_0$.
The $\ell=1$ states decouple so that the $\ell\geq 1$ and $\ell \geq 2$ curves are the same. The $\ell\geq 3$ and $\ell\geq 4$ curves only differ slightly. The zoom-ins of the kinks are also shown.}
	\label{fig:kink}
\end{figure}

In Figure \ref{fig:yualpha}, we find that there are two kinks at $(\tilde \gi_0^2,\tilde \bi_1^2)=(4.32,6.53)$ and $(\tilde \gi_0^2,\tilde \bi_1^2)=(2.42,7.16)$, which will be referred to as ``Kink 1'' and ``Kink 2'' respectively; see Figure \ref{fig:kink} for the close-ups.  
Often, a kink indicates a theory with special features, as it delineates two continuous classes of EFTs. (From the perspective of the convex cone of the $s^2$ coefficients, already, kinks correspond to UV states that are irreps of the spacetime and internal symmetries of the EFT \cite{Zhang:2020jyn}.)
To probe the nature of these kinks, in Figure \ref{fig:kink}, we also calculate the bounds on $\beta_1^2$ and $\gamma_0^2$ with some of the UV spin states suppressed. By $\ell\geq \ell_0$, we assume that there are no UV states with spin $\ell<\ell_0$. We do not plot the $\ell\geq 1$ case because the $\ell=1$ states decouple from this process so that the $\ell\geq 1$ curve is the same as the $\ell\geq 2$ curve. 
Also, in Figure \ref{fig:kink}, we also find that the $\ell\geq 3$ and $\ell\geq 4$ curves only differ slightly.

The reason why the bounds on $\beta_1$ and $\gamma_0^2$ receive no contributions from the $\ell=1$ partial waves and are insensitive to the $\ell=3$ partial waves is a result of spin selection rules, thanks to the fact that the EFT only includes spin-0 and spin-2 modes. To see this, note that the amplitude discontinuity in the dispersion relations schematically goes like
\begin{equation}
	{\rm Disc } \mc{M}^{\mathbb{1}\mathbb{2}\mathbb{3}\mathbb{4}}(\mu,t)\sim \sum_{\ell} d^{\ell}_{h_{12},h_{43}}(\arccos(1+2t/\mu)) c^{\mathbb{1}\mathbb{2}}_{\ell,\mu} (c^{ \mathbb{\bar{3}} \mathbb{\bar{4}} }_{\ell,\mu})^*
	\nonumber
\end{equation}
For an odd $\ell$, we have $c^{\mathbb{1}\mathbb{2}}_{\ell,\mu}=0$ if $\mathbb{1}=\mathbb{2}$, due to the Bose symmetry. On the other hand, for $\mathbb{1}\neq\mathbb{2}$, $\mathbb{3}\neq\mathbb{4}$ and $\ell=\text{odd}$, the Wigner functions $d^{\ell}_{h_{12},h_{43}}$ vanish if $\ell < {\rm max}\{|h_{12}|,|h_{43}|\}$. This is the case for $\ell=1$, because, for a scalar-tensor EFT, when $h_i\neq h_j$, we have $|h_i-h_j|=2$ or $4$. Therefore, the bounds receive no contributions from the $\ell=1$ partial waves. For the $\ell=3$ case, we still have $\ell < {\rm max}\{|h_{12}|,|h_{43}|\}$ if we consider pure graviton scattering in which we have $|h_i-h_j|=4$. So, for $\ell=3$, the Wigner functions in most of the dispersion relations vanish. Additionally, the bounds on $\gamma_0$ and $\beta_1$ turn out to be insensitive to the rest dispersion relations from the non-pure graviton scatterings. Therefore, the bounds are insensitive to the $\ell=3$ partial waves.

Since the $\ell=1$ states decouple, the $\ell\geq 2$ curve in Figure \ref{fig:kink} shows that the horizontal boundary of the all-$\ell$ bound on $\beta_1^2$ and $\gamma_0^2$ cannot be reached if we assume that there are no scalar degrees of freedom in the UV, while the vertical boundary does not have contributions from the heavy scalars.

As with the all-$\ell$ case, the bound on $\beta_1^2$ and $\gamma_0^2$ with $\ell \geq 2$ still mainly results from the four-graviton sum rules and is insensitive to the value of $\alpha$. However, this changes if the UV theory only has higher spin states $\ell\geq 3$, and then the bounds on $\beta_1^2$ and $\gamma_0^2$ are significantly reduced in all directions, as shown in Figure \ref{fig:kink}. Furthermore, for $\ell\geq 3$, we find that the dimension $\alpha$ is fixed to be $\mathcal{O}(1/(M_P^2\Lambda^2))$ and can no longer reach the all-$\ell$ upper bound $\mathcal{O}(1/\Lambda^4)$. To understand this, we can look in the graviton decoupling limit, where all the forward limits of the dispersive sum rules can be used. Notice that the lowest order $st$ null constraint $a^{0000}_{3,1}=a^{0000}_{1,3}$ gives
\begin{equation}\label{null_guanjian}
	0=16\pi \sum_{\ell\geq 4,\text{even};X}(2\ell +1) \int_{\Lambda^2}^{\infty}\frac{\text{d}\mu}{\pi}\frac{\ell(\ell+1)(\ell^2+\ell-8)}{2\mu^5}|c^{00}_{\ell,\mu}|^2\,,
\end{equation}
where we have imposed $c^{\mathbb{12}}_{\ell,\mu}=0$ for $\ell<3$ as intended and used the fact that $c^{00}_{\ell,\mu}=0$ for odd $\ell$. When $\ell\geq 4$, the right-hand side of \eref{null_guanjian} is non-negative for all $\ell$ and $\mu$. Therefore, \eref{null_guanjian} implies $c^{00}_{\ell,\mu}=0$ for all $\ell$ and $\mu$, which leads to $a^{0000}_{k,n}=0$ for $k+n\geq 3$. This means that the scalar self-interaction operators with dimension 8 or higher must vanish. Therefore, all these operators must be suppressed by appropriate powers of $M_P$ away form the decoupling limit, which is consistent with the numerical results that $\alpha\sim \mathcal{O}(1/(M_P^2\Lambda^2))$ in Figure \ref{fig:kink}. In the language of Section \ref{sec:dimAna}, this suggests that the correspondence $c^{00}_{\ell,\mu}\Leftrightarrow \Lambda/M_P$ is the only option. The numerical results are essentially the same if the UV theory only has higher spin states $\ell\geq 4$, as shown in Figure \ref{fig:kink}.

On the other hand, a theory with only $\ell \geq 5$ does not exist. The reason is exactly the same as why a pure scalar theory with only $\ell\geq 3$ does not exist. Notice that, in the presence of gravitons, the lowest order null constraint for 2-to-2 scalar scattering in the forward limit is $0=\big\langle \partial_t^4 F^{0000}_{3,\ell}(\mu,0) \big\rangle$. With the assumption $\ell\geq 5$, it becomes
\begin{equation}
	0=16\pi \sum_{\ell\geq 6,\text{even};X}(2\ell +1) \int_{\Lambda^2}^{\infty}\frac{\text{d}\mu}{\pi}\bigg(\frac{\ell^6}{18}+\frac{\ell^5}{6}-\frac{55 \ell^4}{36}-\frac{10 \ell^3}{3}+\frac{233 \ell^2}{36}+\frac{49 \ell}{6}\bigg)\frac{1}{\mu^8}|c^{00}_{\ell,\mu}|^2\,,
\end{equation}
where we have used the fact that $c^{00}_{\ell,\mu}=0$ for odd $\ell$ again. When $\ell\geq 6$, the right hand side is positive for all $\mu$. Therefore we can infer that $c^{00}_{\ell,\mu}=0$, which in turns implies that $1/M_P^2=0$. Therefore, such kind of scalar-tensor theories can not exist.

On the other hand, if the UV theory only has finite spins $\ell<\ell_{\rm M}$, causality bounds will restrict $\beta_1^2$ and $\gamma_0^2$ to be zero. We can easily see this for $\ell<4$ directly from the sum rules. To this end, note that we have $F^{+++-}_{1,\ell}(\mu,t)=(\cdots) d^{\ell,\mu,t}_{4,0}+(\cdots)\partial_td ^{\ell,\mu,0}_{0,-4}+(\cdots)\partial_td ^{\ell,\mu,0}_{4,0}=0$ for $\ell<4$ from their definitions. If the UV partial amplitude has no support for $\ell \geq 4$ spins, we can infer that $F^{+++-}_{1,\ell}(\mu,t)=0$ for all $\ell$. Therefore, we have
\begin{equation}
	-\frac{\gamma_0}{M_P^4}t^2=\bigg\langle F^{+++-}_{1,\ell}(\mu,t)\bigg\rangle =0 \,,~~\text{for all $-\Lambda^2<t<0$}\,,
\end{equation}  
which suggest that $\gamma_0=0$. (The same result can also be obtained by using $F^{+++-}_{2,\ell}$.) Similarly, for $F^{+0-0}_{1,\ell}$, we have $F^{+0-0}_{1,\ell}(\mu,t)=(\cdots) d^{\ell,\mu,t}_{2,2}=0$ for $\ell<2$.
If the UV partial amplitude has no support for $\ell \geq 2$ spins, we have 
\begin{equation}
	-\frac{1}{M_P^2}- \frac{\beta_1^2}{M_P^4}t^2=\bigg\langle F^{+0-0}_{1,\ell}(\mu,t)\bigg\rangle =0 \,,~~\text{for all $-\Lambda^2<t<0$}\,,
\end{equation}
which leads to $\beta_1=0$. Moreover, this also leads to $M_P\to \infty$, which means that this kind of scalar-tensor theory is excluded by causality bounds. For some larger $\ell_{\rm M}$, we have numerically verified that $\beta_1^2$ and $\gamma_0^2$ are also forced to be zero by positivity bounds.

\begin{figure}[h]
\centering
    \includegraphics[scale=0.22]{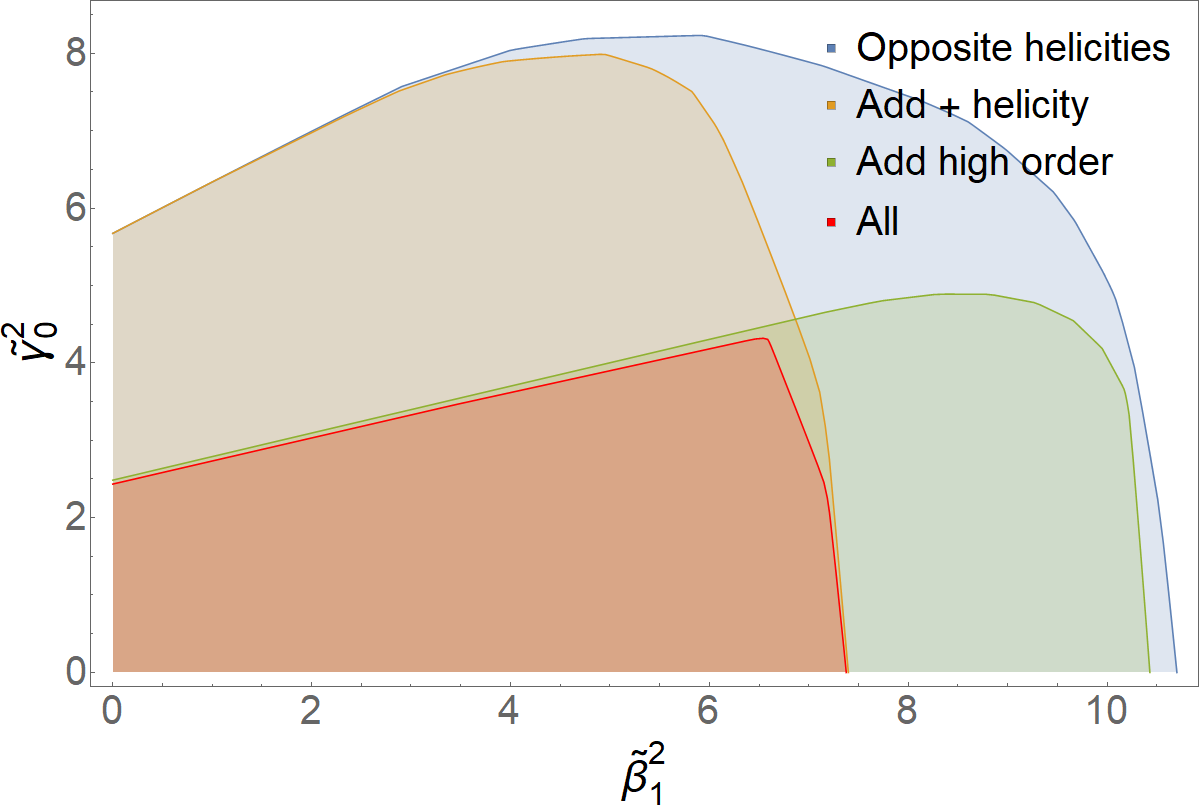}
 \caption{Bounds on $\bi^2_1$ and $\gi^2_0$ using different sum rules. The ``Opposite helicities" bound is obtained using 4 sum rules with $F^{++--}$ and $F^{+-+-}$, {\it i.e.}, \cref{F21211,F21212,F22112,F22113}. The ``Add $+$ helicity" bound is obtained by adding the $F^{++++}$ sum rules, {\it i.e.}, \cref{F22211,F22212}, while the ``Add high order" bound is obtained by adding 3 high order sum rules (\ref{F21213}), (\ref{F21214}) and (\ref{F22114}), compared with the ``Opposite helicities" case. The ``All" bound is the one shown in Figure \ref{fig:yualpha}. Kink 1 is located near the interaction point of ``Add $+$ helicity" and ``Add high order".}
 \label{fig:gi0different}
\end{figure}

It is also instructive to see how presence or absence of certain sum rules impacts the bound on $\gi_0$ and $\bi_1$. Starting from a small set of sum rules with only the graviton scattering with opposite helicities, Figure \ref{fig:gi0different} shows that adding sum rules from $\mc M^{++++}$ significantly strengthens the bound on $\bi_1$, the coefficient of $\phi \mc{G}$, while adding high order sum rules with opposite helicities primarily enhances the bound on $\gi_0$, the coefficient of $\mc{R}^{(3)}$. The former is due to the fact that $\bi_1^2$ also appears in the sum rule with $F^{++++}$, unlike $\gi_0^2$. The latter is because $\gi_0^2$ can manifest in high order sum rules with opposite helicities. It can be observed that Kink 1 is roughly located at the intersection point of the two choices of adding extra sum rules in the optimization.

As mentioned previously, there has been a lot of recent interest in astrophysics to probe the $\phi \mc{G}$ operator in strong and dynamical gravity environments, as it is one of the leading viable scalar-curvature couplings beyond Einstein's gravity. In some of these settings, the $\phi \mc{G}$ operator and the scalar kinetic term are assumed to be the only extra Lagrangian terms, which is a fine-tuned scenario we shall consider in Section \ref{sec:finet}. From the perspective of the causality bounds, fine-tuned cases often have tighter bounds, sometimes incredibly restrictive, as we shall see. Therefore, one deduces conservative conclusions when comparing the generic causality bounds with the observational results for the fine-tuned theories. On the other hand, when constraining the $\bi_1$ coefficient from binary compact star observations, the results, in contrast to the causality bounds, are less sensitive to corrections from higher dimensional operators, as $\phi \mc{G}$ gives the leading contributions in the astrophysical computations. Thus, the proxy model with only $\phi \mc{G}$ should capture the salient astrophysical features of a generic model. With these in mind, we shall use the observational bounds to constrain the cutoff of the scalar-tensor EFT in the following.

\begin{table}[h]
\centering
\setlength{\arrayrulewidth}{0.25mm}
\renewcommand{\arraystretch}{1.2}
\begin{tabular}{|c||c|c|c|c|c|}
\hline
$\Lambda(10^{-10}$eV)        &  BHXB \cite{Yagi:2012gp}  &  ${\rm NS}$ \cite{Saffer:2021gak}     & ${\rm GW}_{\rm BBH}$  \cite{Perkins:2021mhb}     &  ${\rm GW}_{\rm NSBH}$  \cite{Lyu:2022gdr}       & ${\rm GW}_{\rm CB}$  \cite{Lyu:2022gdr} \\
\hline \hline
Conservative    & $1.5$ & $2.2$ & $1.6$  &  $2.1$ &  $2.4$  \\
\hline
Kink 1     & $1.4$ & $2.1$  & $1.6$   &  $2.0$  &  $2.3$  \\
\hline
Fine-tuned  & $0.27$  & $0.39$ & $0.30$  &  $0.38$  &  $0.43$  \\
\hline
\end{tabular}
\caption{Lower bounds on the EFT cutoff $\Lambda$ (in units of $10^{-10}\text{eV}$) from binary compact star observations, by converting $\tilde \beta_1 M_P \log(\Li/m_{\text{IR}})/\Lambda^2<\beta_1^{\rm obs}$ with various choices of the dimensionless $\tilde \beta_1$. ``Conservative'', ``Kink 1'' and ``Fine-tuned'' refer to choosing $\tilde \beta_1$ to be, respectively, its global upper bound, at Kink 1 in Figure \ref{fig:yualpha} and when $\tilde{g}^{T_1}_{4,0}=g^{T_1}_{4,0}\Lambda^6 M_P^2/\log\left(\Lambda/m_{\text{IR}}\right)=0.01$ and $\tilde{g}^{T_1}_{6,0}=g^{T_1}_{6,0}\Lambda^{10} M_P^2/\log\left(\Lambda/m_{\text{IR}}\right)=0.01$.    
The BHXB bound comes from a black hole low mass X-ray binary (A0620-00), while the neutron star (NS) bound is from the mass-radius measurement of pulsar J0740+6620. The other bounds are extracted from constraints from the dephasing of gravitational waves: ${\rm GW}_{\rm BBH}$ is inferred from combining several low mass binary black hole events, ${\rm GW}_{\rm NSBH}$ is from the best neutron star black hole binary event (GW200115) and ${\rm GW}_{\rm CB}$ is extracted from combining several BBH and NSBH events.}
\label{tab:GW}
\end{table}

Having established the sharp causality bounds, we can convert these experimental bounds to the bounds on the cutoff of the theory for a few specific EFTs. A specific EFT has a specific dimensionless $\tilde \beta_1$, and the lower bound on $\Lambda$ can be extracted from
\be
\label{beta1beta1obs}
\f{\tilde \beta_1 M_P \log\(\Li/m_{\text{IR}}\)}{\Lambda^2} = \beta_1<\beta_1^{\rm obs}\,,
\ee
where $\beta_1^{\rm obs}$ is an observational bound and we choose $1/m_{\text{IR}}$ to be the Hubble scale. In Table \ref{tab:GW}, we have surveyed three EFTs: for the ``Conservative'' case we take $\tilde \beta_1$ to be its maximum value in Figure \ref{fig:yualpha}, which is valid regardless of values of other Wilson coefficients; ``Kink 1'' refers to the Kink 1 in Figure \ref{fig:yualpha} (the bounds on $\Lambda$ for Kink 2 being almost the same); the ``Fine-tuned'' case is when we take $\tilde \beta_1$ to be its maximum value when higher order coefficients $g^{T_1}_{4,0}$ and $g^{T_1}_{6,0}$ are set to be relatively small $\tilde g^{T_1}_{4,0}=0.01,~\tilde g^{T_1}_{6,0}=0.01$, which will significantly reduce the upper bound on $|\tilde \beta_1|$ (see Figure \ref{fig:yug4}) and in turn impose much stronger bounds on the EFT cutoff. The observational constraints on $\beta_1^{\rm obs}$ in Table \ref{tab:GW} are obtained as follows. ``BHXB'' refers to a bound from a black hole low mass X-ray binary A0620-00 where the black hole's companion is a K-type main-sequence star, whose matter is accreted into the black hole to produce X-rays \cite{Yagi:2012gp}. ``${\rm GW}_{\rm BBH}$'' \cite{Perkins:2021mhb} , ``${\rm GW}_{\rm NSBH}$'' \cite{Lyu:2022gdr} and ``${\rm GW}_{\rm CB}$'' \cite{Lyu:2022gdr} are bounds from the newly available observational channel of gravitational waves. The BBH one is inferred from combining several most constraining low mass binary black hole events, the NSBH case is from the neutron star black hole binary event (GW200115) and the CB one is extracted from combining several BBH and NSBH events. All the bounds from these binaries are derived from the fact that the scalar Gauss-Bonnet coupling gives rise to corrections to the binary's orbital decay rate due to extra scalar dipole radiation that is of  ``$-1$ PN" order. For X-ray observations from the BHXB, this results in corrections to the period, while for gravitational waves this leads to dephasing in the waveforms. It is also not surprising that the strongest bounds come from the lower mass compact stars, as gravity is the strongest in those environments. On the other hand, ``NS'' refers to a bound from the mass-radius measurement of pulsar J0740+6620, taking the most conservative case with respect to the choice of the equation of state for the neutron star \cite{Saffer:2021gak}. 

To clarify, Table \ref{tab:GW} constrains the cutoffs of the models at a few special places within the causality bounds. These special models serve as benchmarks for causality-bounds-compatible EFT models. Since $\tilde \beta_1$ is known for these specific models, one can convert the experimental bound to the lower bound on the cutoff, via Eq (\ref{beta1beta1obs}). This exercise is related to causality bounds because these specific models are special only from the point of view of the causality bounds. It can be viewed as a succinct survey about how the cutoff may change within the causality bounds, which serves to gauge the strength of the causality bounds in the context of observational bounds.

It is worth noting that the current experimental bounds are parametrically close to the bound from the cosmic censorship that there should be no naked singularity. An intriguing feature of the scalar-Gauss-Bonnet coupling $\phi \mc{G}$ is that it leads to a finite radius singularity at $r^4_S=48\beta_1^2$ \cite{Sotiriou:2014pfa}. So if we require the singularity to be cloaked by the event horizon, for a spherically symmetric black hole with its horizon at $r_h$, this leads to an upper bound on the coupling
\be
\beta_1^2<\f{r_h^4}{48} \,,
\ee
(For generic $f(\varphi) \mc{G}$, we will have $(\d f(\varphi(r_h))/\d \varphi)^2<{r_h^4}/{48}$.) This will impose stronger bounds on the cutoff $\Lambda$ for smaller black holes for which $r_h$ is smaller. Taking $r_h$ to be $\sim$km would lead to $\Lambda\gtrsim \mc{O}(1)\times 10^{-10}$eV if $\tilde\beta_1$ saturates its upper bound.

\subsection{More generic Gauss-Bonnet couplings}

\begin{figure}
	\begin{minipage}[b]{0.5\columnwidth}
		\centering
		\subfloat[$\gamma_0=0$]{\includegraphics[width=0.9\linewidth,height=5cm]{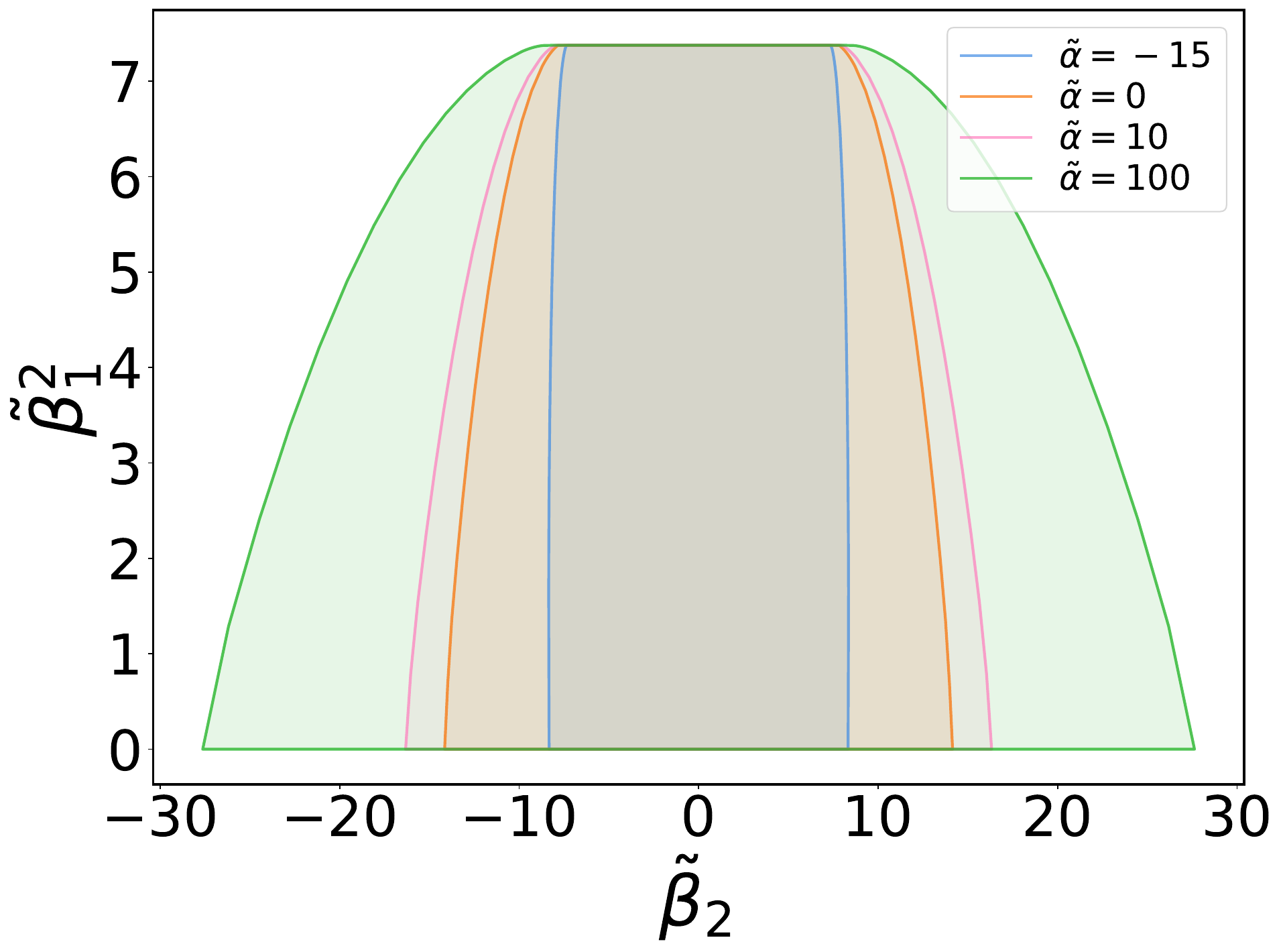}	
		}
		\\
		\subfloat[$\beta_1=0$]{\includegraphics[width=0.9\linewidth,height=5cm]{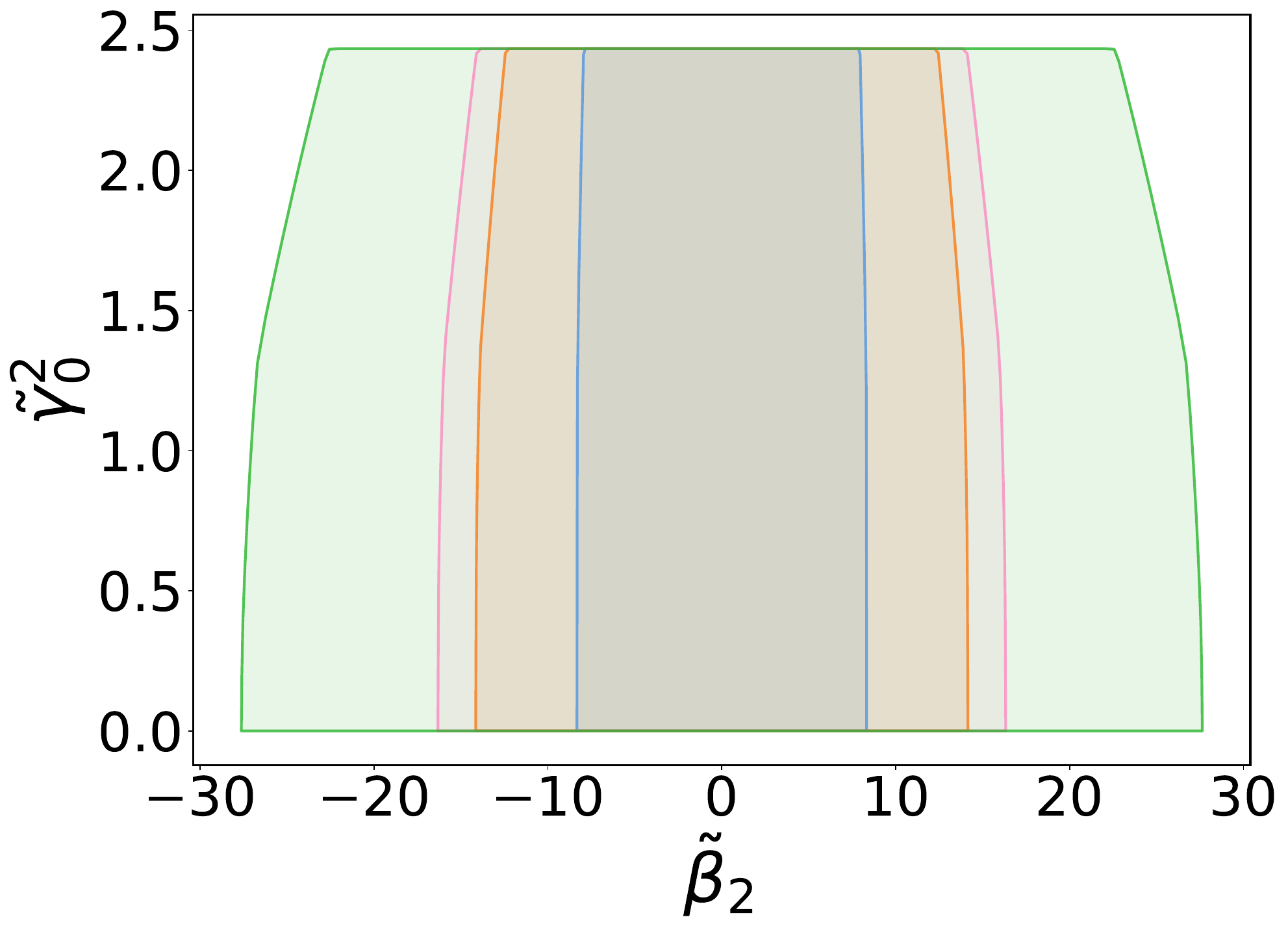}}
	\end{minipage}
	\begin{minipage}[b]{0.5\columnwidth}
		\centering
		\subfloat[$\tilde{\gamma}_0/\tilde{\beta}_1=+ 1$]{\includegraphics[width=0.9\linewidth,height=5cm]{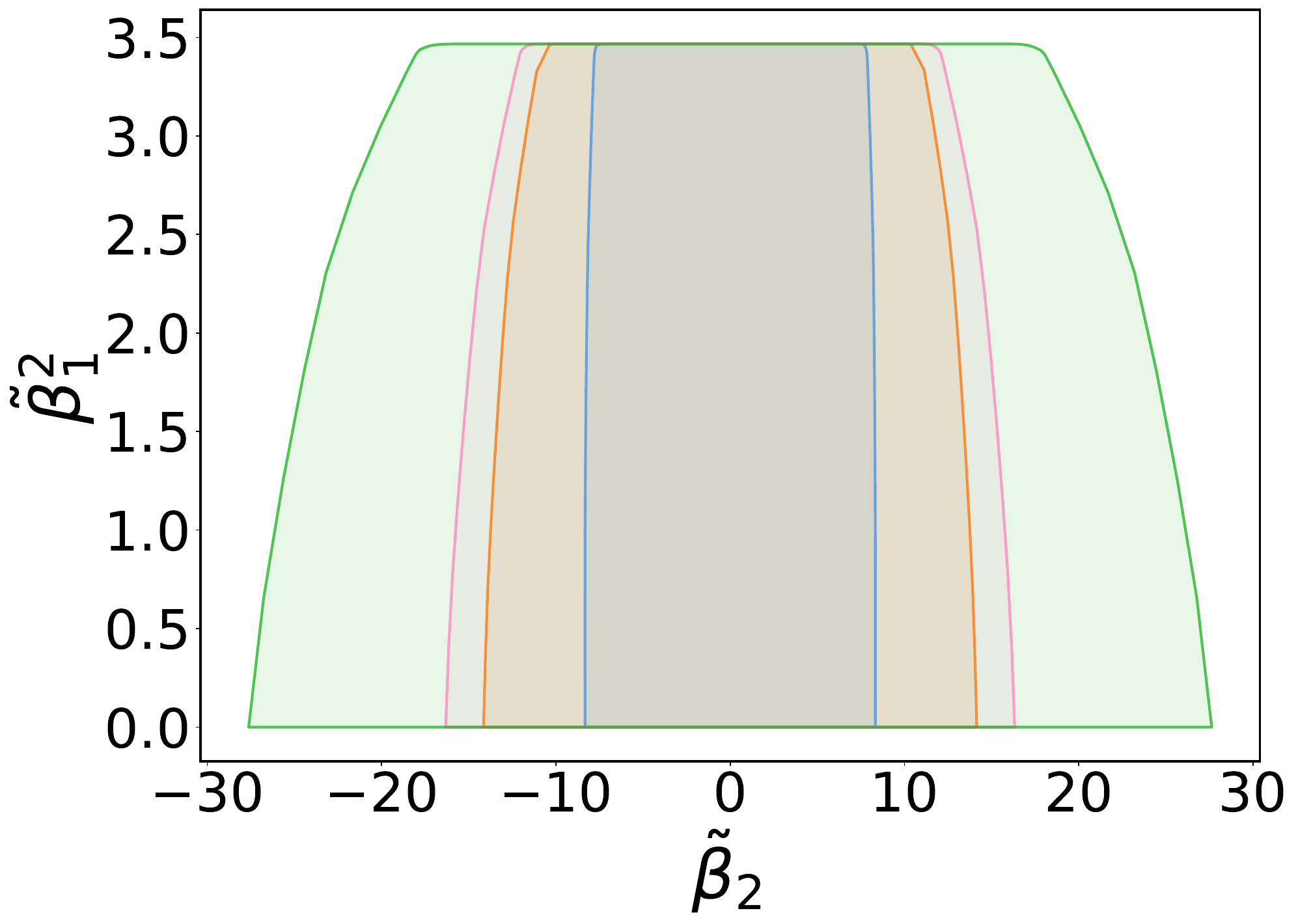}}
		\\
		\subfloat[$\tilde{\gamma}_0/\tilde{\beta}_1=- 1$]{\includegraphics[width=0.9\linewidth,height=5cm]{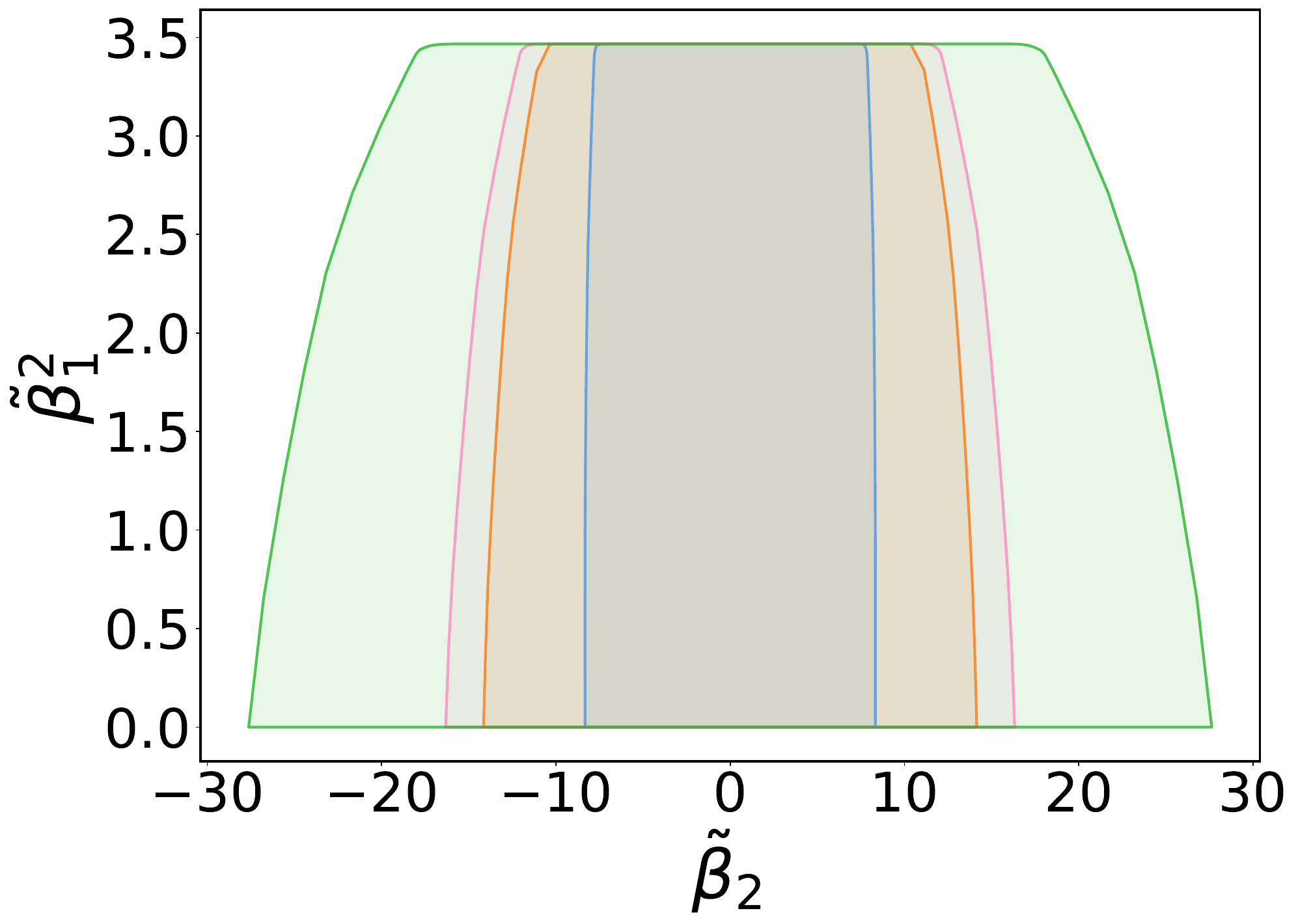}}
	\end{minipage}
	\caption{Bounds on $\beta_2$, $\beta_1^2$ and $\gamma_0^2$ for various $\alpha$. Four representative cross sections are chosen that pass through the $\bi_2$ axis. The dimensionless coefficients are defined as follows: $\tilde{\alpha}={\alpha M_P^2\Lambda^2}/{\log(\Lambda/m_{\text{IR}})}$, $\tilde{\beta}_2=\beta_2 \Lambda^2/\log(\Lambda/m_{\text{IR}})$, $\tilde{\beta}_1^2=\beta_1^2 \Lambda^4/(M_P^2 \log(\Lambda/m_{\text{IR}}))$ and $\tilde{\gamma}_0^2=\gamma_0^2 \Lambda^8/(M_P^4 \log(\Lambda/m_{\text{IR}}))$.}
	\label{fig:d4}
\end{figure}

Beyond the linear coupling $\phi \cal G$, more complex couplings to the Gauss-Bonnet invariant $f(\phi) \cal G$ have also been extensively investigated, due to their ability to generate the interesting mechanism of (spontaneous) scalarization \cite{Silva:2017uqg, Doneva:2017bvd}. Scalarization is a phenomenon where the scalar field can develop a non-trivial profile (non-constant across the space) when the curvature near  compact stars is sufficiently strong in some modified gravity models. All known scalarization mechanisms for the black hole utilize the scalar-Gauss-Bonnet coupling $f(\phi) \cal G$ \cite{Doneva:2022ewd} (for neutron stars other ways are possible). In particular, they usually rely on the $\phi^2\mc{G}$ term to give rise to an effective scalar mass term that is of a tachyonic nature, which is quenched by higher order terms to achieve stable scalarization. Near a compact star, the effective mass term has to be sufficiently negative to develop tachyonic instabilities, as there are positive contributions to the effective potential coming from the mass and angular momentum of the compact star. As the effective mass term is proportional to $\cal G$, this requires the curvature near the compact star to be sufficiently large, so smaller compact stars tend to be scalarized.

In this subsection, we shall constrain the function $f(\phi)$ to the next leading order $\phi^2$. The leading two orders of the $f(\phi)$ expansion can contribute to tree-level 4-leg amplitudes and thus can be effectively constrained with our method. More specifically, we will be concerned with the coefficients in the following Lagrangian terms
\begin{equation}\label{nr14}
\mc{L}\supset \sqrt{-g}\( \frac{\gamma_0}{3!}{\cal R}^{(3)}+\frac{\beta_1}{2!}\phi {\cal G} +\frac{\beta_2}{4}\phi^2 {\cal G} + \frac{\alpha}{2}(\partial \phi)^4 \)
\end{equation}
and investigate how the bound on $\bi_1$ and $\bi_2$ varies with $\ai$ and $\gamma_0$.

Since we have seen in the last subsection that keeping only the quadratic pieces of $\gamma_0$ and $\beta_1$ in the dispersion relations quite accurately captures the bounds, we shall directly adapt that approximation here.  Since $\beta_2$ is only contained in the dispersive sum rule with $F^{++00}_{2,\ell}$, in order to derive a bound on $\beta_2$, we need to specify the value of $\alpha$, as discussed in Section \ref{sec:alpha}. In this subsection, we assume $\alpha \sim \mathcal{O}({1}/{(M_P^2\Lambda^2)})$. (The case of $\alpha\sim \mathcal{O}({1}/{\Lambda^4})$, particularly the scaling of $\beta_2$ with respect to $\ai$, will be explored in Section \ref{sec:xcheck}.)

A few cross sections of the 3D bounds for $\beta_2$, $\beta_1^2$ and $\gamma_0^2$ are shown in Figure \ref{fig:d4}, all of the cross sections passing through the $\bi_2$ axis. While the boundaries of $\beta_1^2$ and $\gamma_0^2$ are insensitive to the value of $\ai$, the limit of $\bi_2$ changes dramatically with it. On the other hand, the limit of $\bi_2$ is insensitive to the values of $\beta_1^2$ and $\gamma_0^2$. We can see that $\beta_2$ is of order $\mathcal{O}({1}/{\Lambda^2})$ and becomes greater when $\alpha$ increases, consistent with the estimate in Section \ref{sec:dimAna}.   Another obvious visual pattern is that the bounds are symmetric with respect to the plane of $\beta_2=0$, up to about 0.01\%. We have also numerically verified that, for $\tilde \alpha\geq -15$, the global minimum and maximum of $\beta_2$ are on the line of $\gamma_0=0$ and $\beta_1=0$.  We have chosen both the cross sections of $\gamma_0\Lambda^2/(\beta_1 M_P)= \pm 1$ because there is a $\gamma_0\beta_1$ term in the sum rules from $\mc{M}^{++-0}$ and its crossing. However, as we can see explicitly in Figure \ref{fig:d4}, the effect of this term is minimal in the results.

The reason why we choose $\tilde \ai$ from $\tilde \alpha=-15$ to plot Figure \ref{fig:d4} is that $\tilde\ai = -16.091$ is its lower bound. A peculiar feature near the lower bound when $\tilde \alpha=-15$ is that the cross sections in Figure \ref{fig:d4} are almost rectangular. From the 3D point of view, the bounds on $\beta_2$, $\beta_1$ and $\gamma_0$ are basically a ``plate'' lying on the  plane of the $\beta_1^2$ and $\gamma_0^2$ directions with a ``thickness'' along the $\beta_2$ direction. This means that the bounds on $\bi_1$ and $\gi_0$ are almost independent of the bounds on $\bi_2$, which is not at all obvious from the sum rules and we have not identified the underlying reason.

\begin{table}[h]
\centering
\setlength{\arrayrulewidth}{0.25mm}
\renewcommand{\arraystretch}{1.2}
\begin{tabular}{|c||c|c|c|c|c|c|c|c|}
\hline
           &\multicolumn{3}{c|}{$\bi_2>0$}   &\multicolumn{5}{c|}{$\bi_2<0$}   \\
\hline
EoS        &  MS1  &  MPA1     & WFF1 &  MS1  &  MPA1  & ENG  & APR4   & WFF1   \\
\hline \hline
$\Lambda(10^{-10}$eV)    & 1.4 & 2.1 & 3.4 & 2.9 & 4.0 & 4.7 & 5.2 & 5.9    \\
\hline
\end{tabular}
\caption{Lower bounds on the EFT cutoff $\Li$ from the NS-WD binary J0348+0432 for various equations of state (EoS) of the neutron star.}
\label{tab:bi21}
\end{table}

\begin{table}[h]
\centering
\setlength{\arrayrulewidth}{0.25mm}
\renewcommand{\arraystretch}{1.2}
\begin{tabular}{|c||c|c|c|}
\hline
NS-WD pair        &  J0348+0432   &  J1012+5307      &  J2222-0137    \\
\hline \hline
$\Lambda(10^{-10}$eV)    & 4.0 & 3.6 & 3.7    \\
\hline
\end{tabular}
\caption{Lower bounds on the EFT cutoff $\Li$ from three NS-WD binaries, assuming
the MPA1 equation of state for the neutron star and $\bi_2<0$. }
\label{tab:bi22}
\end{table}

In the absence of the $\bi_1$ term, the $\bi_2$ term can result in scalarization in compact stars \cite{Silva:2017uqg,Doneva:2017bvd}. As argued in Section \ref{sec:dimAna}, a relatively suppressed $\bi_1$, compared with $\bi_2$, is parametrically natural for a generic UV completion. We will also verify this numerically in Section \ref{sec:xcheck}. The $\bi_2$ coupling has been observationally constrained with binary pulsars, as it can also give rise to dipole scalar radiation, which affects the orbital decay rate of the binaries. Ref \cite{Danchev:2021tew} considered three neutron star-white dwarf (NS-WD) binaries, and has put some upper bounds on $|\bi_2|$ for both $\bi_2>0$ and $\bi_2<0$. Note that for black holes $\bi_2$ needs to be positive in order to have tachyonic instabilities, which is necessary for scalarization to occur, but for neutron stars both signs of $\bi_2$ are possible. Similar to the case of $\bi_1$, we can convert these experimental constraints to bounds on the cutoff $\Li$ by saturating $|\bi_2|$ with the causality bounds for $\ai=0,~\bi_1=0$: $|\tilde \bi_2 | / {\Li^2}  <  | \bi_2^{obs} |$. In Table \ref{tab:bi21} the uncertainties of the $\Li$ bounds are surveyed for various different equations of state of the neutron stars, while in Table \ref{tab:bi22} the $\Li$ bounds extracted from three different NS-WD binaries are compared.

The observational bounds on $\bi_1$ and $\bi_2$ have only been established individually. Assuming the scalar interacts with gravitational strength, the $\bi_1$ term gives rise to the leading order effects in the relevant astrophysical processes, and we may take the observational upper bounds on $\bi_1$ to be valid for all reasonably valued $\bi_2$. This is a rough approximation, which we shall be content with in the absence of any rigorous 2D astrophysical analysis so far.  

Then, if we let the observational bound on $\bi_2$ saturate its causality bound (letting the right end of the thick green line, which lies on the $\tilde\bi_2$ axis, align with the most right end of the causality bound on $\tilde\bi_2$ when $\tilde\bi_1=0$), which fixes the cutoff of the EFT and in turn fixes the causality bound on $\bi_1$ and $\bi_2$, the causality bound on $\bi_1$ and $\bi_2$ may be used to tentatively exclude parameter regions in the $\bi_1$ and $\bi_2$ space and compare with the observational bounds. See Figure \ref{fig:b12} for a comparison with a couple of choices of the observational bounds. We emphasize that this is not intended to be a rigorous comparison. Rather, it is just an attempt to estimate potential interactions between the causality bounds and the observational constraints, which should be updated when suitable astrophysical analyses become available.

\begin{figure}[thbp]
  \centering
  \includegraphics[scale=0.17]{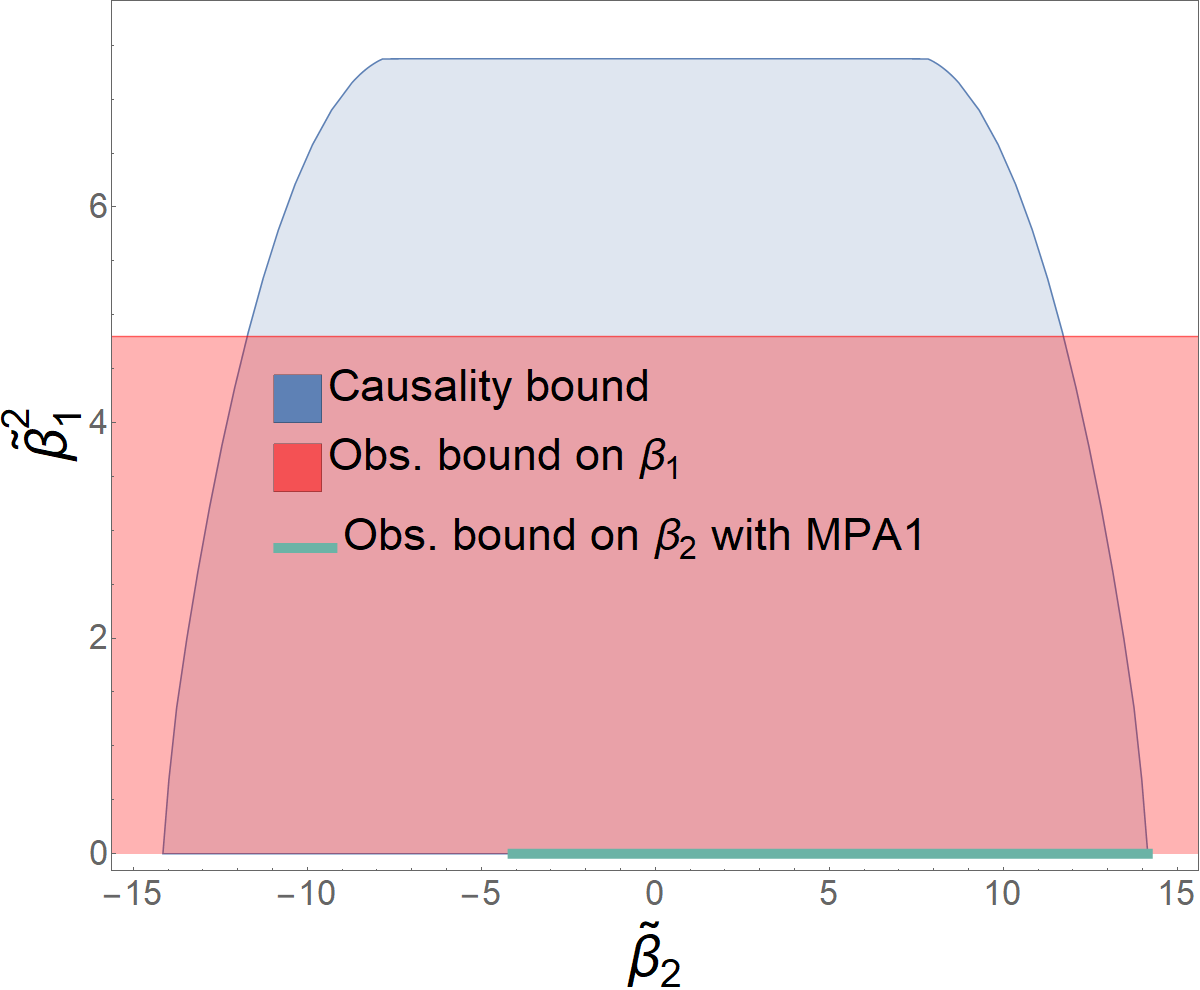}
~~~~
   \includegraphics[scale=0.17]{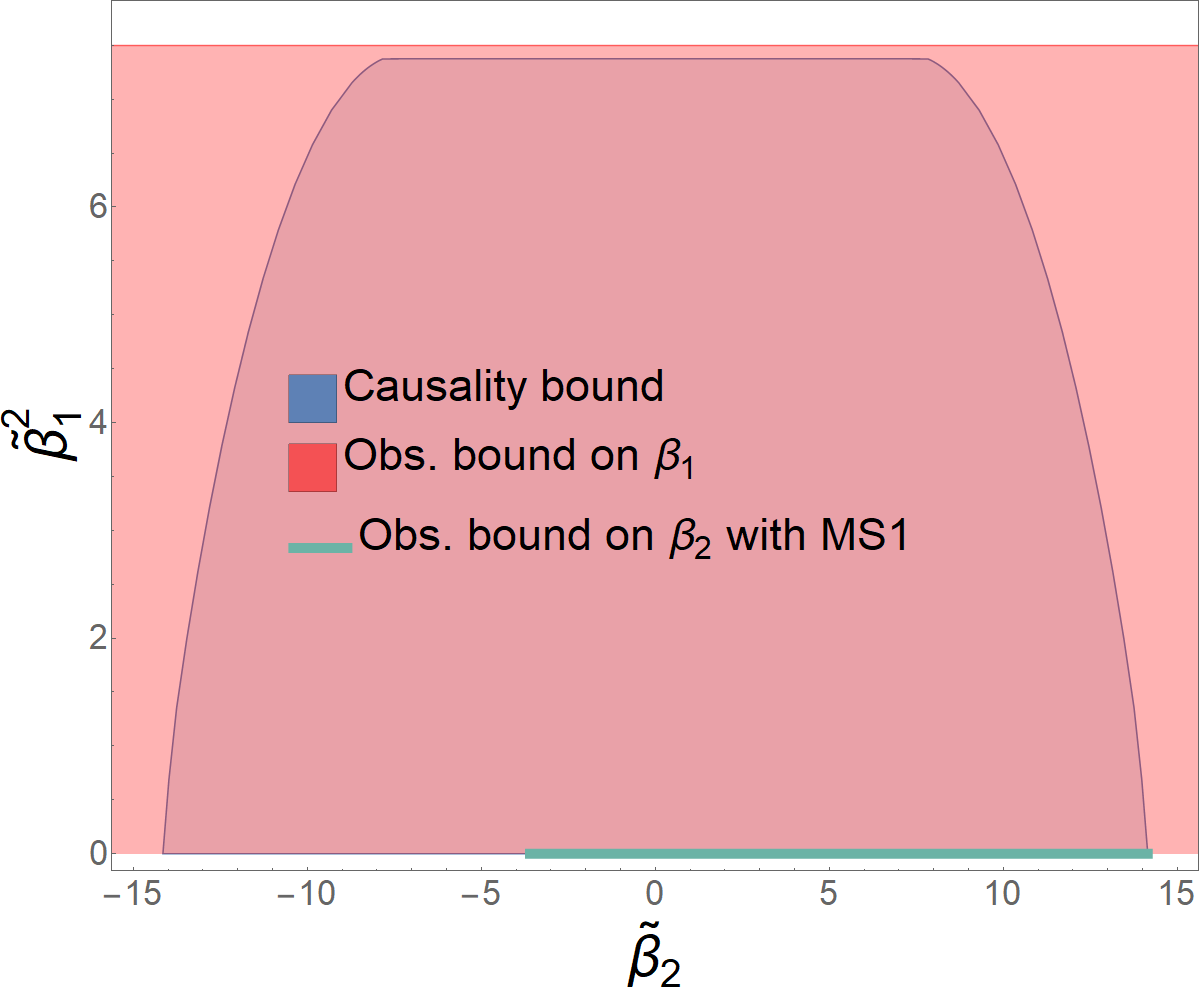}
  \caption{Comparisons of the observational bounds and causality bounds on $\bi_1$ when the observational bound on $\bi_2$ saturates its causality bound. The ``Causality bound" corresponds to the theoretical bounds for the case of $\ai=0$ and $\gi_0=0$ in Figure \ref{fig:d4}. ``Obs.~bound on $\bi_1$'' means the observational upper bound on $\bi_1$ coming from the most stringent gravitational wave constraint of \cite{Lyu:2022gdr}, assuming $\bi_2=0$. ``Obs.~bound on $\bi_2$ with MPA1/MS1'' means the observational upper bound on $\bi_2$ from the orbital decay rate measurements using the MPA1/MS1 equation of state for the neutron stars \cite{Danchev:2021tew}, assuming $\bi_1=0$.  The observational bound on $\bi_2$ using MS1, shown on the right subfigure, is the most conservative one in \cite{Danchev:2021tew}.  We let the observational bound on $\bi_2$ saturate its causality bound, which means that the right end of the thick green line (lying on the $\tilde\bi_2$ axis) is at the most right end of the causality bound on $\tilde\bi_2$ when $\tilde\bi_1=0$.}
    \label{fig:b12}
\end{figure}

\subsection{Other six derivative terms}
\label{sec:boundd6}

\begin{figure}
	\begin{minipage}[b]{0.5\columnwidth}
		\centering
		\subfloat[$\tilde{\gamma}_1$]{\includegraphics[width=0.76\linewidth]{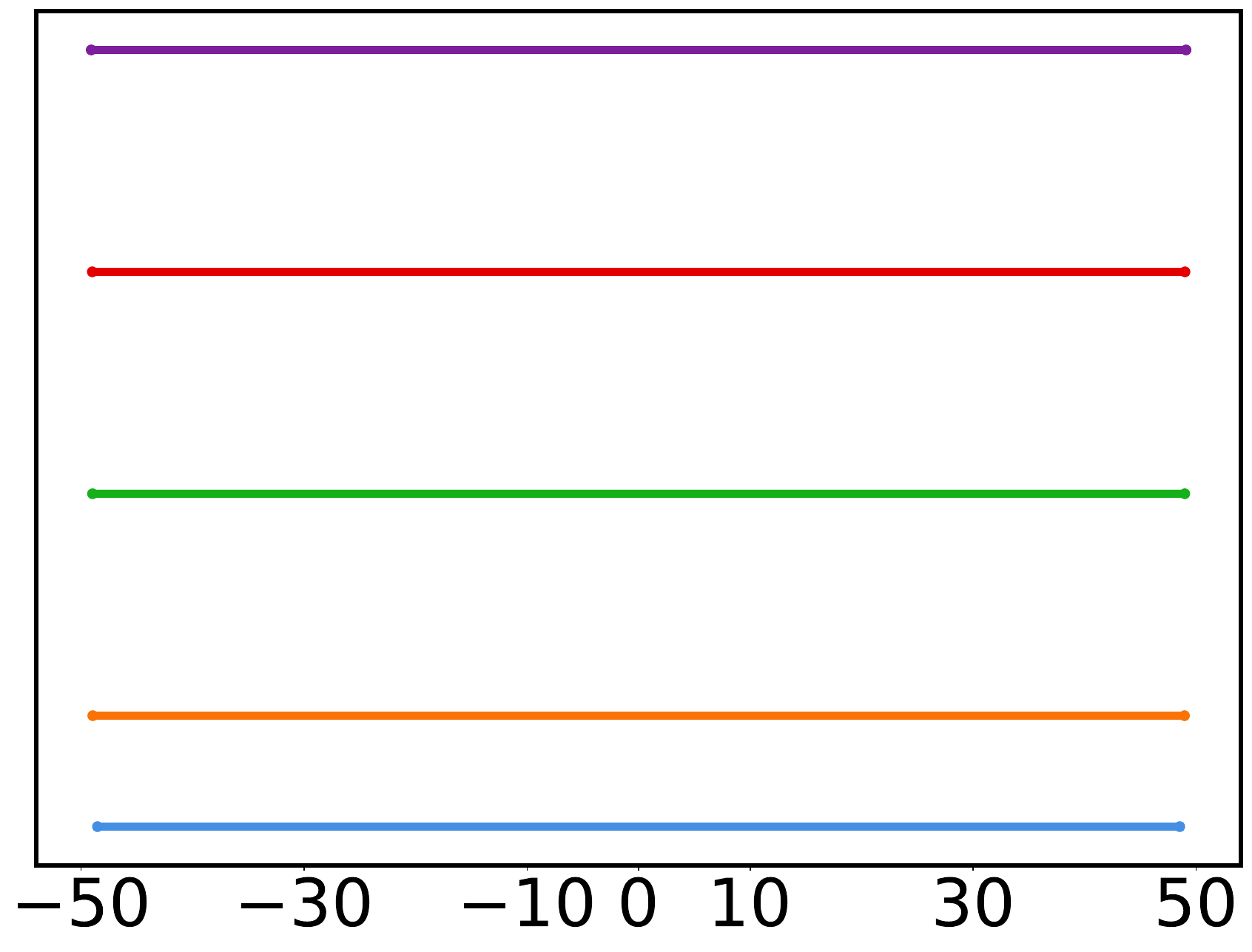}
		\label{fig:d6_a}	
	}
		\\
		\subfloat[$\tilde{\gamma}_2$]{\includegraphics[width=0.76\linewidth]{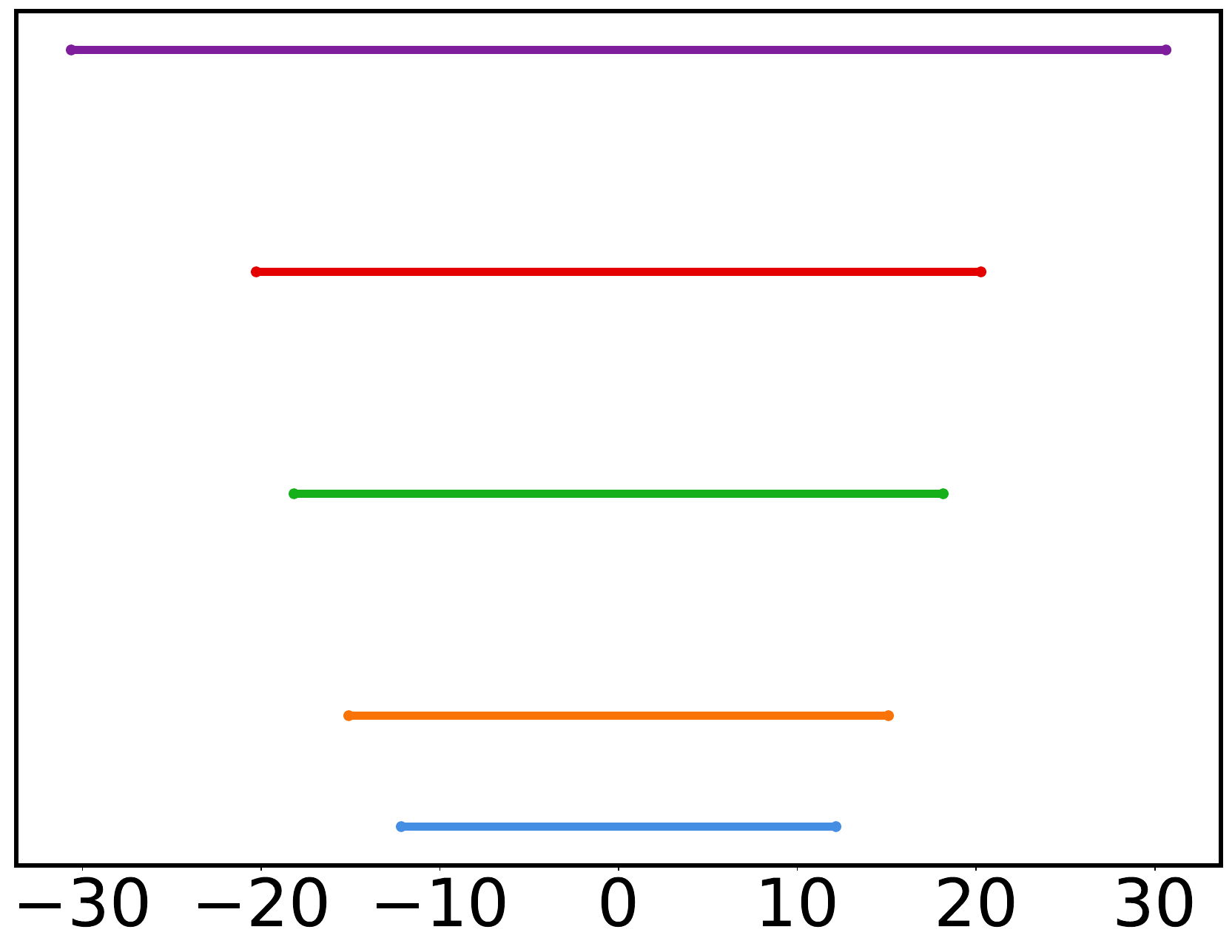}}
		\label{fig:d6_c}
	\end{minipage}
	\begin{minipage}[b]{0.5\columnwidth}
	\centering
	\subfloat[$\tilde{\gamma}_3$]{\includegraphics[width=0.76\linewidth]{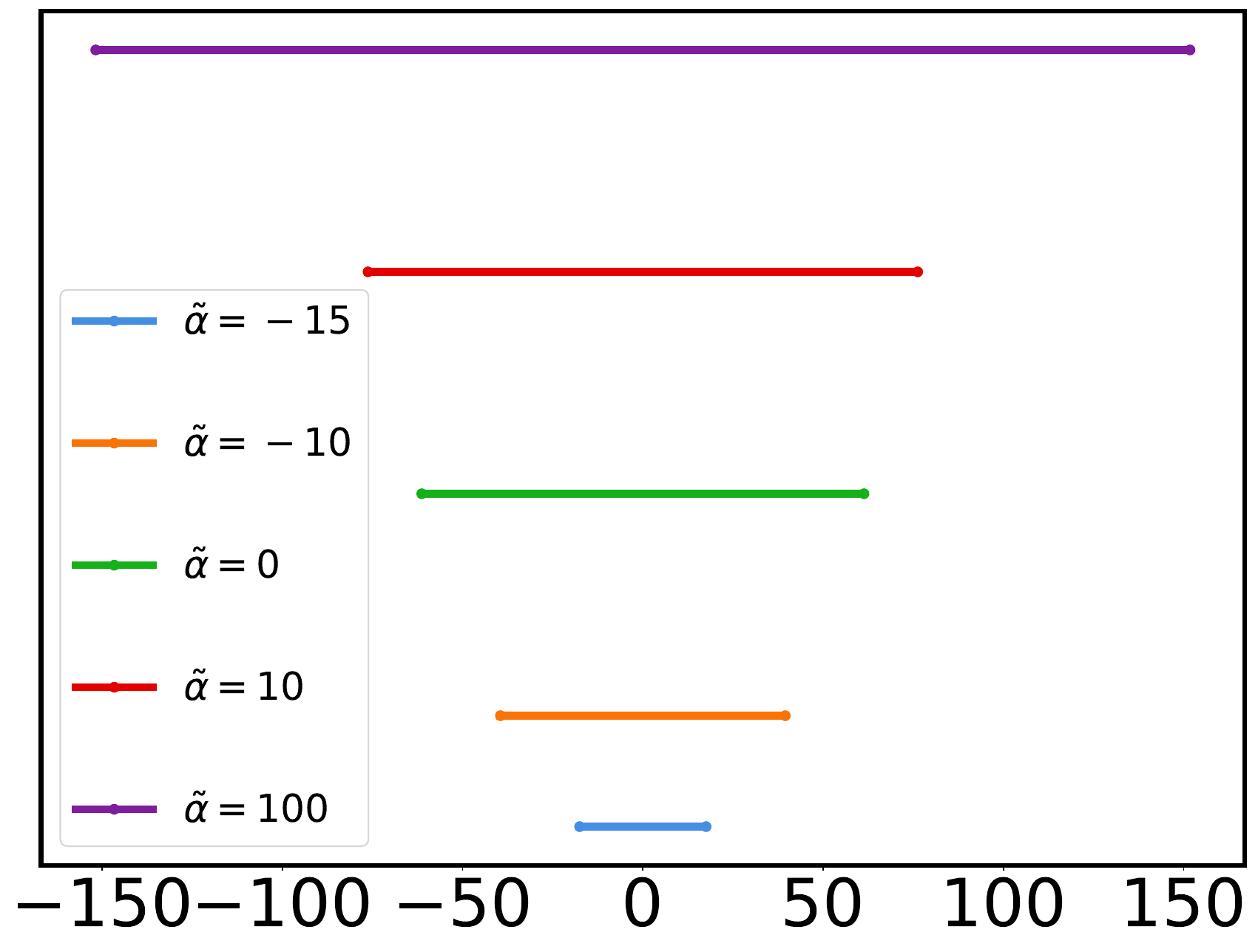}}
	\label{fig:d6_b}
	\\
	\subfloat[$\tilde{\gamma}_4$]{\includegraphics[width=0.76\linewidth]{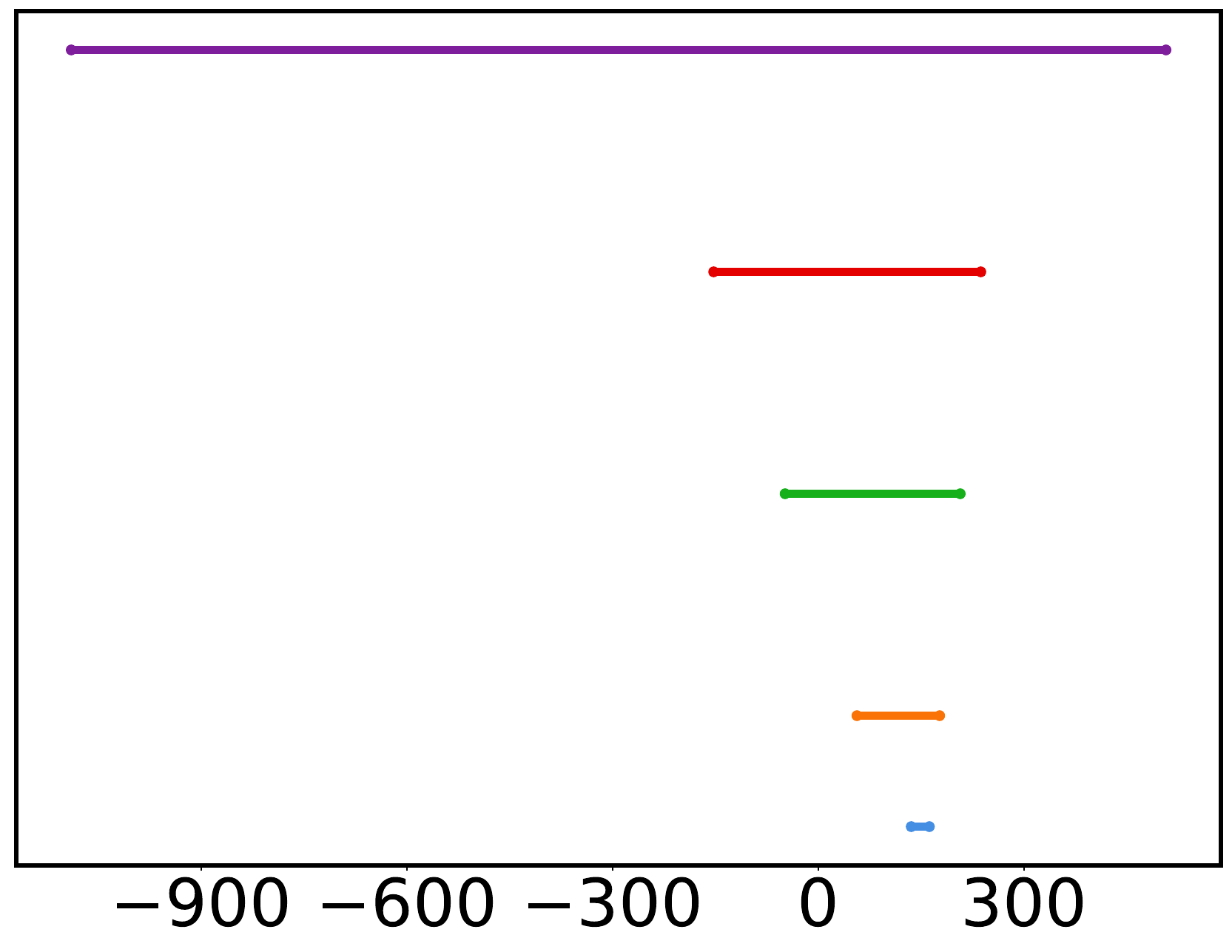}}
	\label{fig:d6_d}
	\end{minipage}
	\caption{Bounds on $\gamma_{1,2,3,4}$ for various values of $\alpha$. As expected, these bounds scale up as $\ai$ increases, that is, as $\alpha$ transits from $\mathcal{O}({1}/{(M_P^2\Lambda^2)})$ and $\mathcal{O}({1}/{\Lambda^4})$. Note that $\tilde{\alpha}={\alpha M_P^2\Lambda^2}/{\log(\Lambda/m_{\text{IR}})}$, $\tilde{\gamma}_1 =\gamma_1 \Lambda^4/(M_P\log(\Lambda/m_{\text{IR}})) $, $\tilde{\gamma}_2 =\gamma_2 \Lambda^4/\log(\Lambda/m_{\text{IR}}) $, $\tilde{\gamma}_3 =\gamma_3 \Lambda^4M_P/\log(\Lambda/m_{\text{IR}}) $ and $\tilde{\gamma}_4 =\gamma_4 \Lambda^4 M_P^2/\log(\Lambda/m_{\text{IR}}) $.  }
	\label{fig:d6}
\end{figure}

In the previous subsections, we have mainly focused on Lagrangian terms with four derivatives, except for $\gamma_0$, which is a term with six derivatives. In this subsection, we shall compute the positivity bounds on all other six derivative terms: $\gamma_1$, $\gamma_2$, $\gamma_3$ and $\gamma_4$. After all, as argued in Section \ref{sec:dimAna}, if the scalar interactions are constrained to be comparable with the gravitational interactions (for example, when $\ai\sim 1/(M_P^2\Lambda^2)$), all the six derivative terms should be all suppressed by $1/\Lambda^4$ (cf.~\eref{action22}).

Numerically, for $\tilde\ai=\alpha M_P^2 \Lambda^2/\log(\Lambda/m_{\text{IR}})\geq -15$, we find that $\gamma_1$, $\gamma_2$ and $\gamma_3$ reach their global bounds ({\it i.e.}, the loosest bounds) approximately when $\gamma_0=0$ and $\beta_1=0$, and the bounds on $\gamma_4$ are actually insensitive to the values of $\gamma_0^2$ and $\beta_1^2$. To see how the bounds change with $\ai$, we shall present the bounds on $\gamma_1$, $\gamma_2$, $\gamma_3$ and $\gamma_4$ for $\tilde\ai=\{-15,-10,0,10,100\}$, with $\gamma_0=0$ and $\beta_1=0$, as shown in Figure \ref{fig:d6}.

In Section \ref{sec:dimAna}, we have argued that $\gamma_1$ must be $\mathcal{O}({M_P}/{\Lambda^4})$ and insensitive to the value of $\ai$. This is what we see with the SDP computations: In Figure \ref{fig:d6_a}, we see that the bounds on $\gamma_1$ depend very weakly on the value of $\alpha$. Despite this, as mentioned in Section \ref{sec:alpha}, because $\gamma_1$ only appears in the sum rules of $F^{+++0}_{1,\ell}$ and $F^{+++0}_{2,\ell}$, we can not use our setup to numerically find the bounds on $\gamma_1$ without specifying $\ai$. Although hardly visible in Figure \ref{fig:d6_a}, the bound on $\gamma_1$ does become weaker very slowly when $\alpha$ increases. In fact, as will be shown in Section \ref{sec:xcheck}, the value of $\gamma_1$ will be of the same order even when $\ai$ is very large, for example, $\alpha\sim\mathcal{O}({1}/{\Lambda^4})$. On the other hand, the bounds on $\gamma_{2,3,4}$ become weaker significantly when $\alpha$ increases, so there is a sizable difference between the case of $\alpha\sim\mathcal{O}({1}/{(M_P^2\Lambda^2)})$ and $\alpha\sim\mathcal{O}({1}/{\Lambda^4})$, which is again consistent with the analysis in Section \ref{sec:dimAna}.

\subsection{Coefficients for large $\ai$}
\label{sec:xcheck}

In the previous subsections, we have seen that the bounds on some coefficients depend on the size of $\ai$. For the explicit bounds on these coefficients, we have chosen $\alpha\sim\mathcal{O}({1}/{(M_P^2\Lambda^2)})$, in which case the scalar self-interaction, along with other interactions involving the scalar, is comparable with the spin-2 interactions. In this subsection, we shall also explore the possibility that $\alpha\sim\mathcal{O}({1}/{\Lambda^4})$ when the scalar self-interaction is strong, close to its upper limit. As already argued in Section \ref{sec:dimAna}, the EFT operators scale differently in this case. In this subsection, we shall confirm these estimates with explicit calculations.

For concreteness, we will concentrate on the following three Wilson coefficients:
\begin{itemize}
	\item[$\gamma_1$:] the coefficient of the $\phi R^3$ term in the Lagrangian, which appears in the $F^{+++0}_{(1,2),\ell}(\mu,t)$ sum rule and thus is of order $\mathcal{O}({M_P}/{\Lambda^4})$ for any value of $\ai$, according to the dimensional analysis in Section \ref{sec:dimAna}; 	
	\item[$\beta_2$:] the coefficient of $\phi^2 {\cal G}$, which only appears in the sum rule with $F^{++00}_{2,\ell}$ and thus is of order $\mathcal{O}({1}/{\Lambda^2})$ when $\alpha\sim\mathcal{O}({1}/{(M_P^2\Lambda^2)})$ or $\mathcal{O}({M_P}/{\Lambda^3})$ when $\alpha\sim\mathcal{O}({1}/{\Lambda^4})$;
	\item[$g^{M_4}_{0,2}$:] the coefficient of the $s^2 u^2$ term in the amplitude $\mathcal{M}^{+0-0}$, which appears in the forward-limit sum rule $g^{M_4}_{0,2}=\langle2 |c^{+0}_{\ell,\mu}|^2/\mu^5 \rangle$ and thus should be of order $\mathcal{O}({1}/{(M_P^2 \Lambda^6)})$ and insensitive to $\ai$, meaning that the coefficient of a 8 derivative operator $\phi^2 R^2\pd^4$ in the Lagrangian should scale as $\mathcal{O}({1}/{\Lambda^6})$ (this is an expected example that does not follow \eref{OphiRscale} with $\tilde N_\phi=\lfloor N_\phi/2\rfloor$, which would suggest an incorrect scaling of $\mathcal{O}({M_P}/{\Lambda^7})$).
\end{itemize}

\begin{figure}
	\begin{minipage}[b]{0.5\columnwidth}
		\centering
		\subfloat[]{\includegraphics[width=0.92\linewidth]{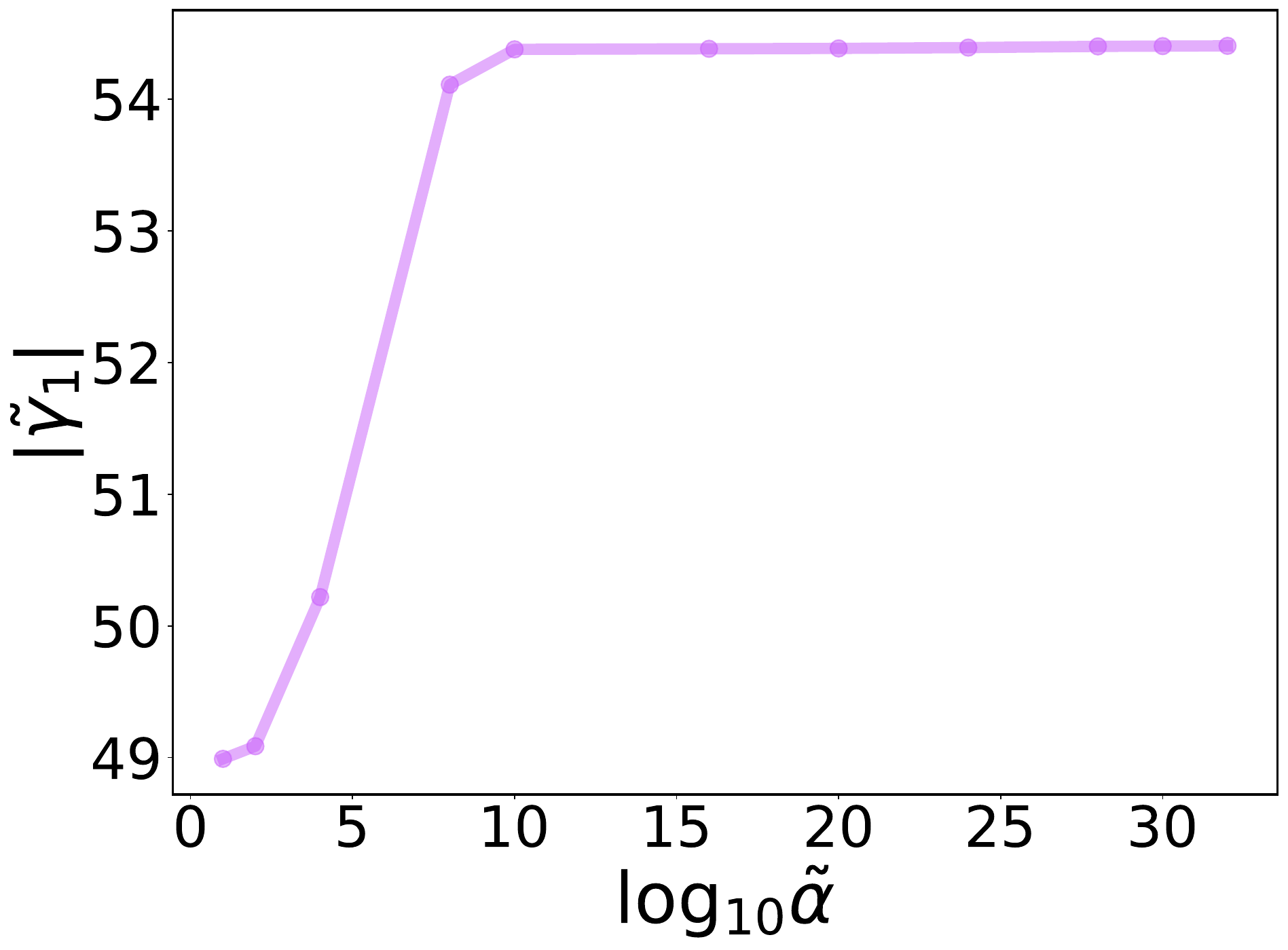}
			\label{fig:p10e_d61}
		}
	\end{minipage}
	\begin{minipage}[b]{0.5\columnwidth}
		\centering
		\subfloat[]{\includegraphics[width=0.92\linewidth]{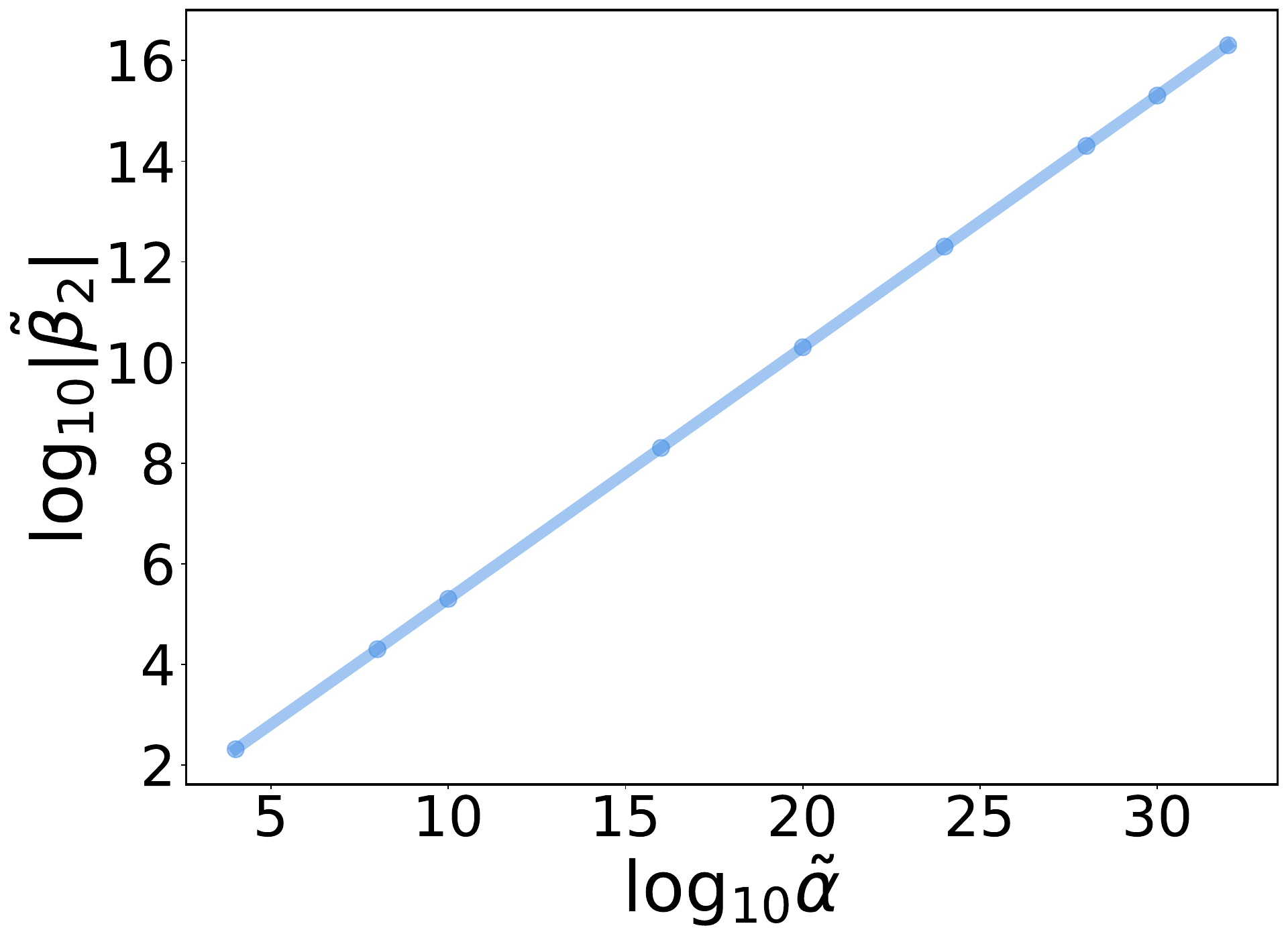}
			\label{fig:p10e_d41}
		}
	\end{minipage}
	\caption{Upper bounds on $|{\gamma}_1|$ and $|{\beta}_2|$ for large values of $\ai$. The upper bound on $|\gamma_1|$ is insensitive to the value of $\alpha$, while the upper bound on $|\beta_2|$ increases rapidly with $\alpha$. The line in subfigure (b) is nearly linear with a slope of $1/2$, which is consistent with the argument in Section \ref{sec:dimAna}. Note that $\tilde{\gamma}_1=\gamma_1 \Lambda^4/(M_P\log(\Lambda/m_{\text{IR}}))$, $\tilde{\beta}_2=\beta_2 \Lambda^2/\log(\Lambda/m_{\text{IR}})$ and $\tilde{\alpha}=\alpha \Lambda^2 M_P^2/\log(\Lambda/m_{\text{IR}})$.}
	
\end{figure}

We shall proceed by probing the bounds on these coefficients with a number of different hierarchies between $M_P$ and $\Lambda$, up to a fiducial big ratio of $M_P/\Lambda=10^{16}$. That is, we shall compute the bounds for these coefficients with $\alpha$ up to $\alpha\sim 10^{32}\log({\Lambda}/{m_{\rm IR}})/{(M_P^2\Lambda^2)}$. As we will see in the following, the bounds on coefficients such as $\gamma_1$ are insensitive to the changes in $\ai$, even for large $\ai$, whereas the bounds on coefficients such as $\beta_2$ increase significantly as $\ai$ increases.

\begin{figure}
\centering
\includegraphics[width=0.5\linewidth]{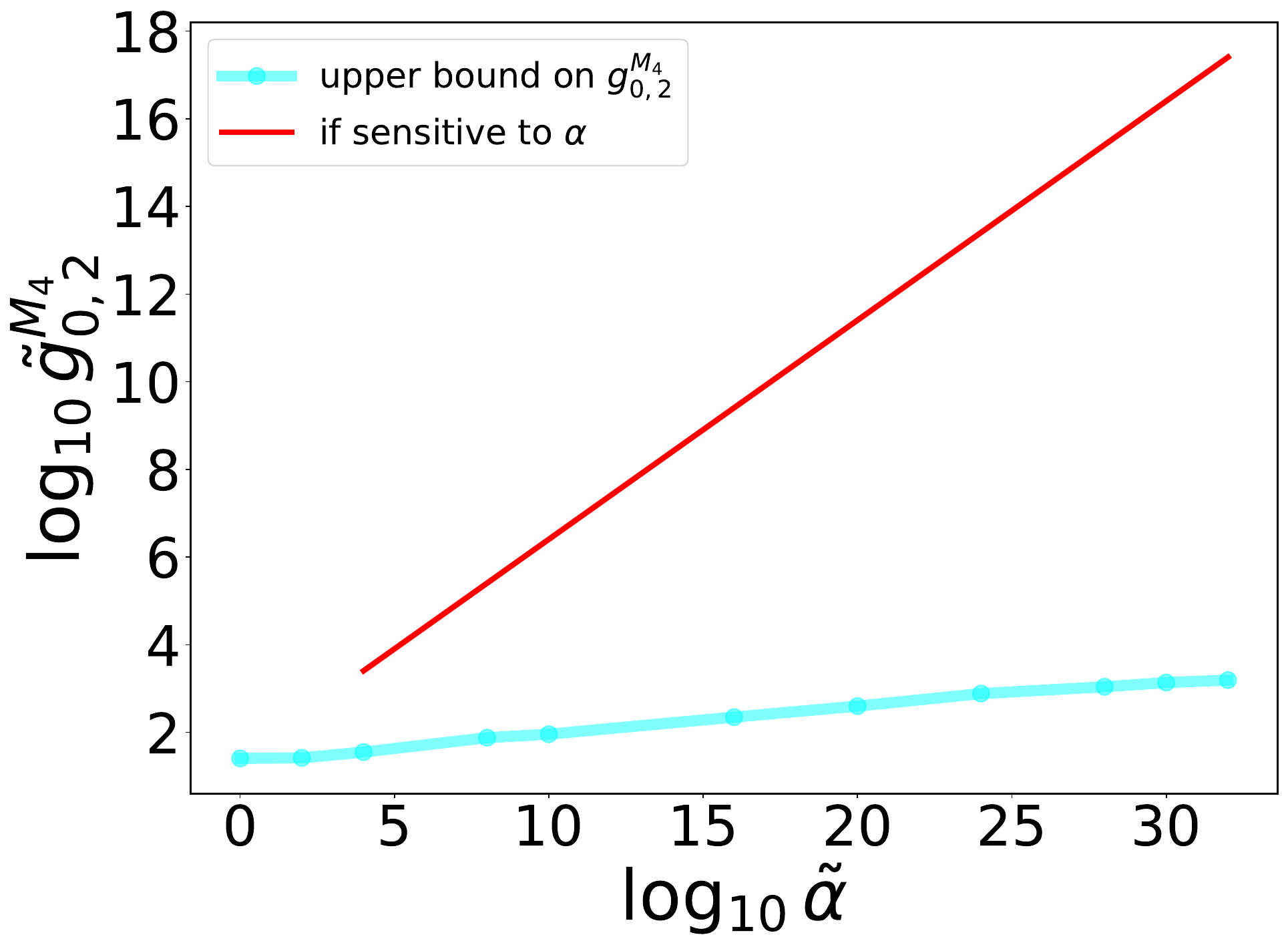}
\caption{Upper bound on $g^{M_4}_{0,2}$ for large $\alpha$. The upper bound on $\tilde{g}^{M_4}_{0,2}$ increases more rapidly than that on $\tilde{\gamma}_1$, but much slower than that on $\tilde{\beta}_2$. We have defined that $\tilde{\alpha}=\alpha \Lambda^2 M_P^2/\log(\Lambda/m_{\text{IR}})$ and $\tilde{g}^{M_4}_{0,2}=g^{M_4}_{0,2}M_P^2\Lambda^6/\log(\Lambda/m_{\text{IR}})$. We see that, in contact to $\beta_2$, its slope is much less than $1/2$ (the case of the red line), meaning that it is insensitive to $\ai$.}
\label{fig:p10e_d81_log}
					
\end{figure}

First, let us see how the upper bound on $|\gamma_1|$ varies for different $\ai$ when $\gamma_0=0$ and $\beta_1=0$. The choice of $\gamma_0=0$ and $\beta_1=0$ makes the obtained bounds approximately the global upper bounds on $|\gamma_1|$ for all $\gamma_0$ and $\beta_1$.  As we see in Figure \ref{fig:p10e_d61}, the upper bound on $|\gamma_1|$ remains stable  around ${M_P}/{\Lambda^4}$ even when $\ai$ has changed for many orders of magnitude, completely consistent with the dimensional analysis in Section \ref{sec:dimAna}. Note that the dimensional analysis in Section \ref{sec:dimAna} suggests that $\gi_1$ is insensitive to $\ai$, because Eq.~(\ref{gi1disp}) does not contain $\hat{c}_{\ell, \mu}^{00}$. In deriving sharp bounds on $\gi_1$, we will use Eq.~(\ref{gi1disp}) along with other dispersion relations. Although the dispersion relations containing $\gamma_1$ do not contain $c^{00}_{\ell,\mu}$, the rest dispersion relations do contain $c^{00}_{\ell,\mu}$. More specifically, $\gamma_1$ is contained in the dispersion relations with $F^{+++0}_{1,\ell}$ and $F^{+++0}_{2,\ell}$. When we add them into the numerical procedure and impose the positivity condition $B_{P_X,\ell}(\mu)\succeq 0$, it is necessary for the $(+0,+0)$ element of $B_{P_X,\ell}(\mu)$ to be positive, which means that the dispersion relation with $F^{+0-0}_{k,\ell}$ must be included. For $k\geq 2$, including $F^{+0-0}_{k,\ell}$ will in turn contribute to the $(00,+0)$ and $(+0,00)$ elements of $B_{P_X,\ell}(\mu)$. Hence, it is necessary for the $(00,00)$ element of $B_{P_X,\ell}(\mu)$ to be positive, so we need to include $F^{0000}_{k,\ell}$ in the SDP. $F^{0000}_{k,\ell}$ does contain $c^{00}_{\ell,\mu}$, which eventually leads to $\gi_1$ having some dependence on $\ai$. These very indirect links also mean that the dependence of $\gi_1$ on $\ai$ is very weak, which is exactly what we see in Figure \ref{fig:p10e_d61}.

For the upper bounds on $|\beta_2|$, we again look at the direction along $\gamma_0=0$ and $\beta_1=0$, which gives approximately the global upper bounds on $|\beta_2|$. In Figure \ref{fig:p10e_d41}, we see that the upper bound on $|\beta_2|$ scales with the square root of $\ai$, accurate to several decimal places for large $\ai$,
\be
\label{biaiscaling}
|\tilde \bi_2^{(\rm up)}| \propto \tilde \ai^{\f12} \,,
\ee
precisely as what is argued in Section \ref{sec:dimAna}. To see why this is consistent with the dimensional analysis in Section \ref{sec:dimAna}, note that a large $\ai$ of order $\ai\sim \Lambda^{-4}$ can be viewed as originating from a large hierarchy between $M_P$ and $\Lambda$: $\ai\sim \Lambda^{-4} = (M_P/\Lambda)^2 (M_P\Lambda)^{-2}$. So the horizontal axis in Figure \ref{fig:p10e_d41} can be viewed as depicting different values of $(M_P/\Lambda)^2$. In Section \ref{sec:dimAna}, we argued that, switching from the $\hat c^{00}_{\ell,\mu}\Leftrightarrow {\Lambda}/{M_P}$ correspondence to $\hat c^{00}_{\ell,\mu}\Leftrightarrow 1$,  the upper bound on $|\beta_2|$ will be boosted by an extra factor of ${M_P}/{\Lambda}$, to be of order  $|\bi_2|\sim M_P/\Lambda^3= ({M_P}/{\Lambda}){\Lambda^{-2}}$, while for $\ai$ the boost factor is $({M_P}/{\Lambda})^2$. This explains the $1/2$ exponent in the fitted \eref{biaiscaling}.

The fact that ${\beta}_2$ increases significantly with $\ai$ has interesting implications for the scalarization models. Notice that a scalarization model should accommodate non-hairy black holes, so the $\bi_1$ coefficient is usually assumed to be negligible, since a sizable $\phi \mc{G}$ coupling generically leads to a hairy black hole \cite{Sotiriou:2013qea}. The fact that the causality bounds allow the $\bi_2$ coefficient to generically have an enhancement of a factor of up to ${M_P}/{\Lambda}$ implies that the $\bi_2$ coupling can be naturally stronger than the $\bi_1$ coupling. This can be achieved by UV models where the scalar interacts with the heavy states stronger than the gravitational force.

Regarding the bounds on $g^{M_4}_{0,2}$, from the sum rule $g^{M_4}_{0,2}=\langle2 |c^{+0}_{\ell,\mu}|^2/\mu^5 \rangle$, we know that the lower bound on $g^{M_4}_{0,2}$ is 0, so let us compute its upper bound. Again, explicitly computation shows that $g^{M_4}_{0,2}$ reaches its global upper bounds when $\gamma_0=0$ and $\beta_1=0$. The dimensional analysis of the sum rule suggests that the bound should be insensitive to the value of $\ai$. Indeed, in Figure \ref{fig:p10e_d81_log}, we see that the upper bound on $g^{M_4}_{0,2}$ only depends on $\ai$ relatively weakly, although more sensitively than the case of $\gamma_1$. This can be seen by comparing with the red line with slope $1/2$, which is for the case if the upper bound were really sensitive to $\alpha$. This example underlies the importance of rigorous calculations if we want to accurately capture the bounds on a specific coefficient.

\subsection{Fine-tuned EFTs}
\label{sec:finet}

Up to now, we have considered generic scalar-tensor EFTs without any {\it a priori} constraints on the Wilson coefficients. The bounds on them purely come from unitary and causality of all possible UV theories, which as we have seen actually defines a power counting scheme for the higher dimensional EFT operators in the Lagrangian. However, for various reasons, one often devises models that fine-tune some of the Wilson coefficients to zero or be suppressed. These reasons may be of a UV nature, where certain UV symmetries or other mechanisms may prohibit the EFT from possessing certain operators; or, there could be some phenomenological considerations to have certain coefficients highly suppressed or tuned to zero so as to make the model fit the observational data. Of course, many results about the leading $\phi\mc{G}$ coupling are insensitive to or independent of the higher order operators, as per the standard EFT power-counting. The same may not be said about other hairy black hole models or the scalarization models. In this subsection, we shall explore the consequences of fine-tunings for a couple of examples in scalar-tensor theory. Notice that {\it a priori} fine-tuning of the Wilson coefficients essentially gives rise to extra sum rules from the perspective of bounding the EFT. For example, restricting some coefficients to zero will lead to some null constraints, which sometimes may result in inconsistencies in the dispersion relations, as we shall see.

By this discussion, we wish to further raise the awareness of the theoretical obstacles one may face if one's model-building replies on fine-tunings of the EFT coefficients, explicitly demonstrating what can go wrong for some inconsiderate model-building with the EFT operators, illustrating the slogan ``not everything goes''. That is, the main purpose of this subsection is not to promote these fine-tuned EFTs or trying to confront them with the observational constraints, rather it is to point out a caveat for potential misuses of fine-tunings in the EFT coefficients. It also acts as a reminder that even following the standard EFT power-counting may not be sufficient, and for a precision result there may be a need to also check with the causality bounds. Alternatively, one may reserve the argument and use the results here and/or the observational bounds on the lower dimensional operators to constrain the higher dimensional operators.

Let us first consider a simple example widely used in discussing hairy black holes and spontaneous scalarization, the $f(\phi) \cal G$ model. In this model, apart from the standard kinetic terms for the scalar and the graviton, all terms except the $f(\phi) \cal G$ operators are tuned to zero:
\begin{equation}\label{ft1}
\mc L =\sqrt{-g}\bigg( \frac{M_P^2}{2}R -\frac{1}{2}(\partial \phi)^2 +\(\frac{\beta_1}{2!}\phi +\frac{\beta_2}{4}\phi^2 +... \) {\cal G}   \bigg)\,.
\end{equation}
In this fine-tuned EFT, we no longer need to subtract the higher order $t$ terms in the left hand side of the original dispersion relations, because the tree level EFT amplitude from this model only contains finite terms. Moreover, the absence of the higher order $t$ terms gives rise to a multitude of extra null constraints. Therefore, in the model, we may simply define sum rules with
\begin{equation}\label{ft2}
\tilde{F}^{\mathbb{1234}}_{k,l}(\mu,t):=\frac{\partial_s^k}{k!}\bigg(\frac{s^2 d^{\ell,\mu,t}_{h_{12},h_{43}}}{\mu^2(\mu-s)}c^{\mathbb{12}}_{\ell, \mu}c^{*\bar{\mathbb3}\bar{\mathbb4}}_{\ell, \mu}+\frac{(-s-t)^2d^{\ell,\mu,t}_{h_{14},h_{23}}}{\mu^2(\mu +s+t)}c^{\mathbb{14}}_{\ell, \mu}c^{*, \bar{\mathbb3}\bar{\mathbb2}}_{\ell, \mu}\bigg)\bigg|_{s\to 0}\,.
\end{equation}
Taking this at its face value, it is easy to find {\it inconsistencies}. Let us look at the forward limit of the sum rule with $\tilde{F}^{++--}_{4,\ell}(\mu,0)$:
\begin{equation}\label{ft5}
0=\bigg \langle \tilde{F}^{++--}_{4,\ell}(\mu,0)\bigg \rangle=\bigg \langle \frac{1}{\mu^5}(|c^{++}_{\ell,\mu}|^2+|c^{+-}_{\ell,\mu}|^2)\bigg\rangle\,.
\end{equation}
Clearly, this implies that every 3-leg partial wave amplitudes $c^{++}_{\ell,\mu}$ and $c^{+-}_{\ell,\mu}$ must vanish for all $\mu$. However, this clearly contradicts with the sum rule from $\tilde{F}^{++--}_{2,\ell}(\mu,t)$:
\begin{equation}\label{ft6}
-\frac{1}{M_P^2}\frac{1}{t}=\bigg\langle \frac{1}{\mu^3}d^{\ell}_{0,0}(1+2t/\mu)|c^{++}_{\ell,\mu}|^2+\frac{1}{(\mu+t)^3}d^{\ell}_{4,4}(1+2t/\mu)|c^{+-}_{\ell,\mu}|^2\bigg\rangle=0\,,
\end{equation}
because it forces $M_P$ to be infinitely large or the Newton's gravitational constant to vanish, which prevents the existence of a coupled scalar-tensor theory. This tells us that the $f(\phi) \cal G$ model (\ref{ft1}) should not be taken at its face value. Instead, we should allow some nonzero values for the higher order operators.

\begin{figure}[tbp]
	\centering
	\includegraphics[scale=0.34]{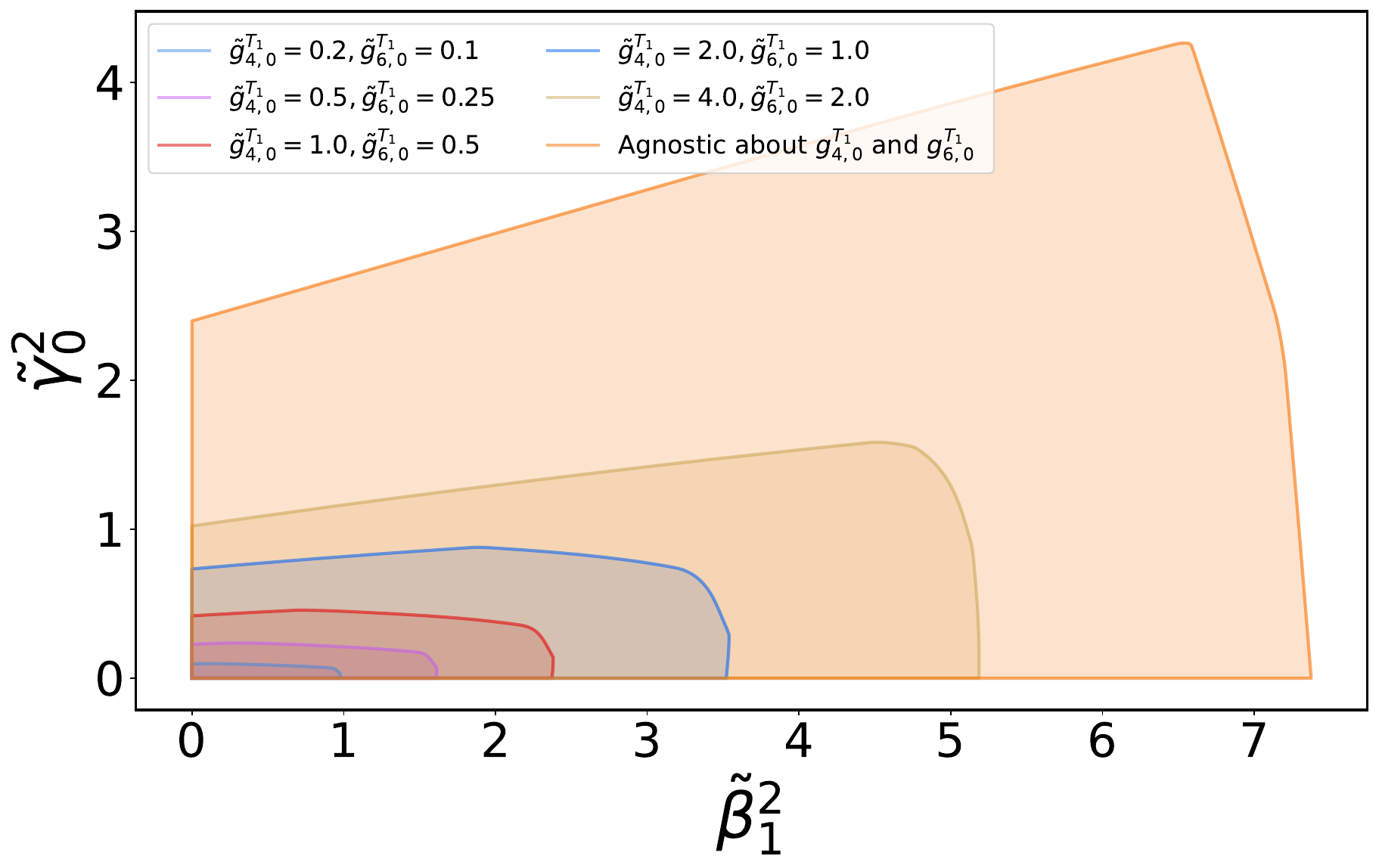}
	\caption{Bounds on $\gamma_0^2$ and $\beta_1^2$ for various $g^{T_1}_{4,0}$ and $g^{T_1}_{6,0}$. Causality bounds require $g^{T_1}_{4,0}$ and $g^{T_2}_{2,0}$ to be nonzero for $\gamma_0$ and $\beta_1$ to have non-vanishing values. We have defined that $\tilde{\beta}_1^2=\beta_1^2 \Lambda^4/(M_P^2 \log(\Lambda/m_{\text{IR}}))$, $\tilde{\gamma}_0^2=\gamma_0^2 \Lambda^8/(M_P^4 \log(\Lambda/m_{\text{IR}}))$, $\tilde{g}^{T_1}_{4,0}=g^{T_1}_{4,0}\Lambda^6 M_P^2/\log\left(\Lambda/m_{\text{IR}}\right)$ and $\tilde{g}^{T_1}_{6,0}=g^{T_1}_{6,0}\Lambda^{10} M_P^2/\log\left(\Lambda/m_{\text{IR}}\right)$. }
	\label{fig:yug4}
\end{figure}

Generically, our formalism provides us with the following criterion to test whether a fine-tuned scalar-tensor model is compatible with unitarity and causality: if we can deduce either $c^{00}_{\ell,\mu}=0$ or $c^{+0}_{\ell,\mu}=0$ or $c^{++}_{\ell,\mu}=c^{+-}_{\ell,\mu}=0$ in the sum rules, then the scalar-tensor theory is inconsistent, in the sense that its Planck mass is forced to be infinitely large. To see this, notice that we have the following sum rules
\bal
		-\frac{1}{M_P^2}+2\alpha t -\gamma_4 t^2&=\bigg\langle F^{0000}_{1,\ell}(\mu,t)\bigg\rangle\,,\\
		-\frac{1}{M_P^2}-\frac{\beta_1^2}{M_P^4}t^2&=\bigg\langle F^{+0-0}_{1,\ell}(\mu,t)\bigg\rangle\,,\\
		-\frac{1}{M_P^2}\frac{1}{t}&=\bigg\langle F^{++--}_{1,\ell}(\mu,t)\bigg\rangle\,.
\eal
Using the explicit expressions of the Wigner d-functions, we can see that $F^{0000}_{1,\ell}$ only contains $|c^{00}_{\ell,\mu}|^2$, $F^{+0-0}_{1,\ell}$ only contains $|c^{+0}_{\ell,\mu}|^2$ and $F^{++--}_{2,\ell}$ only contains a sign-definite combination of $|c^{++}_{\ell,\mu}|^2$ and $|c^{+-}_{\ell,\mu}|^2$. Thus, we can infer that ${1}/{M_P^2}$ must go to zero if $c^{00}_{\ell,\mu}=0$ or $c^{+0}_{\ell,\mu}=0$ or $c^{++}_{\ell,\mu}=c^{+-}_{\ell,\mu}=0$.

Therefore, care should be taken to completely switch off coefficients that are allowed by the symmetries of the EFT.  For the $f(\phi) \cal G$ model to be consistent with the causality bounds, we need to abandon the rigid definition of (\ref{ft1}) and switch back on some other operators in the Lagrangian, for example, the $\gamma_0$ term or some other higher dimensional terms. To determine how large the extra coefficients need to be in order to be consistent with causality and unitarity, we can run our numerical programs. We will see that the bounds on the coefficients of $f(\phi)$ shrink as we tune the higher dimensional coefficients to be smaller. For example, in Figure \ref{fig:yug4}, we can see how the bounds on $\gamma_0^2$ and $\beta_1^2$ reduce as $g^{T_1}_{4,0}$ and $g^{T_1}_{6,0}$ go toward zero along the surface $g^{T_1}_{4,0}=2\Lambda^4 g^{T_1}_{6,0}$. It is interesting to see that these higher order terms in the Lagrangian can have such dramatic effects on the bounds on the lower order Wilson coefficients, merely assuming that there exists an analytic UV model, even though the higher orders may be negligible phenomenologically.

\begin{figure}
\centering
\includegraphics[width=0.45\linewidth]{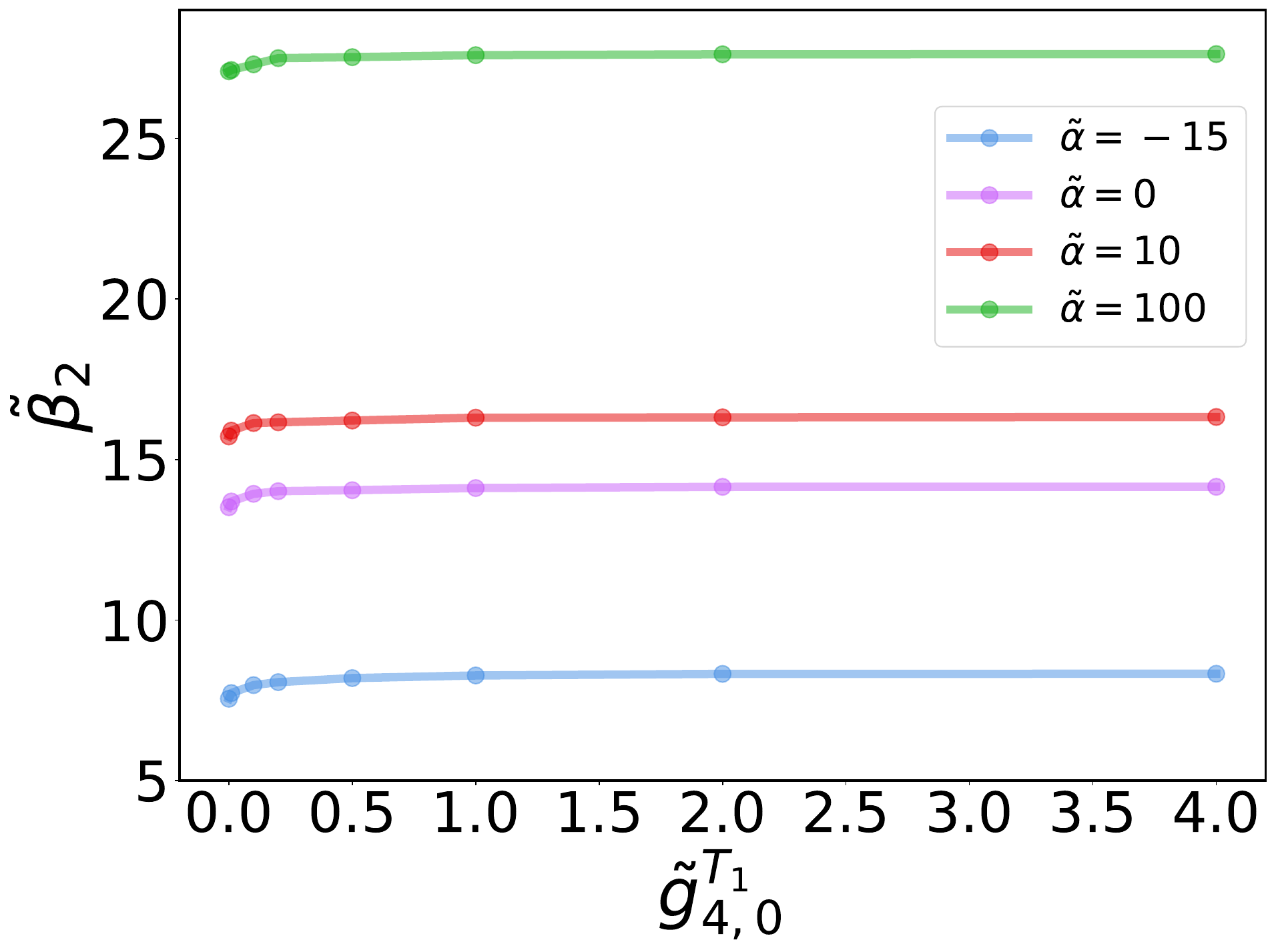}
~~~~~~~~
\includegraphics[width=0.45\linewidth]{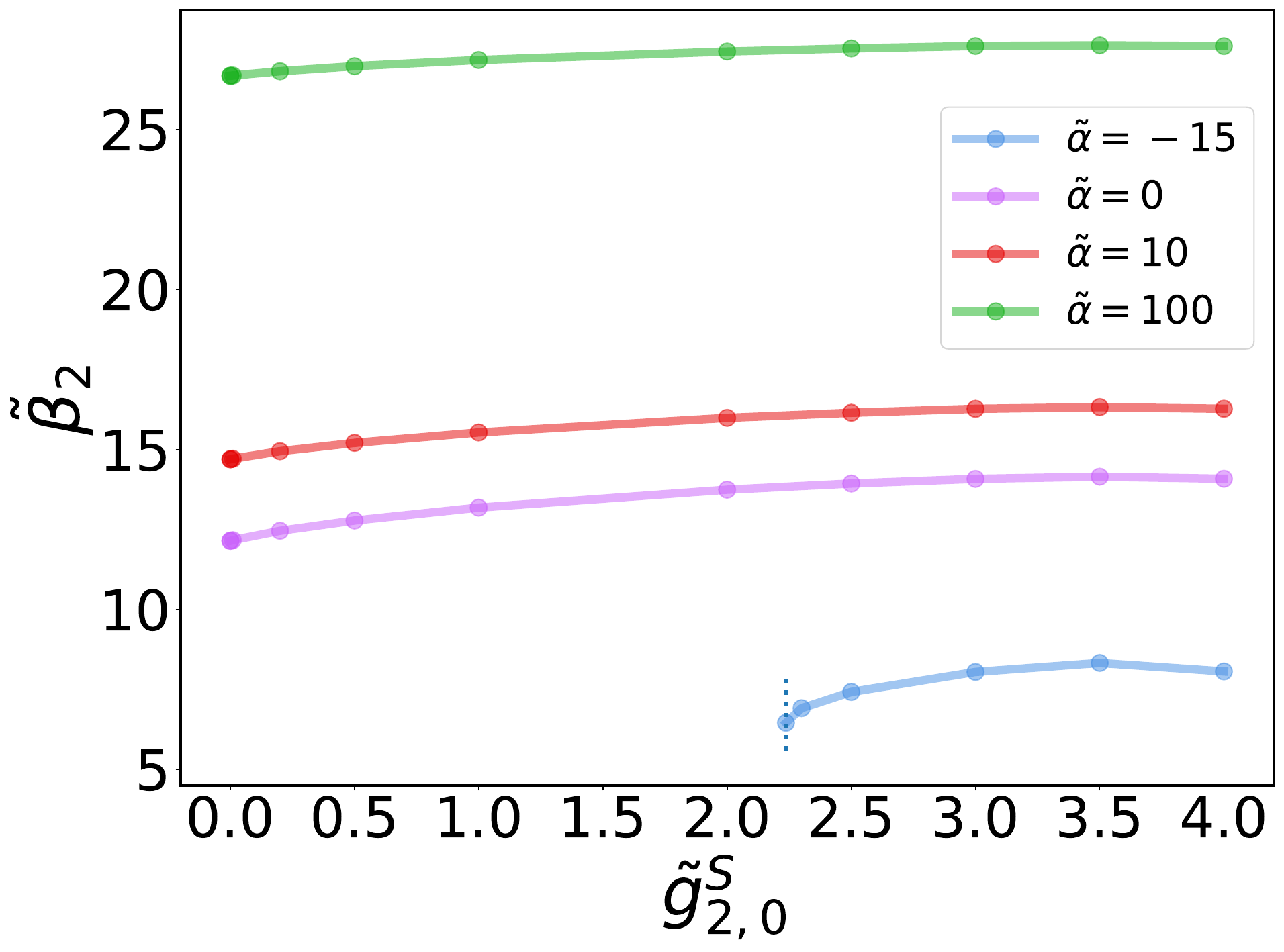}
	\caption{Insensitivity of the upper bound of $\tilde{\beta}_2$ to higher order coefficients. The short dotted line in the right subfigure denotes the lower bound on $\tilde{g}^{S}_{2,0}$ when $\tilde{\alpha}=-15$. We have defined that $\tilde{\alpha}=\alpha M_P^2 \Lambda^2/\log (\Lambda/m_{\text{IR}})$, $\tilde{\beta}_2=\beta_2  \Lambda^2/\log (\Lambda/m_{\text{IR}})$, $\tilde{g}^{T_1}_{4,0}=g^{T_1}_{4,0} M_P^2\Lambda^6 /\log (\Lambda/m_{\text{IR}})$ and $\tilde{g}^{S}_{2,0}=g^{S}_{2,0} M_P^2\Lambda^6 /\log (\Lambda/m_{\text{IR}})$. Note that $\tilde{\alpha}=-15$ is almost the lower causality bound of $\tilde{\alpha}$. }
	\label{fig:ft23}
\end{figure}

On the other hand, the bound on the $\beta_2$ coupling is insensitive to the higher order Wilson coefficients; see Figure \ref{fig:ft23}. Note that as $g^{T_1}_{4,0}$ and $g^{S}_{2,0}$ approach zero, the upper bound on $\beta_2$ decreases only slightly. Not visible in Figure \ref{fig:ft23}, we have also verified this numerically as both $g^{T_1}_{4,0}$ and $g^{S}_{2,0}$ approach zero.

We can understand the difference in sensitivity for $\bi_1$ and $\bi_2$ without actually solving the SDP.
Let us look at the example of how $g^{T_1}_{4,0}$ affects the bounds on $\bi_1$ and $\bi_2$. For $g^{T_1}_{4,0}$, its forward-limit sum rule is positive definite on the right hand side: $g^{T_1}_{4,0} = \< \tilde{F}^{++--}_{4,\ell}(\mu,0)\>=\<  (|c^{++}_{\ell,\mu}|^2+|c^{+-}_{\ell,\mu}|^2)/\mu^5\>\geq 0$. Note that this positive structure is important for the arguments below and, fortunately, this kind of forward-limit sum rules come by quite often. In the SDP, the $g^{T_1}_{4,0}$ coupling enters $B_{P_X,\ell}$ in \eref{BposCon} as 
\be
\label{Bineqapp}
B_{P_X,\ell}  \sim  (\cdots) +  y_*  \mc{O}\({\mu^{-n_*}}\)   + y^{T_1}_{4,0} \mc{O}\({\mu^{-5}}\)  \succeq 0
\ee
where $y^{T_1}_{4,0}$ is the decision variable associated with $g^{T_1}_{4,0}$ in the optimization process and $y_*$ is the decision variable associated with a Wilson coefficient $\bi_*$ that we are concerned with. For every viable set of decision variables, acting $\<...\>$ on \eref{Bineqapp}, we get a condition on the Wilson coefficients
\be
\label{biineqapp}
 [\cdots] +  y_* \bi_* + y^{T_1}_{4,0}  g^{T_1}_{4,0} \geq 0
\ee

First, suppose that $g^{T_1}_{4,0}$ is suppressed and becomes smaller, and let us see how it affects the bounds on the coefficient $\bi_*$. Owing to the smallness of  $g^{T_1}_{4,0}$, $y^{T_1}_{4,0}$ can be very large and still does not significantly affect the inequality (\ref{biineqapp}), and $g^{T_1}_{4,0}$ becoming smaller will allow $y^{T_1}_{4,0}$ to be larger. In the small  $g^{T_1}_{4,0}$ limit, \eref{biineqapp} can be approximated by $[\cdots] +  y_* \bi_* \geq 0$, which leads to the upper bound on $\bi_*$: $\bi_* \leq |[\cdots]/y_*| $. Then, thanks to the positivity of the $g^{T_1}_{4,0}$ sum rule, when $g^{T_1}_{4,0}$ becomes smaller, linear matrix inequality (\ref{Bineqapp}) will allow $y_*$ to take more values, which in turn means that the bounds on $\bi_*$ will become tighter. This is what we have seen for both $\bi_1$ and $\bi_2$ in Figure \ref{fig:yug4} and Figure \ref{fig:ft23}, albeit for $\bi_2$ the effect is very small.

The reason why $\bi_1$ is sensitive to the value of $g^{T_1}_{4,0}$ is linked to the fact that $n_*=4$ for $\bi_1$. To see this, let us first consider the large $\mu$ region of linear matrix inequality (\ref{Bineqapp}). In this region, both the $y_*$ and $y^{T_1}_{4,0}$ terms ({\it i.e.,} the $\bi_1$ and $g^{T_1}_{4,0}$ terms) are negligible, compared to the leading $\mc{O}(\mu^{-3})$ term in $B_{P_X,\ell}$. So the large $\mu$ region does not significantly constrain $y_*$. On the other hand, in the small $\mu$ region,  the positive $y^{T_1}_{4,0}$ term can be significant due to the $\mc{O}\({\mu^{-5}}\)$ scaling, which also leads to weak constraints on $y_*$. Therefore, a loosely constrained $y_*$ results in a strong bound on $\bi_1$. Furthermore, as $g^{T_1}_{4,0}$ becomes smaller, $y^{T_1}_{4,0}$ is allowed to take larger values, which leads to stronger bounds on $\bi_1$. This is what we saw in Figure \ref{fig:yug4}.

On the other hand, for $\bi_2$, we have $n_*=3$ in \eref{Bineqapp}. In this case, the argument for the small $\mu$ region is similar to that of $\bi_1$. However, in the large $\mu$ region, the $y_*$ term has the leading $\mc{O}(\mu^{-3})$ scaling, so linear matrix inequality (\ref{Bineqapp}) now does impose significant constraints on $y_*$, which leads to loose bounds on $\bi_2$. This explains why $\bi_2$ is insensitive to $g^{T_1}_{4,0}$.

\begin{figure}
\centering
\includegraphics[width=0.48\linewidth]{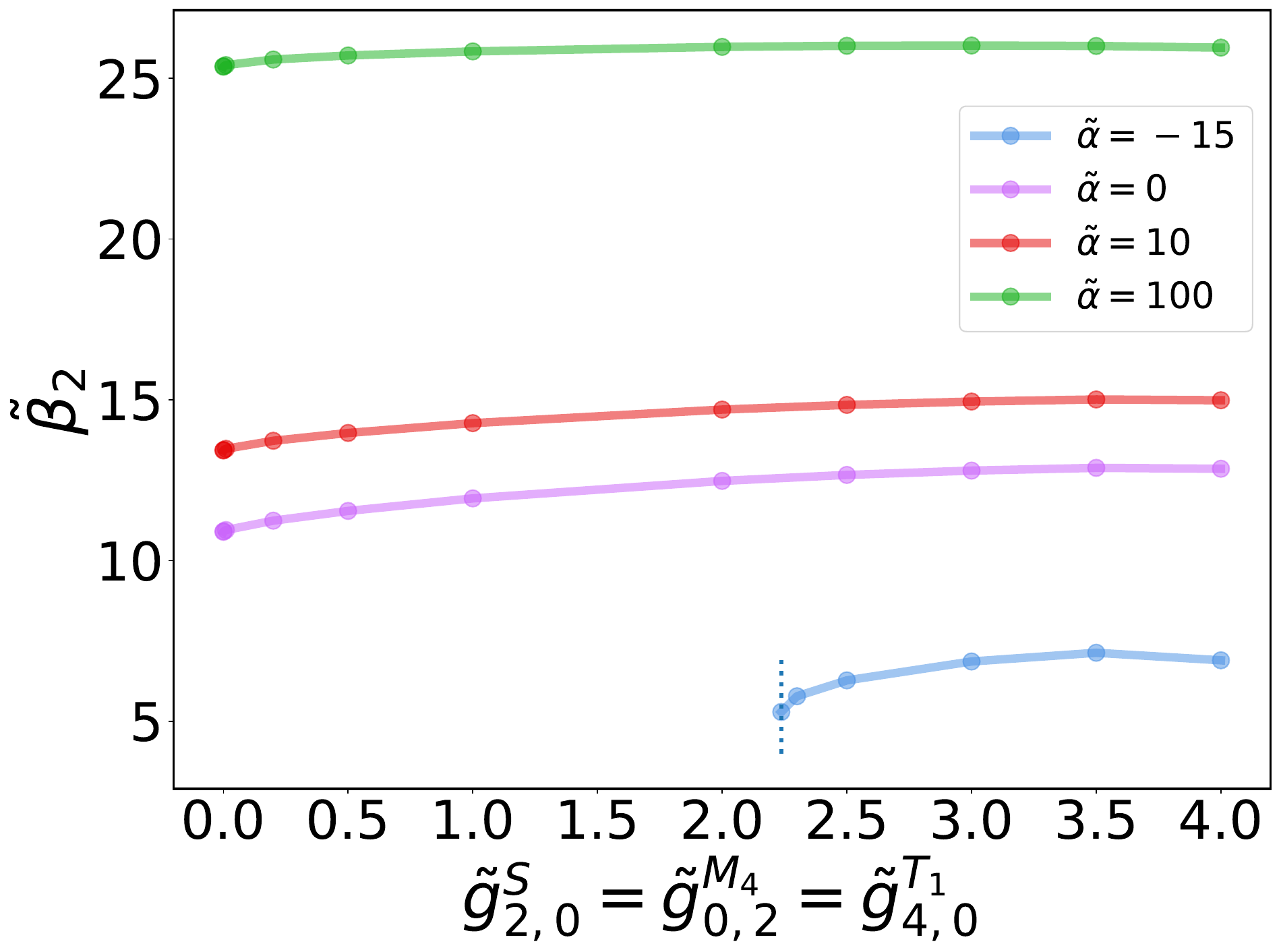} 
	\caption{Entanglement in $B_{P_X,\ell}$ does not necessarily leads to strong correlations between the coefficients. We take $g^{S}_{2,0}$, $g^{M_4}_{0,2}$ and $g^{T_1}_{4,0}$ to approach zero along $\tilde{g}^{S}_{2,0}=\tilde{g}^{M_4}_{0,2}=\tilde{g}^{T_1}_{4,0}$. The short dotted line denotes the lower bound on $\tilde{g}^{S}_{2,0}$ when $\tilde{\alpha}=-15$. Note that $\tilde{\beta}_2=\beta_2 \Lambda^2/\log(\Lambda/m_{\text{IR}})$, $\tilde{g}^{T_1}_{4,0}=g^{T_1}_{4,0} M_P^2\Lambda^6 /\log (\Lambda/m_{\text{IR}})$, $\tilde{g}^{M_4}_{0,2}=g^{M_4}_{0,2}M_P^2\Lambda^{6}/\log(\Lambda/m_{\text{IR}})$ and $\tilde{g}^{S}_{2,0}=g^{S}_{2,0} M_P^2\Lambda^6 /\log (\Lambda/m_{\text{IR}})$.}
	\label{fig:ft44}
\end{figure}

One may be tempted to suggest that the difference between $\bi_1$ and  $\bi_2$ in sensitivity to $g^{T_1}_{4,0}$ is due to the distribution of relevant terms in the $B_{P_X,\ell}$ matrix. One may observe that the $g^{T_1}_{4,0}$ sum rules give rise to nonzero contributions to the blue region in the lower right corner of the $B_{P_X,\ell}$ matrix, as depicted below:
\bc
\includegraphics[width=0.35\linewidth]{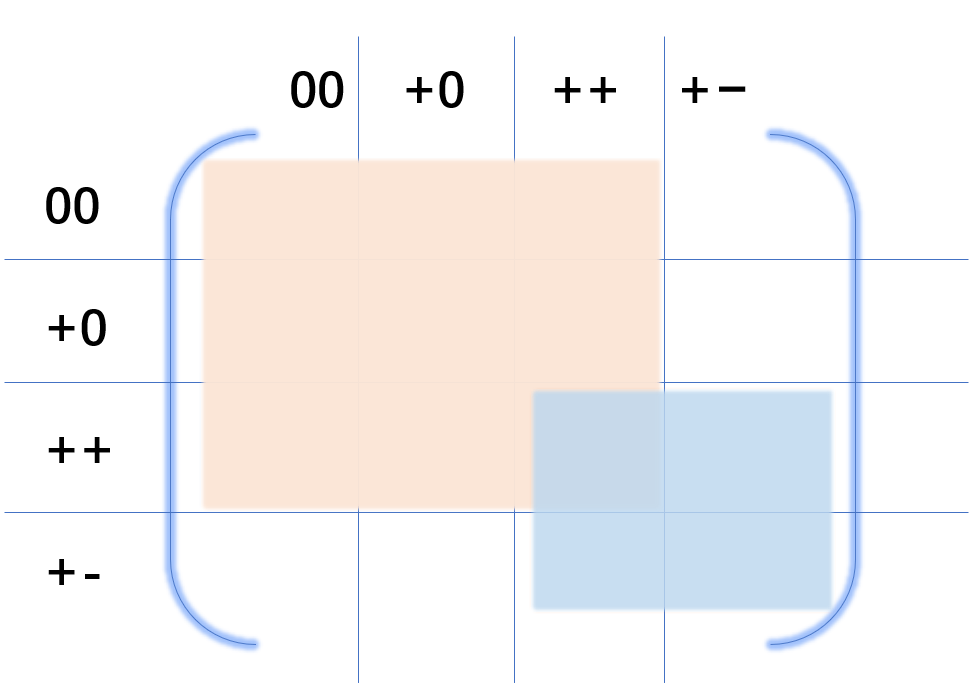}
\ec
The main $\bi_1$ sum rules occupy the same region in $B_{P_X,\ell}$, while the $\bi_2$ sum rule occupies the brown region in the upper left corner, which only slightly overlaps with the $g^{T_1}_{4,0}$ block. This means that $\bi_1$ and $g^{T_1}_{4,0}$ are more entangled in the $B_{P_X,\ell}$ matrix, which might suggest that the mixing in $B_{P_X,\ell}$ is the main reason for $\bi_1$ to be more sensitive to $g^{T_1}_{4,0}$. However, this might not be the case here. We find that, while being separated in $B_{P_X,\ell}$ generally leads to insensitivity between the coefficients, being mixed in $B_{P_X,\ell}$ does not necessarily leads to strong correlations between the coefficients. For example, for the case of Figure \ref{fig:ft44}, the three higher order coefficients occupy the whole $B_{P_X,\ell}$ matrix, and yet we still find that $\bi_2$ is insensitive to these coefficients.

\acknowledgments

We would like to thank Yue-Zhou Li, Ning Su, Shi-Lin Wan, Hao Xu and Yang Zhang for helpful discussions. SYZ acknowledges support from the Fundamental Research Funds for the Central Universities under grant No.~WK2030000036, from the National Natural Science Foundation of China under grant No.~12075233 and 12247103, and from the National Key R\&D Program of China under grant No. 2022YFC220010. We acknowledge use of the GNU parallel package \cite{tange_2022_7239559}.

\appendix

\section{Generic 4-leg amplitudes for scalar-tensor theory}
\label{sec:4pointAmp}

In this appendix,  we shall derive the generic forms of the tree-level amplitudes for scalar-tensor theory.  The amplitudes can be written as functions of $s, t, u$, with certain symmetries among the Mandelstam variables, and also need to satisfy the helicity structure of the scattering particles.  For a tree-level amplitude, there are only two types of contributions, one being two 3-leg vertices connected by a propagator and the other type being a 4-leg contact vertex. 

For massless particles, the on-shell 3-leg amplitudes, with the momenta extended to be complex, are uniquely fixed by the momentum conservation and the little group scaling up to an overall constant \cite{Benincasa:2007xk}
\be
\label{threePres}
\!{\cal M}(1^{h_{1}} 2^{h_{2}} 3^{h_{3}})
\propto\!
\begin{cases}
\langle 12\rangle^{h_{3}-h_{1}-h_{2}}\langle 23\rangle^{h_{1}-h_{2}-h_{3}}\langle 31\rangle^{h_{2}-h_{3}-h_{1}}, & h \leq 0 ,
\\
{[12]^{h_{1}+h_{2}-h_{3}}[23]^{h_{2}+h_{3}-h_{1}}[31]^{h_{3}+h_{1}-h_{2}},} & h \geq 0 ,
\end{cases}
\ee
where $h_i$ is the helicity of particle $i$ and $h\equiv h_1+h_2+h_3$. A 4-leg amplitude can be obtained by glueing one leg of a 3-leg amplitude with one leg of another 3-leg amplitude with opposite helicity. 
Alternatively, we can simply enumerate the Lagrangian terms with lowest few mass dimensions and compute the leading few orders of amplitudes from those terms. This allows us to enumerate all possible pole contributions to the amplitudes from double 3-leg vertex insertions. Then, the rest of the terms can be enumerated in a fashion similar to how \eref{threePres} is obtained, as we shall see shortly. For the latter approach, note that the Lagrangian terms that can give rise to 3-leg vertices are given by
\be
\label{L3pointv}
 \mc{L} \supset   \sqrt{-g} \bigg(  \f{M_P^2}2 R -\f12 \nd_\mu \phi \nd^\mu \phi   - \f{\li_3}{3!} \phi^3  + \f{\bi_1}{2!}  \phi {\cal G}  + \f{\gi_0}{3!} {\cal R}^{(3)} \bigg)\,.
\ee
Computing the relevant amplitudes with these terms only, the independent 2-to-2 amplitudes are given by
\bal
\label{poleStart}
\mc{M}^{0000}_{(3)} &=\li_3^2\bigg(\frac{1}{s}+\frac{1}{t}+\frac{1}{u}\bigg)+\frac{1}{M_P^2} \bigg(\frac{su}{t}+\frac{st}{u}+\frac{ut}{s}\bigg)\,,
\\
\mc{M}^{++--}_{(3)} &=\frac{1}{M_P^2}\frac{s^3}{tu}-\frac{\beta_1^2}{M_P^4}s^3+\frac{\gi_0^2}{M_P^6}s^3tu\,,
\\
\mc{M}^{+++-}_{(3)} &= \frac{\gi_0}{M_P^4}stu\,,
\\
\mc{M}^{++++}_{(3)} &= \frac{10\gi_0}{M_P^4}stu-\frac{3\beta_1^2}{M_P^4}stu+\frac{\gi_0^2}{M_P^6}stu(s^2+t^2+u^2)\,,
\\
\mc{M}^{+++0}_{(3)} &=0\,,
\\
\mc{M}^{++0-}_{(3)} &=\frac{\beta_1}{M_P^3}s^2-\frac{\gi_0\beta_1}{M_P^5}s^2tu\,,
\\
\mc{M}^{++00}_{(3)} &=\frac{\li_3\bi_1}{M_P^2}s+\frac{\gi_0}{M_P^4}stu +\frac{\beta_1^2}{M_P^4}s^3\,,
\\
\mc{M}^{+-00}_{(3)} &=\frac{1}{M_P^2}\frac{tu}{s}+\frac{\beta_1^2}{M_P^4}stu\,,
\\
\label{poleEnd}
\mc{M}^{+000}_{(3)} &=\frac{\beta_1}{2M_P^3}(s^2+t^2+u^2)\,,
\eal
where we have included contributions from both the amplitudes with double 3-leg insertions and those from contact 4-leg vertices. The subscript $(3)$ indicates that these contributions are from the above Lagrangian terms containing 3-leg vertices. Note that most of the terms in \eref{L3pointv} can generate both 3-leg and 4-leg vertices, and it is only when both of them are included can the Ward identities be satisfied.  Amplitudes with other helicities can be obtained from the above ones via crossing, using \eref{fullcrossS1} to \eref{onecrossS4}.

Having found all terms including the 3-leg vertices, we now turn to those purely from the contact vertices. Possible such terms can be constructed by considering restrictions from little group scaling, momentum conservation, locality and crossing symmetry. To see how this works, first note that a 4-momentum goes like $p\sim |]\langle|$ and a polarization vector goes like $\epi\sim |]\langle|/[]$ or $|]\langle|/\<\>$. So, by Lorentz symmetry, a 4-leg amplitude term from a contact term must be a product of powers of $[ij]$ and $\<ij\>$ going like
\be
\!\!\!\mc{M}_m^{\mathbb{1234}}\!\! \propto \! [12]^{a_{12}}[13]^{a_{13}}[14]^{a_{14}}[23]^{a_{23}}[24]^{a_{24}}[34]^{a_{34}}\<12\>^{b_{12}}\<13\>^{b_{13}}\<14\>^{b_{14}}\<23\>^{b_{23}}\<24\>^{b_{24}}\<34\>^{b_{34}},\!\!
\ee
where $m$ denotes the number of partial derivatives in the contact vertex and $a_{ij}$ and $b_{ij}$ are integers. From little group scaling and locality, we can infer some constraints on $a_{ij}$ and $b_{ij}$.

Let us see what these constraints are.
Since a momentum in the contact vertex is of dimension 1 and polarization vectors or tensors are dimensionless, we must have
\be
 a_{12}+a_{13}+a_{14}+a_{23}+a_{24}+a_{34}+b_{12}+b_{13}+b_{14}+b_{23}+b_{24}+b_{34}=m \,.
\ee
Also, since  $|i]$ and $|i\rangle$ scale as $|i] \to t_i|i],~|i\rangle\to t_i^{-1} |i\rangle$ and polarization tensors scales $\epi_{h_i} \to t_i^{2h_i}\epi_{h_i}$ under little group scaling, we have
 \bal
 \begin{cases}
 a_{12}+a_{13}+a_{14}-b_{12}-b_{13}-b_{14}=2 h_{1}\,,
 \\
 a_{12}+a_{23}+a_{24}-b_{12}-b_{23}-b_{24}=2 h_{2}\,,
 \\
 a_{13}+a_{23}+a_{34}-b_{13}-b_{23}-b_{34}=2 h_{3}\,,
 \\
 a_{14}+a_{24}+a_{34}-b_{14}-b_{24}-b_{34}=2 h_{4}\,.
 \end{cases}
 \eal
Furthermore, since we are considering a theory that is local and the momenta from the partial derivatives in the local EFT operators contribute non-negatively to the amplitude's dimension, it must be that the minima of $m$ for the following helicity amplitudes are as follows
\bal
\begin{cases}
m \ge 0 {\rm, ~~for~~ } \mc{M}^{0000}_m \,,\\
m \ge 2 {\rm, ~~for~~ } \mc{M}^{+000}_m \,,\\
m \ge 4 {\rm, ~~for~~ } \mc{M}^{++00}_m, ~\mc{M}^{+-00}_m, ~\mc{M}^{++--}_m \,,\\
m \ge 6 {\rm, ~~for~~ } \mc{M}^{+++0}_m, ~\mc{M}^{++-0}_m\,, \\
m \ge 8 {\rm, ~~for~~ } \mc{M}^{++++}_m, ~\mc{M}^{+++-}_m\,.
\end{cases}
\label{Mmconsts}
\eal
The reason why the lower $m$ terms vanish is similar to the well-known fact that in non-Abelian gauge theory non-MHV 2-to-2 amplitudes must vanish. The easiest way to see this for cases other than $\mc{M}_m^{++--}$, as in the case of non-Abelian gauge theory, is to appropriately choose reference momenta to make all the contractions between the polarizations vanish. So the nonzero $\mc{M}_m^{\mathbb{1234}}$ (except for $\mc{M}_m^{++--}$) are those where every Lorentz index in the polarizations is accompanied by one contracting momentum from the vertex partial derivative. As for $\mc{M}_m^{++--}$, we can set the reference momentum in the polarization tensor of the 1st and 2nd particle to be $p_4$  and that of the 3rd and 4th particle to be $p_1$, which will only leave the contraction of polarizations between particle 2 and 3 nonzero, and then the polarizations of particle 1 and 4 must contract with momenta from the vertex to give rise to nonzero terms, leading to $m \ge 4$ for  $\mc{M}^{++--}_m$. We emphasize that although it is easy to see it by choosing some special reference momenta, the constraint (\ref{Mmconsts}) obviously holds independent of the choice of reference momenta. So we still have the freedom to choose the reference momenta.

Indeed, if we choose the reference momentum in the polarization tensor of the 1st and 2nd particle to be $p_4$ and that of the 3rd and 4th particle to be $p_1$ for all the helicity amplitudes, then we further have the following constraints
  \bal
  a_{13}\geq {\rm min}[0,h_3]\,,~~
  a_{14}\geq {\rm min}[0,h_1]+{\rm min}[0,h_4]\,,~~
  a_{24}\geq {\rm min}[0,h_2]\,,
  \\
  b_{13}\geq {\rm min}[0,-h_3]\,,~~
  b_{14}\geq {\rm min}[0,-h_1]+{\rm min}[0,-h_4]\,,~~
  b_{24}\geq {\rm min}[0,-h_2]\,,
  \eal
and all the other $a_{ij}$ and $b_{ij}$ are non-negative. (A caveat is that one should find appropriate reference momenta in the above construction; otherwise there can be spurious terms in the final amplitude. This can be done by going through a few choices of the reference momenta and pick up the most constraining one.) Furthermore, the contact vertices do not give rise to poles of $s,t,u$ in the amplitude, so we also have
\bal
\begin{cases}
a_{12}+a_{34}+b_{12}+b_{34} \geq 0 \,,\\
a_{13}+a_{24}+b_{13}+b_{24} \geq 0 \,,\\
a_{14}+a_{23}+b_{14}+b_{23} \geq 0 \,.
\end{cases}
\eal
With all these constraints established, we can solve these constraints for $a_{ij}$ and $b_{ij}$. Typically, these constraint equations lead to multiple (in fact, many) solutions. For example, for the case of $m=4,h_1=h_2=+2,~h_3=h_4=0$, there are 9 solutions for $a_{ij}$ and $b_{ij}$, while we have 4570 solutions for $m=10,h_1=h_2=h_3=h_4=+2$. However, they all collapse to a small number of cases after converting to expressions in terms of $s,t,u$.

To convert to an expression in terms of $s,t,u$, we can use an explicit choice for the momenta (all momenta chosen as ingoing and thus related to physical ones by $p^{\rm physical}_3=-p_3$ and $p^{\rm physical}_4=-p_4$) and the spinors
\be
p_1 = \oi(1,0,0,1),~ p_2= \oi(1,0,0,-1)\,,~
p_3 = -\oi(1, \sin\thi, 0, \cos \thi)\,,~
p_4 =  -\oi(1,- \sin\thi, 0, - \cos \thi)\,,\nonumber
\ee
\be
|1\> = \sqrt{2\oi}
\left(\begin{array}{c}
0 \\
1
\end{array}\right),~
|2\> = \sqrt{2\oi}
\left(\begin{array}{c}
1 \\
0
\end{array}\right),~
|3\>=i\sqrt{2 \oi}\left(\begin{array}{c}
-\sin \frac{\theta}{2} \\
\cos \frac{\theta}{2}
\end{array}\right),~
|4\>=i\sqrt{2 \oi}\left(\begin{array}{c}
\cos \frac{\theta}{2} \\
\sin \frac{\theta}{2}
\end{array}\right)\,,
\ee
where particle 1, 2, 3 and 4 are moving in the direction of $(0,0)$, $(\pi, \pi)$, $(\thi,\phi)$ and $(\pi-\thi, \phi+\pi)$ with $\phi=0$ respectively. Here $\thi$ and $\phi$ are the polar and azimuthal angles. There is an extra $i$ in the $|3\>$ and $|4\>$ expression because we need to analytically continue $\sqrt{\oi}$ to $i\sqrt{\oi}$ to account for unphysical $p_3$ and $p_4$.\,\footnote{For a massless particle, a generic momentum is given by $p^{\mu}=\oi(1,  \sin \theta \cos \phi,  \sin \theta \sin \phi,  \cos \theta)$, and a generic helicity-spinor can be written as
\be
|p]^{\dot{a}}=\tilde\li^{\dot{a}}=\sqrt{2 \oi}\left(\begin{array}{c}
\cos \frac{\theta}{2} \\
\sin \frac{\theta}{2} e^{i \phi}
\end{array}\right) ,~~~ |p\>_a=\li_{a}=\sqrt{2 \omega}\left(\begin{array}{l}
-\sin \frac{\thi}{2} e^{-i \phi} \\
+\cos \frac{\thi}{2}
\end{array}\right) \,.
\ee
Note that $\varepsilon^{12}=\varepsilon_{21}=+1, ~ \varepsilon^{21}=\varepsilon_{12}=-1,~\varepsilon^{ij} =\varepsilon^{\dot{i}\dot{j}},~\varepsilon_{ij} =\varepsilon_{\dot{i}\dot{j}}$, We have $p^{\rm physical}=\epi_p p$, where $\epi_p=-1$ if the direction of the physical mementum goes against the assumed direction (otherwise $\epi_p=1$), and also $\<pk\>^*=\epi_p\epi_k [kp]$ because there is an extra $i$ in $|j]$ and $|j\rangle$.}  Also, we have $s=-(p_1+p_2)^2=-(p_3+p_4)^2=(2\oi)^2,~t =-{s}(1-\cos \theta)/2=-s\sin^2 ({\thi}/2) ,~
u =-{s}(1+\cos \theta)/2=-s\cos^2 ({\thi}/2)$. We can cast $\oi$ and $\thi$ in terms of $s,t,u$
\be
\oi =\f12 \sqrt{s}, ~~\sqrt{2\oi}=s^{\f14},~~ \sin \f{\thi}2 = \sqrt{\f{-t}{s}},~~ \cos \f{\thi}2 = \sqrt{\f{-u}{s}} ,
\ee
from which we can find that
\bal
\<12\>&=\sqrt{s},\<13\>=-i\sqrt{-t},\<14\>=i\sqrt{-u}, \<23\>=-i\sqrt{-u},\<24\>=-i\sqrt{-t},\<34\>=-\sqrt{s},
\\
[12]&=-\sqrt{s},\,[13]=i\sqrt{-t},\,[14]=-i\sqrt{-u},\, [23]=i\sqrt{-u},\,[24]=i\sqrt{-t},\,[34]=\sqrt{s}.
\eal
Substituting these replacements into the large numbers of expressions in terms of $[ij]$ and $\<ij\>$ and imposing appropriate crossing symmetries, we can see that they collapse to a small number of functions of $s,t,u$.

In the end, we find that the results are consistent with simply taking the $f_i(s,t,u)$ functions in \eref{gstartdef} - \eref{gfinaldef} to be generic polynomials of $s, t, u$ that share the symmetries of the corresponding amplitudes, except for $\mc{M}^{++++}$. In the $\mc{M}^{++++}$ case, letting $f_{T_3}(s,t,u)$ be generic polynomials of $x,y$ would give rise to a couple of spurious terms, which should vanish according to the analysis above. Therefore, including the contributions from the 3-leg vertices (\eref{poleStart} to \eref{poleEnd}), we can parametrize the 4-leg amplitudes as follows:

\bal
\mc{M}^{0000} &=\li_3^2\(\frac{1}{s}+\frac{1}{t}+\frac{1}{u}\)-\li_4+\frac{1}{M_P^2} \(\frac{su}{t}+\frac{st}{u}+\frac{ut}{s}\)+\sum_{n\ge0, m\ge0}g^{S}_{m,n}x^ny^m \,,
\\
\mc{M}^{++--} &=\frac{1}{M_P^2}\frac{s^3}{tu}-\frac{\beta_1^2}{M_P^4}s^3+\frac{\gi_0^2}{M_P^6}s^3tu+\sum_{n\ge4, m\ge0}g^{T_1}_{n,m}s^n(tu)^m\,,
\\
\mc{M}^{+++-} &=\frac{\gi_0}{M_P^4}y+\sum_{n\ge0, m\ge2}g^{T_2}_{m,n}x^ny^m
\\
\mc{M}^{++++} &=\(\frac{10\gi_0}{M_P^4}-\frac{3\beta_1^2}{M_P^4}\)y+\sum_{\substack{n\ge0, m\ge0,\\ m+n\ge 2}}g^{T_3}_{m,n}x^ny^m\,,
\\
\mc{M}^{+++0} &=\sum_{n\ge0, m\ge1}g^{M_1}_{m,n}x^ny^m\,,
\\
\mc{M}^{++0-} &=\frac{\beta_1}{M_P^3}s^2-\frac{\gi_0\beta_1}{M_P^5}s^2tu+\sum_{n\ge3, m\ge1} g^{M_2}_{n,m}s^n(tu)^m\,,
\\
\mc{M}^{++00} &=\frac{\li_3 \bi_1}{M_P^2}s+\frac{\gi_0}{M_P^4}stu +\sum_{n\ge2, m\ge0}g^{M_3}_{n,m}s^n(tu)^m\,,
\\
\mc{M}^{+-00} &=\frac{1}{M_P^2}\frac{tu}{s}+\frac{\beta_1^2}{M_P^4}stu+\sum_{n\ge0, m\ge2}g^{M_4}_{n,m}s^n(tu)^m\,,
\\
\mc{M}^{+000} &=\frac{\beta_1}{2M_P^3}x+\sum_{n\ge0, m\ge1}g^{M_5}_{m,n}x^ny^m\,,
\eal
where $x=s^2+t^2+u^2,~y=s t u$. All other amplitudes can be obtained by crossing.

\section{Explicit sum rules with $st$ symmetry imposed}
\label{sec:expSum}

Here we explicitly list all the $st$-symmetry imposed sum rules that we use in this paper, for a quick reference. The definition of $F^{\mathbb{1234}}_{k, \ell}(\mu, t)$ is given by
\bal
		F^{\mathbb{1234}}_{k,\ell}(\mu,t)  &= \frac{\partial_s^k}{k!}\bigg(\frac{s^2}{\mu^2(\mu-s)}d^{\ell,\mu,t}_{h_{12}, h_{43}}c^{\mathbb{12}}_{\ell, \mu}c^{*\bar{\mathbb{3}}\bar{\mathbb{4}}}_{\ell, \mu}+\frac{(-s-t)^2}{\mu^2(\mu +s+t)}d^{\ell,\mu,t}_{h_{14}, h_{23}}c^{\mathbb{14}}_{\ell, \mu}c^{*, \bar{\mathbb3}\bar{\mathbb2}}_{\ell, \mu}\bigg)\bigg|_{s\to 0}
		\\
		&~~ -\frac{\partial_t^k}{k!} \bigg(
		\frac{s^3}{\mu^3(\mu-s)}d^{\ell,\mu,t}_{h_{13},h_{42}}c^{\mathbb{13}}_{\ell,\mu}c^{*\bar{\mathbb2}\bar{\mathbb4}}_{\ell,\mu}+
		\frac{(-s)^3}{(\mu+t)^3(\mu+s+t)}d^{\ell,\mu,t}_{h_{14},h_{32}} c^{\mathbb{14}}_{\ell,\mu}c^{*\bar{\mathbb2}\bar{\mathbb3}}_{\ell,\mu}
		\bigg)\bigg|_{t\to 0,s\to t} \,.\nonumber
\eal
The $st$-symmetry imposed sum rules derived from the $stu$-symmetric amplitudes are: 
{{\small
\begin{multicols}{2}
\noindent
	\begin{align}
		-\frac{1}{M_P^2}+2\ai t- \gi_4 t^2&=\Big\langle F^{0000}_{1, \ell}(\mu, t)\Big\rangle   \label{F00001}\\
		-\frac{1}{M_P^2}\frac{1}{t}+2\ai -\gi_4 t+&12g^S_{0,2}t^2 \nonumber\\&=\Big \langle F^{0000}_{2, \ell}(\mu, t)\Big \rangle  \label{F00002}\\
		8g^{S}_{0,2}t-4g^{S}_{1,1}t^2 &= \Big\langle F^{0000}_{3, \ell}(\mu, t)\Big\rangle    \\
		4g^{S}_{0,2}-2g^{S}_{1,1}t+\Big(g^{S}_{2,0}&+48g^{S}_{3,0}\Big)t^2\nonumber  \\&=\Big\langle F^{0000}_{4, \ell}(\mu, t)\Big\rangle  \\
		24g^{S}_{0,3}t-12g^{S}_{1,2}t^2 &= \Big\langle F^{0000}_{5, \ell}(\mu, t)\Big\rangle \label{GF00005}   \\
		8g^{S}_{0,3}-4g^{S}_{1,2}t+\Big(2g^{S}_{2,1}&+160g^{S}_{0,4}\Big)t^2\nonumber  \\&=\Big\langle F^{0000}_{6, \ell}(\mu, t)\Big\rangle \label{GF00006} \\
		\frac{\beta_1}{M_P^3}t-\f{\gi_3}{M_P}t^2&=\Big \langle F^{+000}_{1, \ell}(\mu, t)\Big \rangle \label{F20001} \\
		\frac{\beta_1}{M_P^3}-\f{\gi_3}{M_P}t &=\Big \langle F^{+000}_{2, \ell}(\mu, t)\Big \rangle  \label{F20002}\\
		-4g^{M_5}_{1,1}t^2 &=\Big \langle F^{+000}_{3, \ell}(\mu, t)\Big \rangle  \\
		-2g^{M_5}_{1,1}t+g^{M_5}_{2,0}t^2 &=\Big \langle F^{+000}_{4, \ell}(\mu, t)\Big \rangle \\
		-\f{\gi_1}{M_P^3}t^2&=\Big \langle F^{+++0}_{1, \ell}(\mu, t)\Big \rangle  \label{F22201}\\
		-\f{\gi_1}{M_P^3}t&=\Big \langle F^{+++0}_{2, \ell}(\mu, t)\Big \rangle  \label{F22202}\\
		-4g^{M_1}_{1,1}t^2 &=\Big \langle F^{+++0}_{3, \ell}(\mu, t)\Big \rangle \\
		-2g^{M_1}_{1,1}t+g^{M_1}_{2,0}t^2&=\Big \langle F^{+++0}_{4, \ell}(\mu, t)\Big \rangle  \\
		-\frac{\gi_0}{M_P^4}t^2&=\Big \langle F^{+++-}_{1, \ell}(\mu, t)\Big \rangle  \label{F22211}\\
		-\frac{\gi_0}{M_P^4}t&=\Big \langle F^{+++-}_{2, \ell}(\mu, t)\Big \rangle  \label{F22212}\\
		0&=\Big \langle F^{+++-}_{3, \ell}(\mu, t)\Big \rangle  \\
		g^{T_2}_{2,0}t^2 &= \Big \langle F^{+++-}_{4, \ell}(\mu, t)\Big \rangle  \\
		\Big(\!-\frac{10\gi_0}{M_P^4}+\frac{3\beta_1^2}{M_P^4}\Big)t^2&=\Big \langle F^{++++}_{1, \ell}(\mu, t)\Big\rangle  \label{F22221}\\
		\Big(\!-\frac{10\gi_0}{M_P^4}+\frac{3\beta_1^2}{M_P^4}\Big)t+&12g^{T_3}_{0,2}t^2\nonumber \\&=\Big \langle F^{++++}_{2, \ell}(\mu, t)\Big \rangle \label{F22222} \\
		8g^{T_3}_{0,2}t-4g^{T_3}_{1,1}t^2 &= \Big \langle F^{++++}_{3, \ell}(\mu, t)\Big \rangle  \\
		4g^{T_3}_{0,2}-2g^{T_3}_{1,1}t+\Big(g^{T_3}_{2,0}&+48g^{T_3}_{0,3}\Big)t^2\nonumber \\&= \Big \langle F^{++++}_{4, \ell}(\mu, t)\Big \rangle\, \label{F22224}. 
	\end{align} 
\end{multicols}
}}

\noindent The $st$-symmetry imposed sum rules derived from amplitudes with $su$, $st$ or $ut$ symmetry are:
{{\small
\begin{multicols}{2}
\noindent
	\begin{align}
		\f{\bi_2}{M_P^2}-\frac{\gi_0}{M_P^4}t-g^{M_3}_{2,1}t^2&=\Big\langle F^{++00}_{2, \ell}(\mu, t)\Big\rangle   \label{F22002} \\
		\f{\gi_2}{M_P^2}+\f{\bi_1^2}{M_P^4}-g^{M_3}_{2,1}t-&g^{M_3}_{3,1}t^2\nonumber \\&=\Big\langle F^{++00}_{3, \ell}(\mu, t)\Big\rangle \label{F22003} \\		
		g^{M_3}_{4,0}-g^{M_3}_{3,1}t+(g^{M_3}_{2,2}&-g^{M_3}_{4,1})t^2 \nonumber \\&= \Big\langle F^{++00}_{4, \ell}(\mu, t)\Big\rangle \\
		-\frac{\gi_0}{M_P^4}t^2 &=\Big\langle F^{+0+0}_{1, \ell}(\mu, t)\Big\rangle  \label{F20201}\\
		-\frac{\gi_0}{M_P^4}t - g^{M_3}_{2,1}t^2&=\Big\langle F^{+0+0}_{2, \ell}(\mu, t)\Big\rangle \label{F20202} \\
		0&=\Big\langle F^{+0+0}_{3, \ell}(\mu, t)\Big\rangle \\
		g^{M_3}_{2,2}t^2 &= \Big\langle F^{+0+0}_{4, \ell}(\mu, t)\Big\rangle \\
		-\frac{\beta_1^2}{M_P^4} t+g^{M_4}_{0,2}t^2&=\Big\langle F^{+-00}_{2, \ell}(\mu, t)\Big\rangle  \label{F21002}\\
		g^{M_4}_{1,2}t^2 &=\Big\langle F^{+-00}_{3, \ell}(\mu, t)\Big\rangle\  \\
		g^{M_4}_{2,2}t^2 &= \Big\langle F^{+-00}_{4, \ell}(\mu, t)\Big\rangle\   \\
		g^{M_4}_{3,2}t^2 &=\Big\langle F^{+-00}_{5, \ell}(\mu, t)\Big\rangle\  \label{GF21005}\\
		g^{M_4}_{4,2}t^2 &= \Big\langle F^{+-00}_{6, \ell}(\mu, t)\Big\rangle\   \label{GF21006}\\
		-\frac{1}{M_P^2}-\frac{\beta_1^2}{M_P^4} t^2&=\Big\langle F^{+0-0}_{1, \ell}(\mu, t)\Big\rangle \label{F20101}\\
		-\frac{1}{M_P^2}\frac{1}{t}-\frac{\beta_1^2}{M_P^4} t+&g^{M_4}_{0,2}t^2\nonumber \\&=\Big\langle F^{+0-0}_{2, \ell}(\mu, t)\Big\rangle \label{F20102}\\
		2g^{M_4}_{0,2}t +2g^{M_4}_{1,2}t^2  &= \Big\langle F^{+0-0}_{3, \ell}(\mu, t)\Big\rangle\   \\
		g^{M_4}_{0,2} + g^{M_4}_{1,2}t +\Big(&g^{M_4}_{2,2}-3g^{M_4}_{0,3}\Big)t^2\nonumber \\&= \Big\langle F^{+0-0}_{4, \ell}(\mu, t)\Big\rangle\  \\
		-3g^{M_4}_{0,3}t -3g^{M_4}_{1,3}t^2  &= \Big\langle F^{+0-0}_{5, \ell}(\mu, t)\Big\rangle\   \label{GF20105}\\
		-g^{M_4}_{0,3} - g^{M_4}_{1,3}t +\Big(&-g^{M_4}_{2,3}+6g^{M_4}_{0,4}\Big)t^2\nonumber \\&= \Big\langle F^{+0-0}_{6, \ell}(\mu, t)\Big\rangle\  \label{GF20106}\\
		\frac{\beta_1}{M_P^3}+\f{\bi_1 \gi_0}{M_P^5}t^2 &=\Big\langle F^{0-++}_{2, \ell}(\mu, t)\Big\rangle \label{F01222} \\
		\frac{\gamma_0 \beta_1}{M_P^5} t -g^{M_2}_{3,1}t^2&= \Big\langle F^{0-++}_{3, \ell}(\mu, t)\Big\rangle \\
		-g^{M_2}_{3,1}t -g^{M_2}_{4,1}t^2&= \Big\langle F^{0-++}_{4, \ell}(\mu, t)\Big\rangle \\
		0&=\Big\langle F^{0+-+}_{1, \ell}(\mu, t)\Big\rangle  \label{F02121}\\
		\frac{\gi_0\beta_1}{M_P^5} t^2&=\Big\langle F^{0+-+}_{2, \ell}(\mu, t)\Big\rangle  \label{F02122}\\
		0&= \Big\langle F^{0+-+}_{3, \ell}(\mu, t)\Big\rangle \\
		0&=\Big\langle F^{0+-+}_{4, \ell}(\mu, t)\Big\rangle \\
		0&=\Big \langle F^{+-+-}_{1, \ell}(\mu, t)\Big \rangle \label{F21211}\\
		0&=\Big \langle F^{+-+-}_{2, \ell}(\mu, t)\Big \rangle \label{F21212}\\
		0&= \Big\langle F^{+-+-}_{3, \ell}(\mu, t)\Big \rangle \label{F21213}\\
		0&=\Big\langle F^{+-+-}_{4, \ell}(\mu, t)\Big \rangle  \label{F21214}\\
		0&= \Big\langle F^{+-+-}_{5, \ell}(\mu, t)\Big \rangle  \label{F21215}\\
		0&=\Big\langle F^{+-+-}_{6, \ell}(\mu, t)\Big \rangle \label{F21216}\\
		-\frac{1}{M_P^2}\frac{1}{t}&=\Big\langle F^{++--}_{2, \ell}(\mu, t)\Big \rangle \label{F22112}\\
		-\frac{\beta_1^2}{M_P^4} -\frac{\gi_0^2}{M_P^6} t^2 &=\Big\langle F^{++--}_{3,\ell}(\mu,t)\Big \rangle  \label{F22113}\\
		g^{T_1}_{4,0} - \frac{\gamma_0^2}{M_P^6}t- g^{T_1}_{4,1}t^2&= \Big\langle F^{++--}_{4, \ell}(\mu, t)\Big \rangle  \label{F22114}\\
		g^{T_1}_{5,0}-g^{T_1}_{4,1}t-g^{T_1}_{5,1}t^2&= \Big\langle F^{++--}_{5, \ell}(\mu, t)\Big \rangle  \label{F22115}\\
		g^{T_1}_{6,0}-g^{T_1}_{5,1}t+\Big(g^{T_1}_{4,2}-&g^{T_1}_{6,1}\Big)t^2\nonumber \\ &= \Big\langle F^{++--}_{6, \ell}(\mu, t)\Big \rangle \,.  \label{F22116}
	\end{align}
\end{multicols}
}}
\noindent Note that one can take the forward limit of the sum rules $F^{\mathbb{1234}}_{k,\ell}(\mu,t)$ with $k\geq3$, which are valuable to extract the useful constraints in the finite $\mu$ and large $\ell$ region.

\section{Explicit example of optimization scheme}
\label{sec:Bec_exa}

The purpose of this appendix is to provide the nitty-gritty of using the dispersive sum rules to obtain causality bounds on the Wilson coefficients. We shall demonstrate these by the explicit example of deriving the bound on $\beta_1^2$ and $\gamma_0^2$ (projected to ${1}/{M_P^2}$), using only the dispersive sum rules from four graviton scattering for the sake of simplicity. This example provides a representative illustration of the essential computational steps and subtleties involved, while in the actual figures in Section \ref{subsec:xian} all available dispersive sum rules are used.

Even if we restrict to four graviton scattering, there are already quite a few dispersive sum rules available to constrain $\beta_1$ and $\gamma_0$, which are given by
\bal
\label{exa:f1}
-\frac{\beta_1^2}{M_P^4} -\frac{\gamma_0^2}{M_P^6} p^4&=\Big\langle F^{++--}_{3,\ell}\big(\mu,-p^2\big)\Big\rangle\,,
&\frac{1}{M_P^2}\frac{1}{p^2}&=\Big\langle F^{++--}_{2,\ell}\big(\mu,-p^2\big)\Big\rangle\,,\nonumber\\
g^{T_1}_{4,0}+\frac{\gamma_0^2}{M_P^6}p^2  - g^{T_1}_{4,1}p^4 &=\Big\langle  F^{++--}_{4,\ell}\big(\mu,-p^2\big)\Big\rangle\,,\nonumber\\
0&=\Big\langle F^{+-+-}_{1,\ell}\big(\mu,-p^2\big)\Big\rangle \,,
&0&=\Big\langle F^{+-+-}_{2,\ell}\big(\mu,-p^2\big)\Big\rangle\,,\nonumber\\
0&=\Big\langle F^{+-+-}_{3,\ell}\big(\mu,-p^2\big)\Big\rangle\,,
&0&=\Big\langle F^{+-+-}_{4,\ell}\big(\mu,-p^2\big)\Big\rangle\,,\nonumber\\
-\frac{\gamma_0}{M_P^4}p^4&=\Big\langle F^{+++-}_{1,\ell}\big(\mu,-p^2\big)\Big\rangle\,,
&\frac{\gamma_0}{M_P^4}p^2&=\Big\langle F^{+++-}_{2,\ell}\big(\mu,-p^2\big)\Big\rangle\,, \nonumber\\
0&=\Big\langle F^{+++-}_{3,\ell}\big(\mu,-p^2\big)\Big\rangle\,,
&g^{T_2}_{2,0}p^4&=\Big\langle F^{+++-}_{4,\ell}\big(\mu,-p^2\big)\Big\rangle\,,\nonumber\\
-\(\frac{10\gamma_0}{M_P^4}-\frac{3\beta_1^2}{M_P^4}\)p^4&=\Big\langle F^{++++}_{1,\ell}\big(\mu,-p^2\big)\Big\rangle \,,\nonumber\\
\(\frac{10\gamma_0}{M_P^4}-\frac{3\beta_1^2}{M_P^4}\)p^2+12g^{T_3}_{0,2} p^4 &=\Big\langle F^{++++}_{2,\ell}\big(\mu,-p^2\big)\Big\rangle\,, \nonumber\\
-8g^{T_3}_{0,2}p^2-4g^{T_3}_{1,1}p^4&=\Big\langle F^{++++}_{3,\ell}\big(\mu,-p^2\big)\Big\rangle\,,\nonumber\\
4g^{T_3}_{0,2}+2g^{T_3}_{1,1}p^2+(g^{T_3}_{2,0}+48g^{T_3}_{0,3})p^4&=\Big\langle F^{++++}_{4,\ell}\big(\mu,-p^2\big)\Big\rangle\,.
\eal
where the EFT cutoff has been set to $\Lambda=1$. As mentioned in Section \ref{sec:numdet}, for technical reasons, we also add some forward-limit sum rules (\ref{forlimSum}) to effectively make use of the constraints in the finite $\mu$ and large $\ell$ region:
\bal\label{exa:disp1}
		-\frac{2\gamma_0^2}{M_P^6}&=\Big\langle \pd_t^2 F^{++--}_{3,\ell}(\mu,0)\Big\rangle\,,&
		-\frac{\gamma_0^2}{M_P^6}&=\Big\langle \pd_t F^{++--}_{4,\ell}(\mu,0)\Big\rangle\,,&
		-\frac{\beta_1^2}{M_P^4}&=\Big\langle F^{++--}_{3,\ell}(\mu,0)\Big\rangle\,,\nn
		8g^{T_3}_{0,2}&=\Big\langle \pd_t F^{++++}_{3,\ell}(\mu,0)\Big\rangle\,,&
		0&=\Big\langle \pd^3_t F^{++--}_{3,\ell}(\mu,0)\Big\rangle\,,
\eal
where for illustration purposes only one forward-limit null constraint is included. It is usually beneficial to include more forward-limit sum rules, as we do for the plots in Section \ref{sec:bounds}. The reason for using forward-limit sum rules, as mentioned, is that the Wigner d-functions, when integrated over $p$ against the weight functions, tend to vanish in this region, and the added forward-limit sum rules provide terms that are polynomials of $\ell$ in the large $\ell$ limit and that are dominant in this region. Ultimately, the reason why adding forward-limit sum rules is helpful is that we only parametrize each of the weight functions with a few parameters, while the actual functional spaces are infinite dimensional.

Similar considerations also apply when choosing the forward-limit sum rules. The two dispersion relations with $\gamma_0$ in \eref{exa:disp1} are formally independent, but they are actually linked by the $st$ crossing. Nevertheless, we use both of them, as the enforcing of the $st$ crossing is not complete due to the finite dimensional truncation of the weight functions. On the other hand, we do not use the forward-limit sum rule $g^{T_3}_{2,0}=\langle F^{++++}_{4,\ell}(\mu,0)/4 \rangle$ because it is formally the same as the sum rule with $\pd_t F^{++++}_{3,\ell}(\mu,0)$, already guaranteed by the $su$ symmetry of the dispersion relations.

Our goal is to extract as much information as possible from these sum rules. To that end, we integrate both sides of the sum rules (\ref{exa:f1}) over various weight functions $\phi^{\mathbb{1234}}_k(p)$ and sum both sides of the sum rules (\ref{exa:disp1}) over weight parameters $z^{\mathbb{1234}}_{k,n}$, which leads to
\bal\label{exa:sum2}
		&\bigg\{
		\int_{0}^{1} \d p \,\phi^{++--}_{2}(p)\frac{1}{p^2}\bigg\}\frac{1}{M_P^2}\nn&+
		\bigg\{\int_{0}^{1} \text{d}p\bigg(-\phi^{+++-}_{1}(p)p^4+\phi^{+++-}_2(p)p^2-10\phi^{++++}_1(p)p^4 + 10\phi^{++++}_2(p)p^2\bigg)\bigg \}\frac{\gamma_0}{M_P^4}\nn&+\bigg\{\int_{0}^{1} \text{d}p
		\bigg(-\phi^{++--}_{3}(p)p^4+\phi^{++--}_{4}(p)p^2 \bigg)-2z^{++--}_{3,2}-z^{++--}_{4,1}\bigg\}\frac{\gamma_0^2}{M_P^6}\nn&+
		\bigg\{\int_{0}^{1} \text{d}p\bigg(-\phi^{++--}_3(p)+3\phi^{++++}_1(p)p^4-3\phi^{++++}_2(p)p^2\bigg)-z^{++--}_{3,0}\bigg\}\frac{\beta_1^2}{M_P^4}\nn
		&+\bigg\{\int_{0}^{1} \text{d}p\,\phi^{++--}_4(p)\bigg\}g^{T_1}_{4,0}
        +\bigg\{\int_{0}^{1} \d p\,\phi^{++--}_4(p)p^4\bigg\}g^{T_1}_{4,1}
        +\bigg\{\int_{0}^{1} \text{d}p\,\phi^{+++-}_4(p)p^4\bigg\}g^{T_2}_{2,0}\nn
        &+\bigg\{\int_{0}^{1} \text{d}p\bigg(12\phi^{++++}_2(p)p^4-8\phi^{++++}_3(p)p^2+4\phi^{++++}_4(p)\bigg)+8z^{++++}_{3,1}\bigg\}g^{T_3}_{0,2}\nn
		&+\bigg\{\int_{0}^{1} \text{d}p\bigg(-4\phi^{++++}_3(p)p^4+2\phi^{++++}_4(p)p^2\bigg)\bigg\}g^{T_3}_{1,1}+
		\bigg\{\int_{0}^{1} \text{d}p\bigg(\phi^{++++}_4(p)p^4\bigg)\bigg\}(g^{T_3}_{2,0}+48g^{T_3}_{0,3})
		\nn
		&=\bigg\langle
		\int_{0}^{1}\text{d}p \bigg(\sum_{\mathbb{1234},k}\phi^{\mathbb{1234}}_k(p)F^{\mathbb{1234}}_{k,\ell}\left(\mu,-p^2\right) \bigg) +\sum_{\mathbb{1234},k,n} z^{\mathbb{1234}}_{k,n} \pd_t^n F^{\mathbb{1234}}_{k,\ell}\left(\mu \right)		
		\bigg\rangle
		\\
		&:=\left\langle\(\mathcal{C}_{P_X,\ell,\mu}\)^T
		B_{P_X,\ell}(\mu )
		\mathcal{C}_{P_X,\ell,\mu}\right\rangle\,,
\eal
where the last equality implicitly defines the $B_{P_X,\ell}(\mu )$ matrices mentioned in Section \ref{sec:genStra}. The weight functions and parameters are so-called decision variables in the optimization problem. For some appropriate chosen decision variables, these matrices can be made semi-positive:
\begin{equation}\label{exa:cons1}
	B_{P_X,\ell}(\mu ) \succeq 0,~~~\text{for  $P_X=\pm 1$, all possible $\ell$ and all $\mu \geq \Lambda^2$}\,,
\end{equation}
which in turn results in the right hand side of \eref{exa:sum2} being semi-positive. This gives rise to a bound on the Wilson coefficients appearing on the left hand side of \eref{exa:sum2}. However, our goal here is more specific: we want to derive a bound on $\beta_1^2$ and $\gamma_0^2$, projected onto ${1}/{M_P^2}$. So we do not want other Wilson coefficients to be involved on the left hand side of \eref{exa:sum2}. We can achieve this by imposing the following constraints on the weight functions:
\begin{equation}\label{exa:cons5}
	\begin{aligned}
		&\int_{0}^{1} \text{d}p \bigg(\phi^{++--}_4(p)\bigg)=0\,,~~~\int_{0}^{1} \text{d}p \bigg(\phi^{++--}_4(p)p^4\bigg)=0\,,~~~
		\int_{0}^{1} \text{d}p \bigg(\phi^{+++-}_4(p)p^4\bigg)=0\,,\\
		&\int_{0}^{1} \text{d}p \bigg(12\phi^{++++}_2(p)p^4-8\phi^{++++}_3(p)p^2+4\phi^{++++}_4(p)\bigg)+8z^{++++}_{3,1}=0\,,\\
		&\int_{0}^{1} \text{d}p \bigg(-4\phi^{++++}_3(p)p^4+2\phi^{++++}_4(p)p^2\bigg)=0\,,~~~~~~
		\int_{0}^{1} \text{d}p \bigg(\phi^{++++}_4(p)p^4\bigg)=0\,.\\
	\end{aligned}
\end{equation}
That is, we are deriving bounds on $\beta_1$ and $\gamma_0$ while being agnostic about all other Wilson coefficients, except for $1/M_P^2$. (This is in contrast to the bounds with some other Wilson coefficients fixed, which can often be stronger.) Then, the causality bounds that we can derive are given by
\bal
\label{exa:ineq1}
		&\int_{0}^{1} \text{d}p \bigg(\phi^{++--}_{2} \frac{1}{p^2}\bigg)\frac{1}{M_P^2}+
		\int_{0}^{1} \text{d}p\bigg(-\phi^{+++-}_{1} p^4+\phi^{+++-}_2 p^2-10\phi^{++++}_1 p^4 + 10\phi^{++++}_2 p^2\bigg)\frac{\gamma_0}{M_P^4}
		\nn&\hspace{50pt}+\bigg\{\int_{0}^{1} \text{d}p
		\bigg(-\phi^{++--}_{3} p^4+\phi^{++--}_{4} p^2 \bigg)-2z^{++--}_{3,2}-z^{++--}_{4,1}\bigg\}\frac{\gamma_0^2}{M_P^6}
		\nn&\hspace{50pt}+\bigg\{\int_{0}^{1} \text{d}p\bigg(-\phi^{++--}_3 +3\phi^{++++}_1 p^4-3\phi^{++++}_2 p^2\bigg)-z^{++--}_{3,0}\bigg\}\frac{\beta_1^2}{M_P^4} \geq 0 \,,
\eal
for all sets of weight functions $\phi^{\mathbb{1234}}_k(p)$ and parameters $z^{\mathbb{1234}}_{k,n}$  satisfying \eref{exa:cons1} and \eref{exa:cons5}. If a set of weight functions $\phi^{\mathbb{1234}}_k(p)$ and parameters $z^{\mathbb{1234}}_{k,n}$ satisfy condition \eref{exa:cons1} and \eref{exa:cons5}, so do the scaled set of $\lambda \phi^{\mathbb{1234}}_k(p)$ and $\lambda z^{\mathbb{1234}}_{k,n}$ with $\lambda>0$, leading to an arbitrary normalization of \eref{exa:ineq1}. Considering that we project all of our bounds onto ${1}/{M_P^2}$, one is tempted to fix the normalization of \eref{exa:ineq1} by setting
\begin{equation}
\label{phinomicon}
	\int_{0}^{1}\text{d}p \phi^{++--}_2(p)\frac{1}{p^2}=1\,.
\end{equation}
However, this is only a formal/schematic assignment, which can not be implemented numerically. As discussed in Section \ref{sec:numdet}, the integration on the left hand side of Eq.~(\ref{phinomicon}) is actually divergent, so we need to introduce an IR cutoff $m_{\text{IR}}$ to regulate it, that is, we actually choose to integrate from $m_{\text{IR}}$ to 1 against all the weight functions in the numerical implementation. As we parametrize $\phi^{++--}_2(p)$ by
\be
\phi^{++--}_2(p)=(1-p)^2\sum_{i=1}x^{++--}_{2,i}p^i
\ee
and the $x^{++--}_{2,1}$ term, {\it i.e.,} the $t$-channel contribution, must be present to yield positivity bounds, the leading term on the left hand side, which is logarithmic divergent as $m_{\text{IR}}\to 0$, comes from the $x^{++--}_{2,1}$ term. For phenomenological interesting cases, the $\log(\Lambda/m_{\text{IR}})$ term is usually $\mathcal{O}(10^2)$, which dominates the left hand side of \eref{phinomicon}.

Thus, in the numerical implementation, we can choose the normalization to be
\begin{equation}\label{exa:cons3}
	x^{++--}_{2,1}=1\,.
\end{equation}
Also, because of the large logarithmic term,  it is a good approximation to neglect the linear term ${\gamma_0}/{M_P^4}$ in the sum rules. (These two approximations are justified numerically in more details in Section \ref{subsec:xian}.) After these considerations, the inequality (\ref{exa:ineq1}) becomes
\bal
		\frac{\log{\Lambda/m_{\text{IR}}}}{M_P^2}&+\bigg\{\int_{0}^{1} \text{d}p
		\bigg(-\phi^{++--}_{3} p^4+\phi^{++--}_{4} p^2 \bigg)-2z^{++--}_{3,2}-z^{++--}_{4,1}\bigg\}\frac{\gamma_0^2}{M_P^6}
		\nn&+\bigg\{\int_{0}^{1} \text{d}p\bigg(-\phi^{++--}_3 +3\phi^{++++}_1 p^4-3\phi^{++++}_2 p^2\bigg)-z^{++--}_{3,0}\bigg\}\frac{\beta_1^2}{M_P^4} \geq 0
\label{LambdamIRphi0}
\eal
where the decision variables $\phi^{\mathbb{1234}}_k(p)$ and $z^{\mathbb{1234}}_{k,n}$ must satisfy linear conditions \eref{exa:cons1}, \eref{exa:cons5} and \eref{exa:cons3}.

To carve out the boundary of the causality bound in a 2D parameter space, we choose a fixed point within the convex bound region, use the optimization scheme to find the end points of a ray at a fixed angle from the fixed point, and scan over all angles to  get the boundary. Although only $\gamma_0^2$ and $\beta_1^2$ appear in the inequality (\ref{LambdamIRphi0}), to use the same parametrization as Section \ref{subsec:xian},  we parametrize $\gamma_0$ and $\beta_1$ as
\begin{equation}\label{exa:per}
		\frac{\gamma_0}{M_P^2}=r \cos \theta \,,~~
		\frac{\beta_1}{M_P}=r \sin \theta \,.\\
\end{equation}
Then the inequality (\ref{LambdamIRphi0}) becomes
\begin{equation}
		\log\frac{\Lambda}{m_{\text{IR}}}\geq \bigg\{\cdots \bigg\}\frac{\gamma_0^2}{M_P^4}
		+\bigg\{\cdots \bigg\}\frac{\beta_1^2}{M_P^2} = \( \bigg\{\cdots \bigg\}\cos^2\thi
		+\bigg\{\cdots \bigg\}\sin^2\thi \)  r^2 \,.
\end{equation}
Therefore, for every fixed $\theta$, we solve the following SDP problem
\bal
		&\text{maximize:}~~~
		\bigg\{
		\int_{0}^{1} \text{d}p\bigg(\phi^{++--}_{3}(p)p^4-\phi^{++--}_{4}(p)p^2 \bigg)+2 z^{++--}_{3,2}+z^{++--}_{4,1} \bigg\}\cos^2\theta \\&\hspace{40pt}+
		\bigg\{\int_{0}^{1} \text{d}p\bigg(\phi^{++--}_3(p)-3\phi^{++++}_1(p)p^4+3\phi^{++++}_2(p)p^2\bigg)+ z^{++--}_{3,0}\bigg\}\sin^2 \theta
		\,,\\&\text{subject to:}~~~ \text{Eqs.~(\ref{exa:cons1}, \ref{exa:cons5}, \ref{exa:cons3}) for all functions $\phi^{\mathbb{1234}}_k(p)$ and parameters $z^{\mathbb{1234}}_{k,n}$ ,}
\eal
to get the lowest upper bound on $r^2$ at the given $\theta$, which can be implemented by the {\tt SDPB} package. Scanning $\theta$ from $0$ to ${\pi}/{2}$ for sufficiently many angles, the optimal results on $r^2$ from different angles will depict the boundary of the bounds on $\beta_1^2$ and $\gamma_0^2$.

Before ending, we would like to comment on whether more constraints can be added to get more information, using the current example. This seems to be possible at first glance but actually unachievable. For example, we know from amplitudes $\mathcal{M}^{+-+-}$ and $\mathcal{M}^{++--}$ that the coefficients $a^{+-+-}_{n,2}=0$, $a^{++--}_{2,n}=0$ with $n\geq 3$. Thus, besides the $st$ null constraints $a^{++--}_{2,n}=a^{+-+-}_{n,2}$ with $n\geq 3$, it seems that we can use naively stronger constraints $a^{+-+-}_{n,2}=a^{++--}_{2,n}=0$. However, the information $a^{+-+-}_{n,2}=0$ is actually already contained in the dispersive sum rules and thus does not give rise to extra null constraints. This can be seen from \eref{4.3.8}:
	\begin{equation}
			\sum_{k=3}a^{+-+-}_{k,2}s^k=
			\bigg\langle  \frac{\partial_t^2}{2!} \bigg(
			\frac{s^3 d^{\ell,\mu,t}_{4,-4}c^{+-}_{\ell,\mu}c^{*-+}_{\ell,\mu}}{\mu^3(\mu-s)}+
			\frac{(-s)^3d^{\ell,\mu,t}_{4,-4}c^{+-}_{\ell,\mu}c^{*-+}_{\ell,\mu}}{(\mu+t)^3(\mu+s+t)}
			\bigg)\bigg|_{t\to 0} \bigg \rangle =\langle 0 \rangle\,,
	\end{equation}
That is, the dispersion relations automatically enforce this extra information. Another example is that, as a result of the structure of 3-leg vertices in the theory, $\beta_1^2$ appears in both the sum rules from $\mathcal{M}^{++--}$ and $\mathcal{M}^{++++}$, which leads to a constraint when equating the expressions for  $\beta_1^2$. This constraint does not come from crossing symmetry. However, again, we do not need to explicitly impose this constraint, as we have used the dispersion relations from both $\mathcal{M}^{++--}$ and $\mathcal{M}^{++++}$ in our optimization programs. Using several dispersion relations involving a Wilson coefficient will give the same result as using one of these dispersion relations and the constraints from these dispersion relations.

\bibliographystyle{JHEP}
\bibliography{refs}

\end{document}